\documentclass[toc]{PoS-PDP}

\title{$\alpha'$-corrections and heterotic black holes}

\ShortTitle{$\alpha'$-corrections and heterotic black holes}

\author{Predrag Dominis Prester\\
        Department of Physics, University of Rijeka\\
        Omladinska 14, HR-51000, Croatia\\
        E-mail: \email{pprester@phy.uniri.hr}}

\abstract{We review some recent results on $\alpha'$-exact calculations of the entropy and 
near-horizon geometry of black holes in heterotic string theory.}

\FullConference{Black Holes in General Relativity and String Theory\\
		 August 24-30 2008\\
		 Veli Lo\v{s}inj, Croatia}

\begin{document}

\section{Introduction}
\label{sec:intro}

One of the big tests and challenges for every prospective theory of quantum gravity is to explain
the nature of black hole entropy, which is in standard Einstein gravity given by Bekenstein-Hawking
formula
\begin{equation} \label{BHentf}
S_{\rm bh} = S_{\rm BH} = \frac{A_h}{4 G_N} \,,
\end{equation}
where $A_h$ denote proper area of the black hole horizon. Indeed, one of the most important
successes of string theory so far is that it is indeed able to provide such statistical explanation by 
direct counting of microstates, at least for simpler class of \emph{extremal} black holes composed 
of some number of strings, branes and other non-perturbative objects wrapped around some 
internal cycle of compactification manifold. There is now large set of extremal black holes for which 
microscopic statistical derivation of (\ref{BHentf}) was done. Restriction to extremal black holes is 
for technical reasons - direct microscopic calculations are presently tractable only in a
regime of small coupling (where black holes are not present), so one has to use either 
supersymmetric non-renormalization properties (for BPS states) or attractor mechanism (for 
non-BPS states) to compare statistical entropy with (\ref{BHentf}). As we compare objects defined 
in different regimes of coupling (weakly coupled strings/branes with strongly coupled black holes) 
we need a way to compare the results. Supersymmetry (BPS case) and/or attractor mechanism 
is providing us with this, but both concepts are intrinsically connected with extremal black holes. 
Also, microscopic countings are frequently more tractable for BPS configurations. It is believed
that by improving our knowledge of nonperturbative string theory we shall be able to extend
these results to more realistic nonextremal black holes. 

In string theory classical gravity description is given by low-energy/curvature effective actions 
which have infinite number of higher-derivative corrections, parameterized by the string length 
parameter $\alpha'$ (so called classical, or "stringy", corrections).\footnote{There are also 
quantum corrections parameterized by string coupling $g_s$, which we shall not discuss. Let us 
just mention that by changing a duality frame one sometimes exchanges classical and quantum corrections, so our discussion is not "purely classical".}
A consequence is that classical black hole entropy is not given by simple Bekenstein-Hawking 
area formula (\ref{BHentf}), but by more complicated Wald formula (discussed in section 
\ref{ssec:wald}). This is giving us an opportunity to make precision tests of correspondence 
between black holes and weakly coupled strings/branes. The most interesting examples are
the cases for which we know microscopic result \emph{exactly} in $\alpha'$. Of course, to
perform full calculation on the gravity side, a priori a knowledge of the complete low energy 
effective action is required, and as is well-known only a limited knowledge beyond 6-derivative 
($\alpha'^2$) order is available at the moment.

In this review we shall show how one can go, on the gravity side, beyond perturbative calculation
of black hole entropy and obtain $\alpha'$-exact results for the extremal black hole entropies and
near-horizon geometries. We concentrate on simple cases of extremal black holes in heterotic
string theory with four charges in $D=4$ dimensions, and those with three charges in $D=5$
dimensions. The main part of the review is devoted to the calculation taking into account 
\emph{the full heterotic effective action}, originally developed in \cite{Prester:2008iu}. We also 
review the calculations based on $R^2$-truncated effective actions (supersymmetric and 
Gauss-Bonnet), and, in the case of small black holes in general $D$, Lovelock-type action. In 
addition, some results are presented which were not included in original papers.

\section{Black holes - solutions of effective actions}
\label{sec:bhsol}

\subsection{Low energy effective action in heterotic string theory - lowest order}
\label{ssec:loleea}

Low energy effective action (LEEA) of heterotic string theory in ten dimensions is a $\mathcal{N}=1$
supergravity theory (having 16 real supersymmetry generators). We shall be interested in purely 
bosonic solutions of this LEEA, so we can restrict ouselves to bosonic sector which contains the 
following fields: dilaton $\Phi^{(10)}$, metric tensor $G_{MN}^{(10)}$, 2-form gauge field 
$B_{MN}^{(10)}$ and ($SO(32)$ or $E_8 \times E_8$) Yang-Mills gauge field $A_M^{(10)}$. For 
simplicity, we shall additionaly restrict ourselves to backgrounds for which Yang-Mills field 
$A_M^{(10)}$ vanishes (later we shall explain how one can use T-duality to reconstruct solutions
with non-vanishing Yang-Mills field). In this case, LEEA in the \emph{lowest order} in string length 
parameter $\alpha'$ and string coupling constant $g_s$, is given by
\begin{equation} \label{10dl0}
\mathcal{A}^{(10)}_{0} = \frac{1}{16\pi G_N^{(10)}} \int d^{10}x \, e^{-2\Phi^{(10)}} \left[
 R^{(10)} + 4 (\partial \Phi^{(10)})^2 - \frac{1}{12} H^{(10)}_{MNP} H^{(10)MNP} \right] \;,
\end{equation}
where $M,N,\ldots=0,1,\ldots,9$, $G_N^{(10)}$ is 10-dimensional Newton constant and 
$H^{(10)}_{MNP}$ is a 3-form gauge field strength corresponding to 2-form $B_{MN}^{(10)}$
\begin{equation} \label{hb0}
H^{(10)}_{MNP} = \partial_M B^{(10)}_{NP} + \partial_N B^{(10)}_{PM} + \partial_P B^{(10)}_{MN}
 \;.
\end{equation}
As the fundamental objects of heterotic string theory in ten dimensions are 1-dimensional 
(elementary string or F1-brane) and 5-dimensional (NS5-brane), interesting effectively 
0-dimensional objects (candidates for black holes) can be obtained through compactification, 
followed by wrapping of $d$-branes around $d$-cycles of compactification manifold. The simplest 
choice is to take the space to have the topology of  $M_D \times T^{10-D}$, where $M_D$ is 
$D$-dimensional Minkowski space and $T^{k}$ is $k$-dimensional torus.

\subsection{Small black holes}
\label{ssec:smallbh}

For a start, let us take $D=9$ where compactification manifold is circle $S^1$, parametrized 
with $0 < x_9 < 2\pi\sqrt{\alpha'}$. Following the rules of Kaluza-Klein compactification, we 
obtain that the massless fields in 9-dimensions are dilaton $\Phi$, metric $G_{\mu\nu}$, 2-form 
$B_{\mu\nu}$, modulus (radius of $S^1$) $T$, and two $U(1)$ gauge fields $A_{\mu}^{i}$, $i=1,2$, defined by 
\begin{eqnarray} \label{kk10d9}
&& \Phi = \Phi^{(10)} - \frac{1}{2} \, \ln (G^{(10)}_{99})\, , \qquad
S = e^{-\Phi}\, , \qquad  T = \sqrt{G^{(10)}_{99}}\, , 
\nonumber \\
&& G_{\mu\nu} = G^{(10)}_{\mu\nu}
  - (G^{(10)}_{99})^{-1} \, G^{(10)}_{9\mu} \, G^{(10)}_{9\nu}\, ,
\nonumber \\
&& A^{(1)}_\mu = {1\over 2} (G^{(10)}_{99})^{-1} \, G^{(10)}_{9\mu}\, ,
\qquad
A^{(2)}_\mu = {1\over 2} B^{(10)}_{9\mu}\, , 
\nonumber \\
&& B_{\mu\nu} = B^{(10)}_{\mu\nu} - 2(A^{(1)}_\mu A^{(2)}_\nu - A^{(1)}_\nu A^{(2)}_\mu)
\end{eqnarray}
where $\mu,\nu,\ldots=0,1,\ldots,8$. Using (\ref{kk10d9}) in (\ref{10dl0}) one obtains the 
effective 9-dimensional action
\begin{eqnarray} \label{ea9d}
\mathcal{A}_0 &=& {1\over 16\pi G_N} \int d^9 x \, \sqrt{-G} \, S \, 
\bigg[ R + S^{-2}\, (\partial_\mu S)^2  -  T^{-2} \, (\partial_\mu T)^2
\nonumber \\
&& - {1\over 12}  (H_{\mu\nu\rho})^2
 - T^2 \, (F^{(1)}_{\mu\nu})^2 - T^{-2} \,  (F^{(2)}_{\mu\nu})^2 \bigg] \, ,
\end{eqnarray}
where $R$ is Ricci scalar computed from 9-dimensional metric $G_{\mu\nu}$, 
$G_N = G_N^{(10)}/(2\pi\sqrt{\alpha'})$ is the effective 9-dimensional Newton constant, and 
$F^{(a)}_{\mu\nu}$ and $H_{\mu\nu\rho}$ are 2-form and 3-form gauge field strengths defined by
\begin{eqnarray} \label{gfs9d}
&& F^{(a)}_{\mu\nu} = \partial_\mu A^{(a)}_\nu - \partial_\nu A^{(a)}_\mu\, , \quad
a=1,2\, , \nonumber \\
&& H_{\mu\nu\rho} = \left[ \partial_\mu B_{\nu\rho}
 + 2 \left( A_\mu^{(1)} F^{(2)}_{\nu\rho} - A_\mu^{(2)} F^{(1)}_{\nu\rho}\right) \right]
 + \hbox{cyclic permutations of $\mu$, $\nu$, $\rho$}\, .
\end{eqnarray}

It was shown in \cite{Peet:1995pe} that action (\ref{ea9d}) has the following asymptoticaly flat 
solutions
\begin{eqnarray} \label{sol9d}
ds^2 &=& G_{\mu\nu} dx^\mu dx^\nu
 = - g_s^{2\gamma}\,(F(\rho))^{-1} \rho^{2\beta} dt^2 + g_s^{2\gamma} \, d\vec{x}^2 \, , 
\nonumber \\
S &=& g_s^{-2}\,(F(\rho))^{1/2} \, \rho^{-\beta} \, , \qquad
T = R \, \sqrt{ \frac{\rho^\beta+2|Q_N|}{\rho^\beta + 2|Q_W|}} \, , 
\nonumber \\
A^{(1)}_{t}  &=& - {g_s^\gamma \over R} \, {Q_N \over (\rho^\beta+2|Q_N|)} \, , \qquad
A^{(2)}_{t} = - R\, g_s^\gamma\,{Q_W\over (\rho^\beta + 2|Q_W|)} \, , \qquad
B_{\mu\nu} = 0 = H_{\mu\nu\rho} \, ,
\end{eqnarray}
where we introduced
\begin{eqnarray} \label{conv9d}
F(\rho) &\equiv& (\rho^\beta + 2|Q_W|) (\rho^\beta+2|Q_N|) \, , \nonumber \\
\rho^2 &\equiv& \vec x^2\, , \qquad \beta \equiv D-3 = 6\,, \qquad 
\gamma \equiv \frac{2}{D-2} = \frac{2}{7}\,.
\end{eqnarray}
We shall keep using $\beta$ and $\gamma$ (without fixing $D=9$) because this solution 
generalizes to arbitrary $D$ (corresponding to compactification on $T^{9-D} \times S^1$).

To fully understand the geometry of solution (\ref{sol9d}) and the physical meaning of
parameters $g_s$, $R$, $Q_N$, and $Q_W$, it is better to pass to a canonical metric tensor 
$G_{E\,\mu\nu}$ (known as Einstein-frame metric)  in which the action (\ref{ea9d}) has 
a more conventional form
\begin{displaymath}
\mathcal{A}_0 = {1\over 16\pi G_N} \int d^9 x \, \sqrt{-G_E} \, ( R_E + \ldots ) \,,
\end{displaymath}
where $R_E$ is Ricci scalar obtained from metric $G_{E\,\mu\nu}$. It is easy to show that
relation between "string-frame" and "Einstein-frame" metrics is given by
\begin{equation} \label{Eframe}
G_{E\,\mu\nu} = S^\gamma \, G_{\mu\nu} \,.
\end{equation}
In the limit $\rho \to \infty$ one gets 
\begin{equation} \label{asym9d}
G_{E\,\mu\nu} \to \eta_{\mu\nu} \,, \qquad S \to g_s^{-2} \,, \qquad T \to R \,. 
\end{equation}
The first relation shows that metric is asymptotically flat, so it describes a point-like object. The 
second relation shows that parameter $g_s$ is effective 9-dimensional string coupling. The last 
relation in (\ref{asym9d})  shows that $R$ is the radius of compactification circle $S^1$ measured 
in string-frame metric, in units of string length parameter $\sqrt{\alpha'}$.

What about $Q_N$ and $Q_W$? First of all, let us show that they are proportional to 
\emph{electric} charges connected to gauge fields $A^{(1)}_\mu$ and $A^{(2)}_\mu$. This follows 
from the asymptotic behavior of corresponding field strengths for $\rho\to\infty$
\begin{eqnarray}
&& F^{(1)}_{\rho t}
 = Q_N \frac{4\,g_s^\gamma}{R} \frac{\beta \rho^{\beta-1}}{(\rho^\beta + 2|Q_N|)^2}
 \to Q_N \frac{4\,g_s^\gamma}{R} \, \frac{D-3}{\rho^{D-2}} \,,
\nonumber \\ 
&& F^{(2)}_{\rho t}
 =  Q_W \frac{R\,g_s^\gamma}{4} \frac{\beta \rho^{\beta-1}}{(\rho^\beta + 2|Q_W|)^2}
 \to Q_W \frac{R\,g_s^\gamma}{4} \, \frac{D-3}{\rho^{D-2}}\, , 
\label{gfasym}
\end{eqnarray}
(where  $D=9$). Combined with the asymptotic flatness of the metric (\ref{asym9d}), this shows 
that electric charges are proportional to $Q_N$ and $Q_W$. 

To understand stringy meaning of charges $Q_N$ and $Q_W$, let us find the mass of our 
configuration. It can be determined in the standard way from asymptotic behavior of canonical 
metric
\begin{equation} \label{egtt}
G_{E\,tt} \simeq - 1 + {16\pi G_N \over (D-2) \Omega_{D-2}}{M\over \rho^{D-3}} \, ,
\end{equation}
where $\Omega_{D-2}$ is the volume of the unit $(D-2)$ sphere
\begin{displaymath}
\Omega_{D-2} = \frac{2 \, \pi^{(D-1)/2}}{\Gamma((D-1)/2)}
\end{displaymath}
Using (\ref{Eframe}) and (\ref{sol9d}) in (\ref{egtt}), and puting $D=9$, we get
\begin{equation}  \label{emass}
M = {(D-3) \, \, \Omega_{D-2} (|Q_N|+|Q_W|) \over 8\pi G_N}
 = \frac{\pi^3}{4 G_N} (|Q_N|+|Q_W|).
\end{equation}
If we define rescaled charges $n$ and $w$ in the following way
\begin{equation} \label{enwcomp}
n = \frac{\pi^3}{4 G_N} g_s^{-\gamma} \sqrt{\alpha'}\, R\, Q_N \,, \qquad
w = \frac{\pi^3}{4 G_N} g_s^{-\gamma} \sqrt{\alpha'}\, \frac{Q_W}{R} \,,
\end{equation}
then the expression for mass (\ref{emass}) becomes
\begin{equation} \label{emass2}
M = \frac{g_s^\gamma}{\sqrt{\alpha'}} \left( {|n|\over R} + |w| \, R \right) \, .
\end{equation}
As we explain later in section \ref{sec:stringy}, if we restrict $n$ and $w$ to integer values 
(\ref{emass2}) is identical to the mass spectrum of a particular subspace of states of elementary 
string wound around the circle $S^1$, with $n$ and $w$ identified with momentum number and 
winding number, respectively.

Let us concentrate now on the behavior of the solution (\ref{sol9d}) in the region $\rho \to 0$.
For this purpose, let us take the limit $\rho\alpha \ll |Q_N|,|Q_W|$. If we introduce rescaled 
coordinates
\begin{equation} \label{e9}
\vec y = g^\gamma
\, \vec x, \qquad r = \sqrt{\vec y^2} = g^\gamma \, \rho\, , \qquad
\tau = g^{-\beta\gamma} \, {(D-3)\Omega_{D-2}\over 4\pi}
t / \sqrt{nw}\, ,
\end{equation}
the solution (\ref{sol9d}) in this limit becomes
\begin{eqnarray} \label{e10}
&& ds^2 = -{r^{12} \over 4} \, d\tau^2 + \, d\vec{y}^2 \, ,
\nonumber \\
&& S = {4 \over \pi^3}\, {\sqrt{|nw|}\over r^6}
\, , \qquad
T = \sqrt{\left| n\over w \right|} \, , \nonumber \\
&& F^{(1)}_{r\tau} = {3\over 2n} \, r^5 \, \sqrt{|nw|}\, , \qquad
F^{(2)}_{r \tau} = {3\over 2w} \, r^5 \, \sqrt{|nw|} \, ,
\end{eqnarray}
Comments on 2-charge solution (\ref{sol9d}):
\begin{enumerate}
\item
From (\ref{e10}) we infer that solution (\ref{sol9d}) describes \emph{black hole} with singular
horizon coinciding with null singularity at $\rho=0$.
\item 
Solution (\ref{sol9d}) is \emph{extremal}: it has vanishing Hawking temperature, and it
can be obtained in a particular limit from multiparameter regular black hole solutions (with the
same charge content) having regular horizons with non-vanishing temperature and entropy, and 
with mass satisfying
\begin{equation} \label{mbps}
M \ge \frac{g_s^\gamma}{\sqrt{\alpha'}} \left( {|n|\over R} + |w| \, R \right) \,,
\end{equation}
where inequality is saturated for solutions (\ref{sol9d}).
\item 
After including fermionic degrees of freedom (of heterotic string theory), it can be shown 
that solution (\ref{sol9d}) for $nw\ge0$ is 1/2-BPS, i.e., it is annihilated by half of supersymmetry 
generators ($16/2=8$). In fact, (\ref{mbps}) already looks like BPS condition, but additional
labor is needed to establish that only when $n$ and $w$ have the same sign solution is 
supersymmetric. This is a consequence of the fact that $N=1$ SUSY in the heterotic string theory
is purely right-handed.
\item
Solution (\ref{sol9d}) was generalized to compactifications on general $k$-dimensional tori $T^k$,
including non-vanishing 2-form $B_{\mu\nu}$ and Yang-Mills field. When all gauge fields are
purely electrically charged, one obtains $(10-k)$-dimensional extremal black holes with the
same properties as (\ref{sol9d}) (singular horizon, 1/2-BPS, etc.)
\item
From (\ref{e10}) we see that near the horizon solution is completely determined by charges $n$ 
and $w$ and is independent of asymptotic values of moduli $g_s$ and $R$. This is an example of 
the \emph{attractor mechanism}. 
\end{enumerate}

We shall be interested primarily in the entropy of black holes. When the gravitational part of an 
action has the simple Einstein form, as is the case for action (\ref{10dl0}) (after passing to 
Einstein-frame metric), then the black hole entropy is given by Bekenstein-Hawking formula
\begin{equation} \label{BHent}
S_{\rm bh} = \frac{A_h}{4 G_N} \,,
\end{equation}
where $A_h$ is the proper area of the black hole horizon measured by Einstein-frame metric.
For the black hole solution (\ref{sol9d}), by using  (\ref{Eframe}) we obviously get that  $A_h$  
\emph{vanishes}, so
\begin{equation} \label{BHsmall}
S_{\rm bh} = 0 \,,
\end{equation}
For the reasons we explain later,  such solutions are called \emph{small} black holes. As it has
two charges, we shall refer to solution (\ref{sol9d}) as \emph{2-charge small black hole}.

\subsection{Large black holes in $D=4$ and $D=5$}
\label{ssec:largebh}

\subsubsection{Compactifications on tori}
\label{sssec:tori}

As we mentioned at the end of section \ref{ssec:loleea}, if we compactify heterotic string theory
on torus $T^k$, where $k\ge5$, we can obtain more complicated point-like objects, e.g., by 
wrapping NS5-branes on such torii, possibly in addition to elementary strings. Now we shall show
that one can indeed find black hole solutions of corresponding effective actions with the charge 
content resembling to states in string theory obtained by wrapping strings and branes.

Let us analyze briefly massless field content in bosonic sector after such Kaluza-Klein 
compactification to $D=10-k$ dimensions. First of all, non-abelian gauge group ($E_8 \times E_8$ 
or $SO(32)$) breaks down to abelian subgroup $U(1)^{16}$. This gives 16 U(1) gauge fields and 
$16 \times k$ scalar fields. From $10$-dimensional metric tensor one obtains $D$-dimensional 
metric tensor, $k$ U(1) gauge fields and $k(k+1)/2$ scalars. From 2-form $B_{MN}$ one obtains 
2-form $B_{\mu\nu}$, $k$ U(1) gauge fields and $k(k-1)/2$ scalars. Altogether, we have dilaton 
$\Phi$, metric $G_{\mu\nu}$, 2-form $B_{\mu\nu}$, $16+2k = 36-2D$ KK U(1) gauge fields, and 
$k(k+16) = (10-D)(26-D)$ scalar fields ($\mu,\nu = 0,\ldots,D-1$). This looks extremely 
complicated, but help comes from the following two observations (details with references can be 
found in \cite{Youm:1997hw})
\begin{enumerate}
\item
Compactification on torus leaves all supersymmetry generators (16 in heterotic theory) unbroken.
If one is searching for BPS solutions (which are preserving some supersymmetry), one can use 
BPS conditions (which are first-order equations) to greatly simplify calculations. 
\item
The tree-level effective action has $O(10-D,26-D)$ symmetry (loop-corrections break this symmetry
to $O(10-D,26-D,\mathbf{Z})$, which is a T-duality group of heterotic theory). This symmetry can
be used to reduce the complexity of the problem. For example, one can use it to put majority
of charges to zero. After solving for such generating solutions, one simply obtains all other solutions
(with arbitrary charge assignments) by applying symmetry transformations on generating solutions. 
For example, in $D=5$ ($k=5$) case, generating solution has only 3 non-vanishing charges, and in 
$D=4$ ($k=6$) case, the number of relevant charges is 5. 
\end{enumerate}

\subsubsection{3-charge large black holes in $D=5$}
\label{sssec:3cbh}

We focus now on the case $k=5$ ($D=5$), and consider a compactification on $T^4 \times S^1$ 
(obviously a special case of $T^5$) in which $T^4$ is completely factorized and flat. In the 
language of the previous paragraph, it means that all KK gauge fields (and corresponding 
charges) which are obtained from $T^4$-indices are taken to be zero. We again take $S^1$ to 
be parametrized by $0 < x_9 < 2\pi\sqrt{\alpha'}$. Non-vanishing dynamical massless fields in 
5-dimensions are then dilaton $\Phi$, metric $G_{\mu\nu}$, 2-form $B_{\mu\nu}$, modulus (radius 
of $S^1$ in $\alpha'$-units) $T$, and two $U(1)$ gauge fields $A_{\mu}^{i}$, $i=1,2$, defined by 
\begin{eqnarray} \label{kk10d5}
&& \Phi = \Phi^{(10)} - \frac{1}{2} \, \ln (G^{(10)}_{99})\, , \qquad
S = e^{-\Phi}\, , \qquad  T = \sqrt{G^{(10)}_{99}}\, , 
\nonumber \\
&& G_{\mu\nu} = G^{(10)}_{\mu\nu}
  - (G^{(10)}_{99})^{-1} \, G^{(10)}_{9\mu} \, G^{(10)}_{9\nu}\, ,
\nonumber \\
&& A^{(1)}_\mu = {1\over 2} (G^{(10)}_{99})^{-1} \, G^{(10)}_{9\mu}\, ,
\qquad
A^{(2)}_\mu = {1\over 2} B^{(10)}_{9\mu}\, , 
\nonumber \\
&& B_{\mu\nu} = B^{(10)}_{\mu\nu} - 2(A^{(1)}_\mu A^{(2)}_\nu - A^{(1)}_\nu A^{(2)}_\mu)
\end{eqnarray}
where now $\mu,\nu,\ldots=0,1,\ldots,4$. Using (\ref{kk10d5}) in (\ref{10dl0}) one obtains the 
effective 5-dimensional action
\begin{eqnarray} \label{ea5d}
\mathcal{A}_0 &=& {1\over 16\pi G_N} \int d^5 x \, \sqrt{-G} \, S \, 
\bigg[ R + S^{-2}\, (\partial_\mu S)^2  -  T^{-2} \, (\partial_\mu T)^2
\nonumber \\
&& - {1\over 12}  (H_{\mu\nu\rho})^2
 - T^2 \, (F^{(1)}_{\mu\nu})^2 - T^{-2} \,  (F^{(2)}_{\mu\nu})^2 \bigg] \, ,
\end{eqnarray}
where $R$ is Ricci scalar computed from 5-dimensional metric $G_{\mu\nu}$, and
$G_N = G_N^{(10)}/(2\pi\sqrt{\alpha'}\mathcal{V})$ is the effective 5-dimensional Newton constant
($\mathcal{V}$ is volume of $T^4$).

$F^{(a)}_{\mu\nu}$ and $H_{\mu\nu\rho}$ are 2-form and 3-form gauge field strengths defined by
\begin{eqnarray} \label{gfs5d}
&& F^{(a)}_{\mu\nu} = \partial_\mu A^{(a)}_\nu - \partial_\nu A^{(a)}_\mu\, , \quad a=1,2\, , 
\nonumber \\
&& H_{\mu\nu\rho} = \left[ \partial_\mu B_{\nu\rho}
 + 2 \left( A_\mu^{(1)} F^{(2)}_{\nu\rho} - A_\mu^{(2)} F^{(1)}_{\nu\rho}\right) \right]
 + \hbox{cyclic permutations of $\mu$, $\nu$, $\rho$}\, .
\end{eqnarray}
As the torus $T^4$ trivially factorizes, and enters only in reparametrization of the Newton constant,
the obtained expressions are the same as those from the begging of section \ref{ssec:smallbh},
the only difference is that now theory is effectively 5-dimensional. We note that truncated action 
(\ref{ea5d}) has reduced supersymmetry (from $\mathcal{N}=4$ to $\mathcal{N}=2$), and has 
T-duality group generated by only one element, given by $T \to 1/T$, 
$A^{(1)}_\mu \leftrightarrow A^{(2)}_\mu$.

Extremal black hole solutions of the action (\ref{ea5d}) were reviewed in \cite{Youm:1997hw}. We 
are primarily interested here in the black hole entropy, which is given by Bekenstein-Hawking
formula (\ref{BHent}). From (\ref{BHent}) it is obvious that one just needs near-horizon behavior
of the solution, so we shall from now on concentrate on it (and avoid writing solutions in the whole
space, which are usually cumbersome). Near-horizon behavior of the mentioned black hole 
solution is given by
\begin{eqnarray} \label{3c5d0}
&& ds^2 \equiv G_{\mu\nu} dx^\mu dx^\nu
 = \frac{\alpha' |m|}{4} \left(-r^2 dt^2 + {dr^2\over r^2}\right) + \alpha' |m|\,d\Omega_3^2 \,, 
\nonumber \\
&& S  = \frac{4 G_N}{\alpha'^{3/2} \pi}\, \frac{\sqrt{|nw|}}{|m|} \,, \qquad 
T = \sqrt{\left| \frac{n}{w} \right|} \,
\nonumber \\
&& F^{(1)}_{rt} = \frac{1}{n} \sqrt{\alpha' |nwm|} \,, \qquad
 F^{(2)}_{rt}  = \frac{1}{w} \sqrt{\alpha' |nwm|} \,, \qquad
 H_{234} = 2\alpha' m \sqrt{g_3} \,,
\end{eqnarray}
and all other components of the fields are zero. Here $d\Omega_3^2$ denotes the metric on the 
unit 3-sphere $S^3$ (with coordinates $x^i$, $i=2,3,4$), and $g_3$ is a determinant of its metric 
tensor.

Solution (\ref{3c5d0}) contains three parameters, $n$, $w$, and $m$, which play the roles of 
charges. First of all, electric charges $n$ and $w$ are appearing in the same way as in the small 
black hole solution (\ref{sol9d}), and so they have the same interpretation in stringy theory, i.e., 
they are momentum and winding number of the elementary string wound around $S^1$ 
circle.\footnote{Indeed, for $m=0$ the solution with near-horizon behavior (\ref{3c5d0}) becomes 
a 5-dimensional generalization of small black hole solution (\ref{sol9d}).} As for $m$, let us first 
show that it is the \emph{magnetic} charge corresponding to 2-form gauge field $B_{\mu\nu}$. 

Generally, an object will be magnetically charged under a $(p-1)$-form gauge field (with 
$p$-form field strength $A_p$) according to formula (up to a constant factor, depending on 
conventions)
\begin{equation} \label{magch}
Q_m = \frac{1}{\Omega_p} \int_{S^p} A_p 
\end{equation}
where integration is over any $p$-sphere which encircles the object, and $\Omega_p$ is the 
volume of the unit $p$-sphere. From (\ref{magch}) directly follows that in $D$ dimensions only 
$d = D-p-2$ dimensional objects can be magnetically charged under $(p-1)$-form gauge field. So, 
in $D=5$ only point-like objects ($d=0$) can be magnetically charged under 2-form
gauge field $B_{\mu\nu}$. Indeed, for solution (\ref{3c5d0}) application of (\ref{magch})
gives
\begin{equation} \label{3cmagch}
Q_m = \frac{1}{\Omega_3} \int_{S^3} H = 2\alpha' N 
\end{equation}
As already mentioned, we expect that the 3-charge black hole solution with near-horizon behavior 
(\ref{3c5d0}) describes configuration in which, beside the elementary string, we also have
NS5-branes wound around $T^4 \times S^1$. We shall show that $m$ is properly normalized to
represent the number of NS5-branes (which we denote $Q_5$).\footnote{This can be also 
understood from the viewpoint in which NS5-brane is dual to elementary string, and duality 
exchanges electric and magnetic charges (that is why elementary string is electrically charged on 
$B_{\mu\nu}$, and NS5-brane magnetically).}

Let us make few comments on the near-horizon solution (\ref{3c5d0}):
\begin{enumerate}
\item
Geometry of solution is AdS$_2 \times S^3$. The isometry group of such geometry is 
$SO(2,1) \times SO(4)$, and this symmetry group is respected by the complete solution, not just
metric. This is what is expected for near-horizon geometry of static spherically symmetric 
\emph{extreme} black hole in $D=5$ (simple example is 5-dimensional Reisner-Nordstrom solution). 
\item
When all charges are finite and non-vanishing, black hole horizon, given by $r=0$, is 
\emph{regular}, with all curvature invariants being finite and well defined on it. Such black 
holes are called \emph{large} extremal black holes.
\item
It is completely determined by charges $n$, $w$ and $m$, and is independent of asymptotic values 
of moduli. This is another example of the attractor mechanism. 
\item
It is by itself exact solution of equations of motion.
\item
For $nw>0$ solution is supersymmetric, but this time is only 1/4-BPS (it breaks 3/4 of 
supersymmetry generators) \cite{Ferrara:1997ci}.
\item
There is a \emph{supersymmetry enhancement} near the horizon. Though solution in full space 
breaks 1/2 of $\mathcal{N}=2$ SUSY present in truncated action (\ref{ea5d}), the near-horizon 
solution is completely supersymmetric. 
\item
As mentioned above, for $m=0$ our 3-charge black hole reduces to 2-charge small black hole of
section \ref {ssec:smallbh}. Taking $m\to0$ in the near-horizon solution  (\ref{3c5d0}) is not
well defined, which is expected because near-horizon geometry of 2-charge small black holes,
given in (\ref{e10}), is not AdS$_2 \times S^3$ but geometry with singular horizon.
\end{enumerate}

From (\ref{3c5d0}) (and using (\ref{Eframe})) we can easily calculate Bekenstein-Hawking black 
hole entropy
\begin{equation} \label{largeBHent}
S_{\rm bh} = \frac{A_h}{4 G_N} = 2\pi \sqrt{|nwm|} \,.
\end{equation} 
As expected for large black holes, the entropy is finite.

\subsubsection{4-charge large black holes in $D=4$}
\label{sssec:4cbh}

To obtain the last example of black holes that we study in this review, one needs to compactify
one more dimension on the circle, which we denote $\widehat{S}^1$. This means that the 
compactification manifold is now $T^4 \times \widehat{S}^1 \times S^1$, and that after
KK procedure the theory is effectively $D=4$ dimensional. As explained in section 
\ref{sssec:tori}, KK procedure leads to effective action with large number of fields, but
which are organized according to $O(6,22)$ symmetry group of T-duality. 

Again, we want to simplify things as much as possible, which means taking as many of the 
degrees of freedom to be zero or trivially constant. As before we take torus $T^4$ to be 
"trivial", and furthermore that 2-form $B_{\mu\nu} \equiv B^{(10)}_{\mu\nu}$ (where now
$\mu,\nu=0,1,2,3$) and KK scalars $G^{(10)}_{89}$ and $B^{(10)}_{89}$ all \emph{vanish}. 
We again take $S^1$ to be parametrized by $0 < x_9 < 2\pi\sqrt{\alpha'}$, and 
$\widehat{S}^1$ by $0 < x_8 < 2\pi\sqrt{\alpha'}$. This leaves us with the following 
massless fields 
\begin{eqnarray} \label{kk10d4}
 &&  {\Phi = \Phi^{(10)} - {1\over 4} \, \ln (G^{(10)}_{99})}
 - {1\over 4} \, \ln (G^{(10)}_{88})
 \, ,\nonumber \\
&&  {S=e^{-2\Phi}}\,, \qquad
  {T = \sqrt{G^{(10)}_{99}}}\, , \qquad
 \widehat{T} = \sqrt{G^{(10)}_{88}}\, , \nonumber \\ 
&& {G_{\mu\nu} = G^{(10)}_{\mu\nu} - (G^{(10)}_{99})^{-1} \,
G^{(10)}_{9\mu}
\, G^{(10)}_{9\nu} - (G^{(10)}_{88})^{-1} \,
G^{(10)}_{8\mu}
\, G^{(10)}_{8\nu}\, , }\nonumber \\ 
&& A^{(1)}_\mu = {1\over 2} (G^{(10)}_{99})^{-1} \, G^{(10)}_{9\mu}\, ,
\qquad   {A^{(2)}_\mu = {1\over 2} (G^{(10)}_{88})^{-1} \, G^{(10)}_{8\mu}\, ,}
\nonumber \\ 
&&  {A^{(3)}_\mu = {1\over 2} B^{(10)}_{9\mu}\, , }
\qquad {A^{(4)}_\mu = {1\over 2} B^{(10)}_{8\mu}}\, ,
\end{eqnarray}
Putting (\ref{kk10d4}) into (\ref{10dl0}) we get the following $D=4$ effective action 
\begin{eqnarray} \label{ea4d}
\mathcal{A}_0 &=& {1\over 16\pi G_N} \int d^4 x \, \sqrt{-G} \, S \, 
 \bigg[ R + S^{-2}\, (\partial_\mu S)^2 - T^{-2} \, (\partial_\mu T)^2
 - \widehat{T}^{-2} \, (\partial_\mu \widehat{T})^2
\nonumber \\
&& - T^2 \, (F^{(1)}_{\mu\nu})^2 - \widehat{T}^2 \, (F^{(2)}_{\mu\nu})^2
 - T^{-2} \, (F^{(3)}_{\mu\nu})^2 - \widehat{T}^{-2} \, (F^{(4)}_{\mu\nu})^2 \bigg] \,,
\end{eqnarray}
where $R$ is Ricci scalar computed from 4-dimensional metric $G_{\mu\nu}$, 
$G_N = G_N^{(10)}/(4\pi^2\alpha'\mathcal{V})$ is the effective 4-dimensional Newton constant
($\mathcal{V}$ is the volume of $T^4$), and $F^{(i)}_{\mu\nu}$ are 2-form strengths of gauge 
fields $A^{(i)}_{\mu}$. Again, truncation of the theory leads to the reduction of the 
supersymmetry (now $\mathcal{N}=2$ instead of $\mathcal{N}=4$) and T-duality (now reduced to 
two generators, $T\to1/T$ and $\widehat{T}\to1/\widehat{T}$).

In $D=4$ dimensions 1-form gauge field is Hodge self-dual, which has a consequence that
point-like configurations (like black holes) can be electrically and magnetically charged
on the same 1-form gauge field. As in the truncated action (\ref{ea4d}) there are four 1-form 
gauge fields, we have altogether 8 charges, 4 electric denoted as
$\{Q^{(i)}\} = \{n,\widehat{n},w,\widehat{w}\}$, and 4 magnetic denoted as
$\{P^{(i)}\} = \{N,\widehat{N},W,\widehat{W}\}$. Correspondingly, we have 8-charge 
black hole solutions. Again, we take advantage of T-dualities and simplify things as possible - it can
be shown that the simplest generating solution has just 4 non-vanishing charges, with one choice 
being $\widehat{n} = \widehat{w} = N = W = 0$.

Extremal black hole solutions of the action (\ref{ea4d}) with such charge content were 
constructed in \cite{Cvetic:1995uj}. Again, we are interested in the near-horizon behavior, so we
present here just $r\to0$ limit of the solution 
\begin{eqnarray} \label{4c4d0}
&& ds^2 \equiv G_{\mu\nu} dx^\mu dx^\nu
 = \frac{\alpha'}{4} |\widehat{N}\widehat{W}| \left(-r^2 dt^2 + {dr^2\over r^2}\right)
  + \frac{\alpha'}{4} |\widehat{N}\widehat{W}|\,\left( d\theta^2 + \sin^2\theta\, d\phi^2 \right) \,, 
\nonumber \\
&& S  = \frac{8 G_N}{\alpha'} \sqrt{\left| \frac{nw}{\widehat{N}\widehat{W}} \right|} \,,
 \qquad  T = \sqrt{\left| \frac{n}{w} \right|} \,, \qquad 
 \widehat{T} = \sqrt{\bigg| \frac{\widehat{W}}{\widehat{N}} \bigg|} \,,
 \\
&& F^{(1)}_{rt} = \frac{\sqrt{\alpha'}}{4\,n} \sqrt{\big| nw\widehat{N}\widehat{W} \big|} ,\quad 
 F^{(3)}_{rt}  = \frac{\sqrt{\alpha'}}{4\,w} \sqrt{\big| nw\widehat{N}\widehat{W} \big|} , \quad
 F^{(2)}_{\theta\phi}  = \frac{\sqrt{\alpha'}}{4}\,\widehat{N} \sin\theta , \quad
 F^{(4)}_{\theta\phi}  = \frac{\sqrt{\alpha'}}{4}\,\widehat{W} \sin\theta
\nonumber
\end{eqnarray}
and all other components of the fields are zero.

What is the meaning of the numbers $n$, $w$, $\widehat{N}$ and $\widehat{W}$? Obviously,
$n$, $w$ are electric charges, and $\widehat{N}$, $\widehat{W}$ are magnetic charges (for this 
just apply (\ref{magch}). What about their stringy interpretation? For $n$ and $w$ everything is
the same as in 2-charge and 3-charge solutions we studied before - they are momentum and
winding numbers of elementary string wound around circle $S^1$. Now, from (\ref{kk10d4}) 
follows that $\widehat{N}$ and $\widehat{W}$ are connected to the other circle $\widehat{S}^1$,
and they are charges of the so called Kaluza-Klein monopole and $H$-monopole, respectively.
If we follow precise definition for $A^{(4)}_{\mu}$ in (\ref{kk10d4}), we see that $\widehat{N}$ 
enters in similar way as the magnetic charge $m$ from section \ref{sssec:3cbh}, which is 
suggesting that $\widehat{W}$ should be directly connected to the number of NS5-branes 
wrapped around $T^4 \times S^1$. However, they are not equal - there is a shift between
NS5-brane charge $\widehat{W}$ and number of NS5-branes $Q_5$, because Kaluza-Klein 
monopole also carries $(-1)$-unit of NS5-brane charge.\footnote{Microscopic explanation of this 
was given in \cite{Bershadsky:1995qy} and macroscopic, in the framework of $R^2$-type SUSY 
effective action, in \cite{Castro:2007ci}. We shall see later how this "charge shift" explicitly 
appears in our macroscopic analysis.}

All in all, it appears that solution (\ref{4c4d0}) describes near horizon geometry, in leading order in
$\alpha'$ and $g_s$, of a configuration in heterotic string theory consisting of an elementary string 
wound $w$ times around $S^1$, $\widehat{N}$ KK-monopoles wrapped around $T^4 \times S^1$ 
(and with "the core" on circle $\widehat{S}^1$), $\widehat{W}$ NS5-branes wrapped around 
$T^4 \times S^1$, and on top of it there is a momentum on $S^1$ with momentum number $n$. In appendix A of \cite{Sen:2005iz} one can find a proof that $n$, $w$, $\widehat{N}$ and 
$\widehat{W}$, as we defined them, are indeed properly normalized, i.e., only as integer numbers 
they have a meaning in string theory interpretation.

Let us make few comments on the near-horizon solution (\ref{4c4d0}):
\begin{enumerate}
\item
Geometry of solution is AdS$_2 \times S^2$, and the whole solution respects symmetry on
isometry group $SO(2,1) \times SO(3)$. This is what is expected for near-horizon geometry of 
static spherically symmetric \emph{extreme} black hole in $D=4$. 
\item
When all charges are finite and non-vanishing, black hole horizon, given by $r=0$, is 
\emph{regular}, with all curvature invariants being finite and well defined on it, so this is another 
example of large extremal black hole.
\item
It is completely determined by charges $n$, $w$, $\widehat{N}$ and $\widehat{W}$, and is 
independent of asymptotic values of moduli. This is yet another example of the attractor mechanism.
\item
It is by itself exact solution of equations of motion.
\item
There are 3 types of solutions: (1) $nw>0$ and $\widehat{N}\widehat{W}>0$ solutions are 
supersymmetric 1/4-BPS (they break 3/4 of the supersymmetry generators) \cite{Ferrara:1997ci}, 
(2) $nw<0$ and $\widehat{N}\widehat{W}<0$ solutions are non-BPS, (3) $nw\widehat{N}\widehat{W}<0$ solutions are non-BPS. In fact, they are examples of 3 possible types of black hole solutions in general
classification in $\mathcal{N}=4$ SUGRA \cite{Cerchiai:2009pi}. 
\item
For $\widehat{N}=\widehat{W}=0$ our 4-charge black hole reduces to 2-charge small black hole of
section \ref {ssec:smallbh}. Taking $\widehat{N}\to0$ and/or $\widehat{W}\to0$ in the 
near-horizon solution  (\ref{4c4d0}) is not well defined, which is expected because 
near-horizon geometry of 2-charge small black holes, given in (\ref{e10}), is not 
AdS$_2 \times S^2$ but geometry with singular horizon.
\end{enumerate}

From (\ref{4c4d0}) (and using (\ref{Eframe})) we can easily calculate Bekenstein-Hawking black 
hole entropy
\begin{equation} \label{4clBHent}
S_{\rm bh} = \frac{A_h}{4 G_N} = 2\pi \sqrt{|nw\widehat{N}\widehat{W}|} \,.
\end{equation} 
As expected for large black hole, the entropy is finite.

\subsubsection{Stringy ($\alpha'$) and quantum ($g_s$) corrections}
\label{sssec:corr}

Full low energy effective action of heterotic string theory is much more complicated then 
(\ref{10dl0}). It has the general form
\begin{equation}
\mathcal{A} = \mathcal{A}_0 + \hbox{ (tree-level higher derivative terms) + (string loop corrections)}
\end{equation}
String loop corrections, parametrized by string coupling constant $g_s$, are coming from
quantum corrections and generally have perturbative and nonperturbative contributions (a thing 
well-known already from ordinary QFT). The higher derivative terms are coming from finite size 
of strings and are parametrized by the square of string length parameter, i.e., $\alpha'$.

As quantum loop-corrections are much more subtle to deal with (quantum effective actions are 
either non-local or not manifestly symmetric on dualities), we shall restrict ourselves to classical 
tree-level analyses. The corresponding effective action is \emph{perturbative} in $\alpha'$
\begin{equation} \label{treeact}
\mathcal{A}_{\rm tree} = \sum_{n=0}^\infty \alpha'^n \mathcal{A}_n
 = \sum_{n=0}^\infty \alpha'^n \int dx^{D} \sqrt{-G} \, \mathcal{L}_n
\end{equation}
and is known incompletely. Only $\mathcal{A}_0$, $\mathcal{A}_1$ and $\mathcal{A}_2$ are 
known fully. As string theory is not expected to be equivalent to any QFT, an expansion in 
(\ref{treeact}) is believed to be infinite. It is obvious from dimension of $\alpha'$ that 
$\mathcal{A}_n$ is composed of $2(n+1)$-derivative terms. As the effective theory contains 
gravity, $\mathcal{A}_n$ can contain powers of Riemann tensor up to $(n+1)$-order and that is 
why the whole $\mathcal{A}_n$ is sometimes called $R^{n+1}$ part of the action. There is a field redefinition scheme in which dependence of the tree-level Lagrangian on the dilaton field $S(x)$ is 
of the form
\begin{equation} \label{treeL}
 \mathcal{L}_{\rm tree} = S \, \mathcal{K}(\partial^m S/S) \,, \qquad m=1,2,\ldots \,.
\end{equation}
This is a manifestation to the fact that effective string coupling is with expectation value of dilaton 
field by a relation which is in our conventions given by $g_{\rm eff}^2 \propto 1/S$.

Let us now go back to the lowest-order near-horizon solutions (obtained from action 
$\mathcal{A}_0$) that we presented in previous sections, and analyse what conditions should be 
applied to the parameters so that perturbative expansions are well-defined. We take first the regular 
large black hole solutions, using 5-dimensional 3-charge black case as an example. Let us 
start with loop-corrections which are parameterized by string coupling constant. From 
solution (\ref{3c5d0}) we copy
\begin{equation} \label{gseff3}
S  = \frac{4 G_N}{\alpha'^{3/2} \pi}\, \frac{\sqrt{|nw|}}{|m|} \,.
\end{equation}
It is obvious that if we are in the regime in which $|nw|\gg m^2$, then $S\gg1$ and so 
$g_{\rm eff}\ll1$, which means that we can ignore loop-corrections. As we do not want to cope 
with loop-corrections, we shall assume that we are in such asymptotic regime. As for 
$\alpha'$-corrections, note that all 2-derivative scalar monomials which appear in $D=5$
effective action (\ref{ea5d}) (i.e., Ricci scalar $R$, $T^2 (F^{(1)})^2$, $T^{-2} (F^{(2)})^2$ 
and $H^2$) when evaluated on the solution (\ref{3c5d0}) are proportional to $1/(\alpha'm)$. It 
is easy to realize that all monomials which contain $2k$ derivatives will be proportional to 
$1/(\alpha'm)^k$. This means that the terms in the expansion (\ref{treeact}) will behave like
\begin{equation}
\alpha'^n \mathcal{K}_n \propto \frac{1}{\alpha'\, m} \frac{1}{m^n}
\end{equation}
We see that expansion in $\alpha'$ is effectively expansion in $1/m$. So, if $m\gg1$ the 
$\alpha'$-corrections will be small and we expect to have well-behaved perturbative expansion. 
The same analyses can be repeated for the case of 4-dimensional 4-charge large black holes 
with near-horizon solution given by (\ref{3c5d0}), leading to the similar conclusions, with the only 
difference that the role of $m^2$ is now played by the product $\widehat{N}\widehat{W}$.

In the case of \emph{small} black hole solutions things are much different. Let us take 2-charge
black hole solution reviewed in section \ref{ssec:smallbh}. We saw that this solution has singular
horizon, which is obvious from near-horizon behavior (\ref{e10}). We see that dilaton is singular on
the horizon/singularity $r=0$, which means that $g_{\rm eff}^2 \propto 1/S = 0$. So, if we are 
hoping that higher-oreder corrections can regularize the solution, it is obvious that loop-corrections 
cannot do this because they vanish on the horizon. What about stringy $\alpha'$-corrections? 
From near-horizon solution follows that $n$-th order terms will behave as 
$\mathcal{K}_n \propto r^{-14(n+1)}$ (in $D=9$),
which shows that higher-order $\alpha'$-terms in the action are more and more singular when 
evaluated on lowest-order solution. So, $\alpha'$-corrections are important. Let us assume that
they can regulate the solution. As we are dealing with extremal black holes, we expect to obtain
AdS$_2 \times S^{D-2}$ near-horizon geometry, with radii of the order of string length, i.e., 
$\ell^2_{A,S} \sim \alpha'$ (indeed, it was shown on explicit examples that inclusion of general 
$R^2$-corrections is leading to such behavior 
\cite{Dabholkar:2004dq,Sen:2004dp,Hubeny:2004ji,Sen:2005kj}). The Ricci scalar is 
$R\sim1/\alpha'$. Repeating the above analyses of behavior of higher-order terms in the 
tree-level Lagrangian we obtain
\begin{equation} \label{smallalph}
\alpha'^n \mathcal{K}_n \propto \frac{1}{\alpha'} \,,
\end{equation}
where we again assumed that $|n/w|\sim1$. Though the solution is regular, we see from 
(\ref{smallalph}) that for small black holes $\alpha'$-expansion is \emph{not} well-defined as
perturbative expansion, and one should find a way to somehow "sum" the complete 
$\alpha'$-dependence. It is easy to understand the reason for this - black holes with horizon 
radius of the order of string length are intrinsically stringy objects, and for such objects we 
do not expect that low energy (or low curvature) expansion is meaningful.

We shall see in the next section that in some cases (including black holes we analyze here) it 
is possible to obtain statistical entropies by counting of microstates in string theory 
\emph{exactly} in $\alpha'$. If we could calculate $\alpha'$-corrections to black hole entropies 
from the gravity side (i.e., by using  low-energy effective action), this would be strong test for the
validity of such stringy description of black holes. On the other hand, if we believe in such
description, we could use the equality of entropies to get some new information on structure of
higher-order terms in effective actions. We shall show here how both of this ideas can be
successfully applied on our examples of extremal black holes in heterotic string theory.

Now when we have understanding of influence for all type of corrections for black holes that we 
study, we can fix the values of parameters. We shall use the convention in which $\alpha'=16$ 
and $G_N=2$ throughout the review, with the exception of section \ref{sec:offshsusy} in which we 
take  $\alpha'=1$ and $G_N=\pi/4$, a convention frequently used 
in the literature for theories in $D=5$.

\section{Stringy description}
\label{sec:stringy}

\subsection{Microstate counting}
\label{ssec:microcount}

We want to find configurations in heterotic string theory which are in the supergravity limit 
described by black hole solutions we analysed in section \ref{sec:bhsol}. The simplest case 
is 2-charge solution, for which we assumed to represent just the elementary string living in
$M_9 \times S^1$ spacetime, which is wound $w$ times around circle $S^1$ and has momentum 
number $n$. The simplicity of this case is that these states are purely perturbative, i.e., they exist 
in the spectrum of a free string. 

Let us now analyze these states in the limit of free string with string coupling $g_s\to0$ such that 
geometry can be considered flat, i.e., $G^{(10)}_{MN} = \eta_{MN}$. We parametrize circle 
$S^1$ with $0\le x_9< 2\pi\sqrt{\alpha'}R$. Perturbative string states are characterized by:
\begin{itemize}
\item
Momentum 9-vector $p^\mu$ in the uncompactified directions (Minkowski space $M_9$).
\item
Right-moving and left-moving momenta in the compact direction ($S^1$) given by
\begin{equation} \label{pRpL}
p_R = \frac{1}{\sqrt{\alpha'}} \left( \frac{n}{R} + wR \right) \,, \qquad
p_L = \frac{1}{\sqrt{\alpha'}} \left( \frac{n}{R} - wR \right) \,.
\end{equation}
\item
Excitations described by independent right- and left-moving oscillators, of which we will need just
the total level numbers $N_R$ and $N_L$ (measured relative to physical vacuum).  
\end{itemize}
Physical states satisfy the following mass-shell conditions
\begin{eqnarray} \label{masshelR}
M^2 &=& p_R^2 + 4 N_R / \alpha' \\
        &=& p_L^2 + 4 N_L / \alpha'
\label{masshelL}
\end{eqnarray}
where $M$ is the mass connected to uncompactified dimensions, defined by 
$M^2 = - p_\mu p^\mu$.

We saw in section \ref{ssec:smallbh}  that for some choices of signatures of charges ($nw>0$) 
black holes are 1/2 BPS states, so let us locate such states in the string spectrum. Heterotic string 
theory has $\mathcal{N}=1$ supersymmetry, which contains 16 generators in 10 dimensions. It 
can be shown that states which satisfy the BPS condition
\begin{equation} \label{halfbps}
M = |p_R| = \frac{1}{\sqrt{\alpha'}} \left| \frac{n}{R} + wR \right|
\end{equation}
are $1/2$-BPS states preserving half of the supersymmetries ($8=16/2$). We see that 
(\ref{halfbps}) is for $nw>0$ exactly equal to mass of BPS small black holes given in 
(\ref{emass2}), which confirms that we are on the right track.\footnote{The factor 
$g_s^{\gamma/2}$ present in (\ref{emass2}) is because in section \ref{ssec:smallbh} we defined 
mass (energy) by using canonical (Einstein-frame) metric, while here we are using string frame 
metric. From (\ref{Eframe}) and (\ref{asym9d}) follows that they asymptotically differ by a factor of
$g_s^\gamma$, which gives to the above difference in mass scale.} 

Putting (\ref{halfbps}) in the condition (\ref{masshelR}) we obtain $N_R=0$, i.e., right-moving 
sector is unexcited for these states. Equating now (\ref{masshelR}) and (\ref{masshelL}), and 
using (\ref{pRpL}), we can write the condition (\ref{halfbps}) equivalently as
\begin{equation} \label{levnw}
N_L = \alpha' (p_R^2 - p_L^2)/4 = nw \,.
\end{equation}
As by definition $N_L\ge0$ it directly follows $nw\ge0$, as expect from supergravity analysis. All 
states with fixed $n$, $w$ and $N_L=nw$ are equal candidates to represent (in the free string 
regime) the BPS 2-charge small black hole with charges $n$ and $w$. Let us calculate the number 
of such states. Closed form expression for general $n$ and $w$ is not known, but we saw in 
section \ref{sssec:corr} that we should be interested in the regime $nw\gg1$ where classical black 
hole solutions can be reliable (quantum corrections are negligible). In this regime it is quite easy to 
get asymptotic expression for number of states
\begin{equation} \label{nos}
\Gamma = e^{4\pi \sqrt{N_L}}\,, \qquad N_L \gg 1 \,,
\end{equation}
which is obviously a huge number. We can assign the statistical entropy to this ensemble of 
states using standard microcanonical definition
\begin{equation} \label{micro}
S_{\rm stat}^{\rm (BPS)} \equiv \ln \Gamma = 4\pi \sqrt{nw}\,, \qquad nw \gg 1 \,.
\end{equation}
It is important to emphasize that though the result (\ref{micro}) is asymptotic in $nw$, it is 
\emph{exact} in $\alpha'$.

What about the cases when $nw<0$ for which we found non-BPS black hole solutions? Though 
we cannot use supersymmetry here\footnote{Note that this sector, in which only right-movers are excited, is basically the same as the corresponding perturbative sector of type II theory (in which
case it is also supersymmetric due to the larger $\mathcal{N}=2$ SUSY). The degeneracy
(for all signs of charges $n$ and $w$) in type II theory is given by (\ref{nosn}).}, we can by 
analogy define ensemble of states with fixed $n$ and $w$, which are unexcited now in left-moving sector, i.e., with $N_L=0$. Putting this in (\ref{masshelL}) we obtain for the mass
\begin{equation} \label{nonbpsm}
M = |p_L| = \frac{1}{\sqrt{\alpha'}} \left| \frac{n}{R} - wR \right| \,,
\end{equation} 
which again agrees with black hole mass formula (\ref{emass2}) for $nw<0$. So we are on the 
right track. Putting (\ref{nonbpsm}) in (\ref{masshelR}) gives us now $N_R=-nw$, which forces 
$nw<0$. We obtained ensemble of states defined by fixing $n$, $w$, and $N_R=-nw$. The 
asymptotic formula for number of such states is again easily calculated and the result is
\begin{equation} \label{nosn}
\Gamma = e^{2\sqrt{2}\pi \sqrt{N_R}}\,, \qquad N_R \gg 1 \,,
\end{equation}
which gives statistical entropy
\begin{equation} \label{micron}
S_{\rm stat}^{\rm (non-BPS)} \equiv \ln \Gamma = 2\sqrt{2}\pi \sqrt{|nw|}\,, \qquad -nw \gg 1 \,.
\end{equation}

Now we pass to microscopic (stringy) description for large black holes. As they generally contain 
non-perturbative objects (like, e.g., $NS5$-branes and KK monopoles) the corresponding 
microstates in string theory are also non-perturbative. This drastically complicates calculation of
statistical entropy by direct counting of string microstates, and only in some special cases (so far 
only BPS where one can use powerful properties of supersymmetry) closed form expressions 
were obtained. Fortunately, for 4-dimensional 4-charge large BPS black holes (discussed in 
section \ref{sssec:4cbh}), and for 5-dimensional 3-charge large BPS black holes (discussed in 
section \ref{sssec:3cbh}) such calculations were done.

For microscopic statistical description of 4-dimensional 4-charge large black holes we take 
heterotic string theory compactified on flat $T^4 \times \widehat{S}^1 \times S^1$, and count 
1/4-BPS micro-configurations consisting of elementary string wound $w$ times around $S^1$, 
$\widehat{N}$ KK-monopoles wrapped around $T^4 \times S^1$, $(\widehat{W}+\widehat{N})$ 
NS5-branes\footnote{It is known that KK-monopole contributes $(-1)$ unit to NS5-brane charge, 
so $\widehat{W}$ denotes total NS5-brane charge of the configuration.} wrapped around 
$T^4 \times S^1$, and a momentum on $S^1$ with momentum number $n$. For 1/4-BPS states 
($nw>0$, $\widehat{N}\widehat{W}>0$) the asymptotic formula for statistical entropy in the regime 
$nw\gg\widehat{N}\widehat{W}$ is given by (see \cite{Sen:2007qy} for a detailed review)
\begin{equation} \label{micro4}
S_{\rm stat}^{\rm (BPS)} = 2\pi \sqrt{nw(\widehat{N}\widehat{W}+4)} \,.
\end{equation}
Again, (\ref{micro4}) is $\alpha'$-exact, and the cumulative effect of $\alpha'$ corrections is
encoded in number 4 inside the square root. This time there is a well defined perturbative 
expansion in $\alpha'$, as expected from our discussion in section \ref{sssec:corr} on properties of corresponding 4-charge large black holes. From (\ref{micro4}) follows that at the lowest order in  
$\alpha'$ we obtain agreement with result for black hole entropy (\ref{4clBHent}). This is one of the 
many examples in which string theory is giving microscopic explanation for black hole 
thermodynamics, in the most direct and straightforward way without any suspicious assumptions.

What about non-BPS states? In the case $nw<0$, $\widehat{N}\widehat{W}<0$ we can use the 
trick to go to the type-II string theory, where those microstates are supersymmetric. As the 
microstates are in NS-NS sector and purely right-moving, the bosonic part of this sector is in 
one-to-one correspondence with that in heterotic theory which means that microcanonical 
entropies are the same. Counting of microstates gives asymptotically for 
$|nw|\gg|\widehat{N}\widehat{W}|$
\begin{equation} \label{micro4n2}
S_{\rm stat}^{\rm (non-BPS)} = 2\pi \sqrt{|nw\widehat{N}\widehat{W}|} \,, \qquad
 nw<0\,, \quad \widehat{N}\widehat{W}<0 \,.
\end{equation}
There are no $\alpha'$-corrections. In the next section we shall give macroscopic explanation for 
this.

As for the non-BPS case with $nw\widehat{N}\widehat{W}<0$, the $\alpha'$-exact direct 
microstate counting was not performed.

For microscopic statistical description of 5-dimensional 3-charge large black holes we take 
heterotic string theory compactified on flat $T^4 \times \times S^1$, and count 1/4-BPS 
micro-configurations consisting of elementary string wound $w$ times around $S^1$, 
$m$ NS5-branes wrapped around $T^4 \times S^1$, and a momentum on $S^1$ with momentum 
number $n$. For 1/4-BPS states ($nw>0$) the asymptotic formula for statistical entropy in the 
regime $nw\gg m$ is given by \cite{Castro:2008ys}
\begin{equation} \label{micro5b}
S_{\rm stat}^{\rm (BPS)} = 2\pi \sqrt{nw(|m|+3)} \,.
\end{equation}
Again, (\ref{micro5b}) is by construction $\alpha'$-exact.

Comments on direct microscopic analysis:
\begin{enumerate}
\item
Microscopic entropies are calculated \emph{exactly} in $\alpha'$, and closed-form expressions
were obtained. 
\item
Large black holes - Lowest order in $\alpha'$ agreeing with Bekenstein-Hawking entropies of corresponding black holes. In principle one can use perturbative analyses to check agreement at 
higher orders, by systematically taking into account higher-derivative terms in supergravity 
effective action.
\item
Small black holes - Microscopic entropy intrinsically non-perturbative in $\alpha'$. To compare it 
with black hole entropy, full tree-level effective action is needed on the gravity side.
\item
Microstate counting is typically performed in the limit of small $g_s$, such that influence of relevant microscopic configurations on space-time geometry can be neglected. On the other hand, in 
supergravity analyses we are dealing with black holes which significantly change the geometry of 
space-time. Obviously, those are two completely different regimes, so why we should be allowed 
to compare the two entropies?\footnote{Mathematically this means the following. Influence of an 
object with mass $M$ on geometry is proportional to $G_N M$. For string configurations typically considered we have $G_N M\propto g_s^a$, with $a>0$. Now, to avoid large quantum effects, we 
take $g_s\ll1$. No effect on geometry means that $G_N M$ should be small compared to the 
string scale, i.e., $G_N M \ll \alpha'^{(D-3)/2}$. But if we have large black holes, for which 
Schwarzschild radius should be much larger then the string length parameter, we have 
$\sqrt{\alpha'} \ll R_{\rm Sch} \propto (G_N M)^{1/(D-3)}$. This is obviously an opposite limit from 
the previous one, so we have two completely different regimes. We note that an "intermediate" 
regime $G_N M \sim \alpha'^{(D-3)/2}$ is fully nonperturbative stringy regime of which very little is 
known so far.} For BPS black holes there is a direct answer - such states are organized in special 
shorter multiplets, and the number of states inside these multiplets cannot change when parameters 
of the theory (such as $g_s$) are changed continuously. Also, if we assume that nothing violent
(like phase transitions) is happening in the process, the number of short multiplets will not change. 
So the total number of states, which gives the entropy, is protected by supersymmetry. For non-BPS 
black holes we cannot use this argument, but we shall show in section \ref{ssec:senef} that attractor mechanism can instead be used to argue that the entropy should not change when we "turn 
effective coupling constant on".
\end{enumerate}

\subsection{AdS/CFT methods}
\label{ssec:adscft}

In the classic paper \cite{Brown:1986nw} Brown and Henneaux showed that gravity in 
$D=3$ dimensions has asymptotic symmetry group containing two independent Virasoro 
algebras,
\begin{eqnarray*}
&& [ L_m,L_n] = (m-n)L_{m+n} + \frac{c_R}{12}m(m^2-1)\delta_{m+n} \,, 
\\ 
&& [ \overline{L}_m,\overline{L}_n] = (m-n)\overline{L}_{m+n}
 + \frac{c_L}{12}m(m^2-1)\delta_{m+n} \,, \\
&& [L_m,\overline{L}_n] = 0
\end{eqnarray*}
where $m,n\in Z$ and $c_R$ and $c_L$ are central charges of corresponding algebras. This 
is exactly what is present in 2-dimensional conformal field theories. It was argued later by 
Maldacena that this is just one example of a more general idea known today as the AdS/CFT 
conjecture \cite{Aharony:1999ti}, which states that $D$-dimensional gravity theory which is
asymptotically AdS should be equivalent to the conformal field theory (without gravity) which is 
"living" on the boundary (in asymptotic infinity) of AdS space. The equivalence is of strong/weak 
type (it connects strongly coupled theory on one side with weakly coupled on the other side) which 
can be extremely useful as it can be used to study strong-coupling behavior by using perturbative calculations. But, because of this it is hard to prove conjecture, as for this one should be able to 
calculate in the regime of strong coupling at least on one side, which is typically not known (that is 
why it is still a conjecture). One possible exception is $D=3$ case, where the dual theories are 
2-dimensional conformal field theories, for which much more is known in strongly coupling regime. 
This is one of the motivations for analyzing cases in which one has asymptotic AdS$_3$ geometry. 
For example, microcanonical entropy (logarithm of number of states) at a level $L_0=\Delta$, 
$\overline{L}=\overline{\Delta}$ is given asymptotically for $\Delta \gg c_R$, 
$\overline{\Delta} \gg c_L$ by the simple Cardy formula
\begin{equation} \label{cardy}
S_{\rm CFT} = 2\pi \sqrt{\frac{c_R \Delta}{6}}
 + 2\pi \sqrt{\frac{c_L \overline{\Delta}}{6}} \,.
\end{equation}
Important property of Cardy formula (\ref{cardy}) is that entropy depends just on central charges, 
and not on the specific details of the conformal theory.

Interestingly, all (non-singular) solutions that we consider in this review contain such 
AdS$_3$ factor in the near-horizon geometry. It can be easily shown that factors 
AdS$_2 \times S^1$, which appear in all these solutions, are \emph{locally} isometric to 
AdS$_3$. They all satisfy
\begin{equation} \label{locads3}
R_{MNPQ} = - \ell_A^{-2} \left( G_{MP} G_{NQ} - G_{MQ} G_{NP} \right)
 \quad \mbox{for} \quad M,N,P,Q \in \{t,r,x^9\} \,.
\end{equation}
meaning that they are locally maximally symmetric. The geometries would be also globally 
isometric to AdS$_3$ if the proper radius of $S^1$  would be infinitely large. We shall assume now
that the radius is large enough so that corresponding finite-size effects are negligible and we can 
take for geometry to have AdS$_3$ factor.\footnote{This means that special limit for the charges is understood, e.g., in our examples it is $|n|\gg|w|$. We shall see below that this is also compatible 
with conditions $\Delta \gg c_R$, $\overline{\Delta} \gg c_L$ generally needed for validity of
Cardy formula (\ref{cardy}).}

For extremal black holes that we analyse, either $\Delta$ or $\overline{\Delta}$ are vanishing, and 
the one which does not vanish is equal to $|n|$. In the cases where all other charges are positive, 
BPS case $n>0$ corresponds to $\Delta=0$ and $\overline{\Delta}=n$, and non-BPS case $n>0$ to 
$\Delta=|n|$ and $\overline{\Delta}=0$. If one could find central charges $c_{R,L}$ 
then the entropy would be simply given by Cardy formula (\ref{cardy}). There are two methods 
which were used in the literature, (1) direct sigma model calculation \cite{Kutasov:1998zh}, 
(2) indirect by using anomaly inflow arguments \cite{Kraus:2005vz,Kraus:2006wn}.

In the first, and historicaly earlier method, one treats heterotic string theory on the relevant compactifications and backgrounds with AdS$_3$ factors, and then heavily relying on explicit 
realizations of $(0,4)$ supersymmetry (present in all cases of interest to us)  and AdS/CFT 
correspondence one is able to obtain relevant central charges \cite{Kutasov:1998zh}. For the 
geometry appearing as near horizon geometry in case of the 4-dimensional 4-charge large black 
holes from section \ref{sssec:4cbh}, when $w>0$, $\widehat{N}\widehat{W}>0$ one obtains
\begin{equation} \label{c4ch}
c_R = 6 w (\widehat{N}\widehat{W}+2) \,, \qquad
c_L = 6 w (\widehat{N}\widehat{W}+4) \,.
\end{equation}
When $w<0$, $\widehat{N}\widehat{W}>0$ the only difference is the left $\leftrightarrow$ right
interchange, which leads to $c \leftrightarrow \overline{c}$, so
\begin{equation} \label{c4ch2}
c_R = 6 |w| (\widehat{N}\widehat{W}+4) \,, \qquad
c_L = 6 |w| (\widehat{N}\widehat{W}+2) \,.
\end{equation}
When used in Cardy formula (\ref{cardy}) this gives in the BPS case ($nw>0$)
\begin{equation} \label{cft4b}
S_{\rm CFT}^{\rm (BPS)} = 2\pi \sqrt{nw(\widehat{N}\widehat{W}+4)} \,,
 \qquad nw>0\,, \quad \widehat{N}\widehat{W}>0 \,,
\end{equation}
which exactly agrees with the expression obtained from direct microstate counting 
(\ref{micro4}). The virtue of AdS/CFT method is that it can give us the result also in the 
non-BPS case in which $n<0$
\begin{equation} \label{cft4n}
S_{\rm CFT}^{\rm (non-BPS)} = 2\pi \sqrt{|nw|(|\widehat{N}\widehat{W}|+2)} \,, \qquad
  nw\widehat{N}\widehat{W}<0 \,.
\end{equation}
Note that this was the case in which direct counting of microstates was not performed, so this 
is a new result. 

As for the other type of non-BPS states defined by $nw<0$, $\widehat{N}\widehat{W}<0$, 
because AdS$_3$ background with $\widehat{N}\widehat{W}<0$ is nonsupersymmetric in the 
heterotic theory, the method apparently cannot be used. 

For the geometry appearing as near horizon geometry in case of the 5-dimensional 3-charge 
large black holes from section \ref{sssec:3cbh}, when $w>0$ one obtains \cite{Kutasov:1998zh}
\begin{equation} \label{c3ch}
c_R = 6 |w k| \,, \qquad c_L = 6 |w| (|k|+2) \,.
\end{equation}
When $w<0$, the only change is $c_R \leftrightarrow c_L$. In (\ref{c3ch}) $k$ denotes the total 
level of affine algebra $\widehat{SL(2)}$ in the right-handed (supersymmetric) sector coming 
from worldsheet symmetries. 

When (\ref{c3ch}) is put in Cardy formula (\ref{cardy}) we obtain in the BPS case ($nw>0$)
\begin{equation} \label{cft3b}
S_{\rm CFT}^{\rm (BPS)} = 2\pi \sqrt{nw(|k|+2)} \,,
\end{equation}
while in the non-BPS case ($nw<0$)
\begin{equation} \label{cft3n}
S_{\rm CFT}^{\rm (non-BPS)} = 2\pi \sqrt{|nwk|} \,.
\end{equation}
BPS entropy (\ref{cft3b}) can be compared with statistical entropy (\ref{micro5b}) obtained by 
direct microstate counting. We see that formulas agree if we take that $k$ is
connected with NS5-brane charge $m$ (which is expected to be equal to the number of 
NS5-branes $Q_5$) through a relation
\begin{equation} \label{kN}
|k| = |m| + 1
\end{equation}
In \cite{Prester:2009mc} it was explicitly shown in supergravity analysis that shift in (\ref{kN}) 
is generated by mixed (gauge-gravity) Chern-Simons term present in heterotic string theory. We 
shall show this in section \ref{ssec:3chbh}.

Later it was shown \cite{Kraus:2005vz,Kraus:2006wn,David:2007ak} that when effective 
3-dimensional theory on AdS$_3$ has $(0,4)$ (or even smaller $(0,2)$ \cite{Kaura:2008us}) 
supersymmetry, central charges are generally determined purely by the coefficients of 
Chern-Simons terms. This method of calculating central charges has two virtues: (i) it is 
general, depending only on symmetries, (ii) as Chern-Simons terms are connected to anomalies 
and correspondingly 1-loop saturated, their coefficients in many cases can be calculated 
exactly (at least in $\alpha'$). In fact, in \cite{Kraus:2005vz} the power of this method 
was demonstrated by calculating central charges (\ref{c4ch}) relevant for the entropy of 
4-dimensional 4-charge black holes. As for the case relevant for 5-dimensional 3-charge black 
holes, i.e., (\ref{c3ch}), such calculations were not performed. Let us mention that 
$\alpha'$-exact gravity calculations \cite{Prester:2009mc} are confirming both results 
(\ref{c4ch}) and (\ref{c3ch}).

\section{Some formalities}
\label{sec:form}

\subsection{Wald entropy formula}
\label{ssec:wald}

As noted in section \ref{sssec:corr}, low energy effective actions of string theories, even on 
tree-level, contain higher-derivatives terms. It is known that in such theories entropy of 
black hole solutions is not any more given by simple Bekenstein-Hawking formula (\ref{BHent}). 
If the theory is \emph{manifestly} diffeomorphism invariant, in which case Lagrangian density 
is of the form
\begin{equation} \label{eqdefgenL}
\mathcal{L} = \mathcal{L}(g_{ab}, R_{\mu\nu\rho\sigma}, \nabla_\lambda R_{\mu\nu\rho\sigma},
 \dots{}, \psi, \nabla_\mu \psi, \dots) \,,
\end{equation}
where $\psi$ denotes matter fields and dots denote higher-order derivatives, then the black
hole entropy is given by Wald formula \cite{Wald}
\begin{equation} \label{Went}
S_{\rm bh} = - 2\pi \int_{\mathcal{H}} d^{D-2}x \sqrt{h}\, E^{abcd} \eta_{ab}\eta_{cd} \,.
\end{equation} 
Here $\mathcal{H}$ is a cross-section of the horizon, $\eta_{ab}$ denotes binormal to 
$\mathcal{H}$, $h=\det(h_{ab})$ is determinant of the induced metric on $\mathcal{H}$, and
\begin{equation} \label{eqdefgenE}
E^{\mu\nu\rho\sigma} = \frac{\partial \mathcal{L}}{\partial R_{\mu\nu\rho\sigma}}
 - \nabla_{\lambda_1} \frac{\partial \mathcal{L}}{\partial \nabla_{\lambda_1}
 R_{\mu\nu\rho\sigma}}
 + \ldots + (-1)^m \nabla_{(\lambda_1} \ldots \nabla_{\lambda_m)}
 \frac{\partial \mathcal{L}}{\partial \nabla_{(\lambda_1} \ldots \nabla_{\lambda_m)}
 R_{\mu\nu\rho\sigma}}
\end{equation} 
The derivative in (\ref{eqdefgenE}) is taken with $g_{\mu\nu}$ and $\nabla_\mu$ fixed.

Two important comments on Wald formula:
\begin{enumerate}
\item
Here it is important to notice that Wald entropy is purely determined from near-horizon 
behavior of black hole solution. Because in higher-derivative theories it is generally not
possible to find exact solutions in the whole space-time, this property is essential if we
are hoping to calculate exact entropies.
\item
There are theories (and heterotic string theory is an example) with Lagrangians containing also 
terms which are \emph{not} manifestly diff-covariant, so called (purely gravitational or mixed) 
Chern-Simons terms. In \cite{Tachikawa:2006sz} a generalization of Wald formula to such 
theories was proposed. However, instead of using this generalised formula, we shall handle 
Chern-Simons terms in a more direct and quicker way developed in \cite{Sahoo:2006pm}.
\end{enumerate}

\subsection{Sen's entropy function method}
\label{ssec:senef}

Let us assume that we have a $D$-dimensional theory with a field content consisting of the metric 
tensor $G_{\mu\nu}$, some number of neutral scalar fields $\phi_s$, and a number of (also 
neutral) $p$-form fields (of which some are U(1) gauge fields with corresponding $(p+1)$-form 
strengths), with the Lagrangian which is \emph{manifestly} gauge and diffeomorphism invariant. 
We are interested in the near-horizon behavior of the rotationally invariant \emph{extremal} black 
holes. One expects that the metric is $AdS_2\times S^{D-2}$, which has $SO(2,1)\times SO(D-1)$ 
as an isometry group, and that the whole background respects this symmetry 
manifestly.\footnote{In \cite{Kunduri:2007vf,Astefanesei:2007bf,Goldstein:2007km,Kunduri:2009ud} 
this was proven for broad class of actions in $D=4$ and $D=5$.} In this case one can apply Sen's 
entropy function formalism \cite{Sen:2005wa} which we now briefly review.\footnote{Method was 
also extended to rotating black holes \cite{Astefanesei:2006dd}.}

The point is that manifest symmetry under $SO(2,1)\times SO(D-1)$ heavily restricts the 
near-horizon behavior of the fields. For example, it follows that the only manifestly 
covariant $p$-forms (which means strengths in case of gauge fields) which are allowed to be 
non-vanishing are 2-form (denoted $F^{I}$) and $(D-2)$-form (denoted $H_m$).\footnote{Obviously
$D$-forms are also allowed, but we shall assume that they are either dualised to scalars or 
written as wedge-products of 2-forms and $(D-2)$-forms.} More completely, 
the near-horizon behavior is constrained to have the following form
\begin{eqnarray} \label{efgen}
&& ds^2 = v_1 \left( -r^2 dt^2 + \frac{dr^2}{r^2} \right) + v_2\,d\Omega_{D-2}^2
 \nonumber \\
&& \phi_s = u_s \;, \qquad s=1,\ldots,n_s 
 \nonumber \\
&& F^{I}_{rt} = f^I \;, \qquad i=1,\ldots,n_F
 \nonumber \\
&& H_{m} = h_m \mathbf{\epsilon}_{S} \; \qquad m=1,\ldots,n_H
\end{eqnarray}
where $v_{1,2}$, $u_s$, $e^I$ and $h_m$ are all constant, and $\mathbf{\epsilon}_{S}$ is an 
induced volume-form on unit sphere $S^{D-2}$. For $F^{I}$ and $H_{m}$ which play the role of 
gauge field strengths, if they are closed forms than it follows that $e^I=f^I$ and $p_m=h_m$ are 
the electric fields and magnetic charges, respectively.\footnote{We see that $D=4$ is a special 
case in which there are only 2-form strengths, but which can carry both electric and magnetic 
charges.}

It can be shown that for background (\ref{efgen}) solving of equations of motion is equivalent 
to extremization of the (algebraic) function $\mathcal{F}$, defined by
\begin{equation} \label{ffun}
\mathcal{F}(\vec{v},\vec{u},\vec{f},\vec{h};\vec{e},\vec{p}) =
 \oint_{S^{D-2}} \sqrt{-G} \, \mathcal{L} \,,
\end{equation}
over $\vec{v}$, $\vec{u}$ and $\vec{f}$. We have divided forms into gauge forms (whose 
corresponding electric field strengths $\vec{e}$ and magnetic charges $\vec{p}$ are taken 
as \emph{fixed}) and non-gauge whose values are variables denoted as $\vec{f}$ and 
$\vec{h}$ (in our examples they will be auxiliary fields). This means that we have to solve a 
system of algebraic equations
\begin{equation} \label{feom}
0 = \frac{\partial\mathcal{F}}{\partial\vec{v}} \,,\qquad
0 = \frac{\partial\mathcal{F}}{\partial\vec{u}} \,,\qquad
0 = \frac{\partial\mathcal{F}}{\partial\vec{f}} \,,\qquad
0 = \frac{\partial\mathcal{F}}{\partial\vec{h}} \,.
\end{equation}
If the system happens to be regular, we can solve it for all unknowns and obtain solutions
for $\vec{v}$ ,$\vec{u}$ and $\vec{f}$ as functions of $\vec{e}$ and $\vec{p}$.

It is more common to express solutions not as functions of electric field strengths but
as a function of electric charges. It can be easily shown that electric charge (in particular
normalization) is given by
\begin{equation} \label{elch}
\vec{q} = \frac{\partial\mathcal{F}}{\partial\vec{e}} \,,
\end{equation}
One of the virtues of the Sen's entropy function method is that it gives straightforwardly
the entropy of black hole. Let us define the entropy function $\mathcal{E}$ as 
Legandre-transform of the function $\mathcal{F}$ with respect to electric field/charge
\begin{equation} \label{efun}
\mathcal{E}(\vec{v},\vec{u},\vec{f},\vec{h},\vec{e};\vec{q},\vec{p}) =
2\pi \left( \vec{q}\cdot\vec{e} - \mathcal{F} \right) \,.
\end{equation}
Then obviously system (\ref{feom}) and (\ref{elch}) is equivalent to extremization of the
entropy function $\mathcal{E}$ with respect to all variables except electric and magnetic
charges which are kept fixed, i.e.,
\begin{equation} \label{efeom}
0 = \frac{\partial\mathcal{E}}{\partial\vec{v}} \,,\qquad
0 = \frac{\partial\mathcal{E}}{\partial\vec{u}} \,,\qquad
0 = \frac{\partial\mathcal{E}}{\partial\vec{f}} \,,\qquad
0 = \frac{\partial\mathcal{E}}{\partial\vec{h}} \,, \qquad
0 = \frac{\partial\mathcal{E}}{\partial\vec{e}} \,.
\end{equation}
By solving this system one obtains $\vec{v}$ ,$\vec{u}$, $\vec{f}$ and $\vec{e}$ as functions 
of charges $\vec{q}$ and $\vec{p}$.

Finally, it was shown in \cite{Sen:2005wa} that the the value of the entropy function at the
extremum gives the same result as black hole entropy calculated from Wald formula 
(\ref{Went}), i.e., 
\begin{equation} \label{efent}
S_{\rm bh}(\vec{q},\vec{p})
 = \mathcal{E} \quad \textrm{(evaluated at the solution of (\ref{efeom}))} \,.
\end{equation}

Comments on the entropy function method:
\begin{enumerate}
\item
It enormously simplifies calculation of near-horizon geometry and entropy, as it turns solving
of system of differential equations into solving of system of algebraic equations.
\item
Manifest gauge and diffeomorphism invariance neccessary. If there are Chern-Simons terms
of any kind (gravitational, gauge, or mixed) additional labor is necessary \cite{Sen:2007qy}.
One idea is to use dimensional reduction to write such terms in the manifestly covariant form.
We shall use this idea in section \ref{sec:full} to handle mixed Chern-Simons term that is 
present in heterotic theory (pure gauge terms, which also exist, will be vanishing in our 
examples).
\item
When one has just the solution with symmetries expected of near-horizon solution, it is not guaranteed that for every such solution there is indeed full black hole solution with such near-horizon behavior. However, for large black holes treated here we know that such correspondence
exist in the lowest-order (because there are explicit complete solutions), and as corrections to 
the near horizon geometry are regular we can expect that this correspondence continues to apply 
at least perturbatively. For small black holes, in which $\alpha'\to0$ limit is singular, we 
cannot be that sure.
\end{enumerate}

\subsection{Field redefinitions}
\label{ssec:fred}

We shall be dealing with tree-level effective action of heterotic string theory which, as discussed  
in section \ref{sssec:corr} has infinite expansion in derivatives (parametrized by $\alpha'$) 
(\ref{treeact}). For such theories, there is no uniquely preferred choice for fields. If we start with 
some set of fields $\phi_i$, we can always make a \emph{field redefinitions} of the type
\begin{equation} \label{fieldred}
\phi_i \to \phi'_i = \phi_i + \sum_{n=1}^\infty \alpha'^n f_i^{(n)}[\phi] \,.
\end{equation}
The Lagrangian, written in transformed fields $\phi'_i$, will have generally different form (except
for the lowest order 2-derivative part which obviously stays unchanged). It can be shown that
perturbative properties are not changed. Important examples are S-matrix and Wald entropy,
which are both perturbatively invariant on field redefinitions. One can use field redefinitions to
make Lagrangian (or some part of it) look "nicer" or more "symmetric" which can simplify some
calculations. In the following sections we shall use the field redefinition freedom.

We emphasize the following points, which are sometimes overlooked in the literature:
\begin{enumerate}
\item
Monomials in Lagrangians are divided into two groups, (a) those whose coefficients are unchanged 
by any field redefinition (\emph{unambiguous terms}), (b) those whose coefficients change under
some field redefinitions \emph{ambiguous terms}). Though by using field redefinitions we can 
\emph{individually} kill every ambiguous term, it is not generally the case that we can kill them 
all \emph{simultaneously}. Ambiguous terms which can be killed simultaneously are called 
\emph{irrelevant}. For some Lagrangians there are ambiguous terms which are \emph{relevant}. It
should be emphasized that relevant ambiguous terms are \emph{equally important} as unambiguous 
terms (which are obviously relevant). Indeed, as shown in \cite{Metsaev:1987zx}, the heterotic 
effective action is giving us a nice example, in which already at first-order in the $\alpha'$-expansion 
(4-derivative terms) relevant ambiguous terms are present.
\item
If we are interested in \emph{non}perturbative results, we have to be more careful when applying
field redefinitions. For example, higher derivatives, typically present in $f_i^{(n)}[\phi]$ in 
(\ref{fieldred}), are generally introducing new degrees of freedom which obviously means that 
transformed Lagrangian will not describe the same theory. In such cases we have to be sure that 
field redefinition is \emph{regular}, not introducing new degrees of freedom or some other 
anomalies. This is why we have to be careful with the choice of field redefinition scheme when we 
treat small black holes (which are $\alpha'$-nonperturbative objects, as we argued 
before).\footnote{Though we emphasize again that $\alpha'$-expansion in effective action is 
not expected to make sense outside the perturbative regime in which corrections are small.}
\end{enumerate}

\subsection{Dualities}
\label{ssec:dual}

In this review we are explicitly analyzing two types of large black holes in heterotic string theory: 
3-charge in 5-dimensions ($T^4 \times S^1$ compactification) and 4-charge in 4-dimensions 
($T^4 \times S^1\times S^1$ compactification).\footnote{And also large black holes in NS-NS 
sector of type-II string theories compactified in the same way.} However, we can use various 
duality relations present in string theory to generalize results to more black holes with more 
charges and/or to different compactifications. This is especially easy for the black hole entropy 
which should be invariant on dualities. 

We have already mentioned one example of this - $T^4 \times S^1$ and 
$T^4 \times S^1\times S^1$ compactifications are special cases of compactifications on torii $T^5$ 
and $T^6$, respectively, which lead to $O(5,21)$ and $O(6,22)$, respectively, T-duality
symmetries of corresponding tree-level effective actions. A consequence of these symmetries is
that if we organize charges in two vectors $\vec{Q}$ (electric charges) and $\vec{P}$ (magnetic
charges) then black hole entropy can be function just of invariant scalar products $Q^2$, $P^2$
and $Q \cdot P$. As 4-dimensional 4-charge black hole solution has $Q^2=nw$, 
$P^2=\widehat{N}\widehat{W}$ and $Q \cdot P=0$, this means that for general large black holes 
satisfying $Q \cdot P=0$ entropies can be constructed from 4-charge formulas just by making
substitutions $nw \to Q^2$ and $\widehat{N}\widehat{W} \to P^2$. With some additional effort (by
adding just one more charge) we could obtain completely general "generating solutions", but this
complicates manipulation of mixed Chern-Simons term. Similar analysis applies to 5-dimensional 
black holes present in $T^5$ compactification. 

We can also apply chains of dualities to connect black holes in heterotic theory on 
$T^4 \times S^1$ and $T^4 \times S^1\times S^1$ with black holes in type-II theories on 
$K3 \times S^1$ and $K3 \times S^1\times S^1$. Let us take first 5-dimensional 3-charge black
holes which are microscopically constituted of elementary string (F1), NS5-branes (NS5), and 
momentum along $S^1$ (P). Then we have the following chain: [Heterotic on $T^4 \times S^1$]
(NS5,P,F1) $\longleftrightarrow$ [IIA on $K3 \times S^1$] (F1,P,NS5) 
$\stackrel{\rm T}{\longleftrightarrow}$ [IIB on $K3 \times S^1$] (P,F1,NS5)
$\stackrel{\rm S}{\longleftrightarrow}$ [IIB on $K3 \times S^1$] (P,D1,D5). We denoted D1- and
D5-branes with D1 and D5, T denotes T-duality on $S^1$, and S is S-duality. Similar chains can 
be constructed for 4-dimensional 4-charge black holes. What is interesting
is that though black hole entropies stay formally invariant, above dualities change interpretations 
of regimes for charges (they exchange strong/weak coupling which affects use of tree-level 
effective action, or small/large radii which, e.g., affects region of validity of Cardy formula).

\section{Full heterotic effective action}
\label{sec:full}

In this section we mainly review the results from \cite{Prester:2008iu,Prester:2009mc}, which we 
present in more complete form (additional solutions are shown). 

\subsection{The action and one conjecture}
\label{ssec:10Daction}

We now start with systematic analyses of $\alpha'$-corrections from the gravity side, which
means starting from the tree-level effective action of (appropriately compactified) heterotic 
string theory, solving for extremal black holes and calculating the corresponding entropies. To simplify the calculations, we concentrate immediately on the near-horizon behavior and use
Sen's entropy function formalism.

Let us start with what is known about the structure of tree-level effective action of 
heterotic string theory in $D=10$ dimensions. It has the general structure
\begin{equation} \label{lalphex}
\mathcal{S}^{(10)} = \int dx^{10} \sqrt{-G^{(10)}} \mathcal{L}^{(10)}
  = \sum_{n=0}^\infty \int dx^{10} \sqrt{-G^{(10)}} \mathcal{L}^{(10)}_n \;,
\end{equation}
Again, as we explained at the beginning of section \ref{ssec:loleea}, we are interested in such configurations for which the only non-vanishing elementary fields are string metric 
$G^{(10)}_{MN}$, dilaton $\Phi^{(10)}$ and 2-form $B^{(10)}_{MN}$. In this case every 
$\mathcal{L}^{(10)}_n$ is a function of the string metric $G^{(10)}_{MN}$, Riemann tensor 
$R^{(10)}_{MNPQ}$, dilaton $\Phi^{(10)}$, 3-form gauge field strength $H^{(10)}_{MNP}$ and the covariant derivatives of these fields. 10-dimensional space-time indices are denoted as 
$M,N,\ldots = 0,1,\ldots,9$. The term $\mathcal{L}^{(10)}_n$ has $2(n+1)$ derivatives, and is multiplied with a factor of $\alpha'^n$.

Ten-dimensional Lagrangian can be decomposed in the following way
\begin{equation} \label{10dtl}
\mathcal{L}^{(10)} = \mathcal{L}^{(10)}_{01}
 + \Delta \mathcal{L}^{(10)}_{\rm CS} + \mathcal{L}^{(10)}_{\rm other} \;.
\end{equation}
The first term in (\ref{10dtl}), explicitly written, is
\begin{equation} \label{l100}
\mathcal{L}^{(10)}_{01} = \frac{e^{-2\Phi^{(10)}}}{16\pi G_{10}} \left[ R^{(10)} +
 4\left(\partial \Phi^{(10)}\right)^2
 - \frac{1}{12} \overline{H}^{(10)}_{MNP} \overline{H}^{(10)MNP} \right] \;,
\end{equation}
where $G_{10}$ is 10-dimensional Newton constant.
3-form gauge field strength is not closed, but instead given by
\begin{equation} \label{hbcs}
\overline{H}^{(10)}_{MNP} = \partial_M B^{(10)}_{NP} + \partial_N B^{(10)}_{PM}
 + \partial_P B^{(10)}_{MN} - 3 \alpha' \overline{\Omega}^{(10)}_{MNP} \;,
\end{equation}
where $\overline{\Omega}^{(10)}_{MNP}$ is the gravitational Chern-Simons form
\begin{equation} \label{gcs}
\overline{\Omega}^{(10)}_{MNP} = \frac{1}{2} \, \overline{\Gamma}^{(10)R}_{\quad\;\;\;\; MQ} \,
 \partial_N \overline{\Gamma}^{(10)Q}_{\quad\;\;\;\; PR} \, 
 + \, \frac{1}{3} \, \overline{\Gamma}^{(10)R}_{\quad\;\;\;\; MQ} \,
 \overline{\Gamma}^{(10)Q}_{\quad\;\;\;\; NS} \,
 \overline{\Gamma}^{(10)S}_{\quad\;\;\;\; PR} \;\; \mbox{(antisym. in $M,N,P$)}
\end{equation}
Bar on the geometric object means that it is calculated using a modified connection
\begin{equation} \label{modcon}
\overline{\Gamma}^{(10)P}_{\qquad MN} = \Gamma^{(10)P}_{\quad\;\; MN}
  - \frac{1}{2} \overline{H}^{(10)P}_{\qquad MN}
\end{equation}
in which $\overline{H}$ plays the role of a torsion. It is believed that Chern-Simons terms appear 
exclusively through Eq. (\ref{hbcs}).

Let us note here in passing that a definition of strength of Kalb-Ramond 2-form $B_{MN}$ given 
in (\ref{hbcs}) is different from the standard one, which is
\begin{equation} \label{hcs}
H^{(10)}_{MNP} = \partial_M B^{(10)}_{NP} + \partial_N B^{(10)}_{PM}
 + \partial_P B^{(10)}_{MN} - 3 \alpha' \Omega^{(10)}_{MNP} \;.
\end{equation}
Here $\Omega^{(10)}_{MNP}$ is gravitational Chern-Simons form corresponding to standard 
Christoffel connection $\Gamma^{(10)M}_{\quad\;\;\;\; NP}$, i.e.,
\begin{equation} \label{cgcs}
\Omega^{(10)}_{MNP} = \frac{1}{2} \, \Gamma^{(10)R}_{\quad\;\;\;\; MQ} \,
 \partial_N \Gamma^{(10)Q}_{\quad\;\;\;\; PR} \, 
 + \, \frac{1}{3} \, \Gamma^{(10)R}_{\quad\;\;\;\; MQ} \,
 \Gamma^{(10)Q}_{\quad\;\;\;\; NS} \, \Gamma^{(10)S}_{\quad\;\;\;\; PR} \;\;
 \mbox{(antisym. in $M,N,P$)}
\end{equation}
We shall discuss in Sec. \ref{ssec:chdef} important difference between these two definitions.

If in (\ref{hbcs}) the Chern-Simons form $\overline{\Omega}^{(10)}_{MNP}$ would be absent, 
then (\ref{l100}) would be equal to (\ref{10dl0}), i.e., we would have 
$\mathcal{L}^{(10)}_{01} = \mathcal{L}^{(10)}_0$ in (\ref{lalphex}). Its presence introduces 
non-trivial $\alpha'$-corrections. Beside, as shown in \cite{Bergshoeff:1989de}, 
supersymmetrization (on-shell completion of $\mathcal{N}=1$ SUSY) of the Chern-Simons term 
introduces a (probably infinite) tower of terms in the effective action (with increasing number 
of derivatives), denoted by $\Delta \mathcal{L}^{(10)}_{\rm CS}$ in (\ref{10dtl}). The first 
two non-vanishing terms (in expansion in $\alpha'$) are
\begin{equation} \label{l101}
\Delta \mathcal{L}^{(10)}_{\mathrm{CS},1}  = \frac{\alpha'}{8}
 \frac{e^{-2\Phi^{(10)}}}{16\pi G_{10}} \overline{R}^{(10)}_{MNPQ} \overline{R}^{(10)MNPQ} 
\end{equation}
and
\begin{equation} \label{l103cs}
\Delta \mathcal{L}^{(10)}_{\mathrm{CS},3} = -\frac{\alpha'^3}{64}
 \frac{e^{-2\Phi^{(10)}}}{16\pi G_{10}}
 \left( 3\, T_{MNPQ}\, T^{MNPQ} + T_{MN}\, T^{MN} \right)
\end{equation}
where
\begin{equation} \label{tten}
T_{MNPQ} \equiv \overline{R}_{[MN}^{(10)\; RS} \, \overline{R}^{(10)}_{PQ]RS} \;,
 \qquad
T_{MN} \equiv \overline{R}_{MP}^{(10)\; QR}\, \overline{R}^{(10)P}_{\qquad NQR} \;.
\end{equation}
Though higher terms present in $\Delta \mathcal{L}^{(10)}_{\rm CS}$ were not explicitly 
constructed, it was argued in \cite{Bergshoeff:1989de} that $\alpha'^n$ contribution should be 
a linear combination of monomials containing $n$ Riemann tensors $\overline{R}_{MNPQ}$ 
calculated from the connection with torsion as given in (\ref{modcon}). This is the key 
information for us. All large black hole near-horizon solutions that we construct and analyze 
here have the property that $\overline{R}_{MNPQ}$ evaluated on them \emph{vanishes}, which 
means that all these terms, including (\ref{l101}) and (\ref{l103cs}), will be irrelevant in 
our calculations (giving vanishing contribution to equations of motion and entropy).

It is well-known that, beside terms connected with Chern-Simons term by supersymmetry, 
additional terms appear in the effective action starting from $\alpha'^3$ (8-derivative) order. 
In (\ref{10dtl}) we have denoted them with $\mathcal{L}^{(10)}_{\rm other}$. One well-known 
example is $R^4$-type (unambiguous) term multiplied by $\zeta(3)$, which appears in all string 
theories. Unfortunately, the knowledge of structure of $\mathcal{L}^{(10)}_{\rm other}$
is currently highly limited, and only few terms have been unambiguously calculated.

From now on, we are going to neglect contributions coming from $\mathcal{L}^{(10)}_{\rm other}$. 
One motivation is following from AdS$_3$/CFT$_2$ correspondence and anomaly inflow 
arguments of \cite{Kraus:2005vz}. There was argued (from 3-dimensional perspective) that for 
geometries having $AdS_3$ factor only Chern-Simons terms are important for calculations of 
central charges (from which one can calculate the black hole entropy). 
$\mathcal{L}^{(10)}_{\rm other}$ neither contains Chern-Simons terms nor is connected by 
supersymmetry to them, it should be irrelevant in such calculations. AdS$_3$/CFT$_2$ argument 
is sometimes used to explain successes of $R^2$-truncated actions (supersymmetric and/or 
Gauss-Bonnet) in calculations of entropies of BPS black holes in $D=4$ and 5. However, 
AdS$_3$/CFT$_2$ argument is relying on supersymmetry and can be confidently used only when
corresponding AdS$_3$ background is supersymmetric (current proofs require $(0,2)$ SUSY). In 
fact, we shall see that results for black hole entropy show that the argument cannot be used in
such non-BPS cases. Also, it would be interesting to have an argument which is not using any 
other information but the structure of effective action in $D=10$. 

In fact, there is such direct argument. If it happens that $\mathcal{L}^{(10)}_{\rm other}$ 
could be written in such a way that every monomial in it contains two powers of 
$\overline{R}_{MNPQ}$, then it would be irrelevant for our calculations and our results would 
be undoubtedly $\alpha'$-exact. The argument is the same as the one we used for 
$\Delta \mathcal{L}^{(10)}_{\rm CS}$ two paragraphs above. Indeed, this property was 
conjectured long time ago, see e.g., \cite{Metsaev:1987zx}. There is a stronger form of the conjecture, which claims that $\mathcal{L}^{(10)}_{\rm other}$ is purely composed of ($G^{MN}$ contracted) products of $\overline{R}_{MNPQ}$, see, e.g., \cite{Kehagias:1997cq}. Though the 
current status of the conjecture appears to be somewhat controversial -- it was disputed in 
\cite{Frolov:2001xr,Peeters:2001ub}, but the most recent detailed calculations 
\cite{Richards:2008sa} of some 8-derivative corrections (some of them recalculating the ones 
from \cite{Frolov:2001xr,Peeters:2001ub}) are giving results \emph{in agreement} with the 
strong form of the conjecture. 

We shall assume that conjecture (at least in weaker form) is correct, which allows us to
neglect $\mathcal{L}^{(10)}_{\rm other}$ part of the Lagrangian, which then is allowing us to
calculate $\alpha'$-\emph{exact} near-horizon solutions and corresponding black hole entropies. 
As in this way we obtain results for the entropies which are in agreement with microscopic calculations, we can say that our results are speaking in favour of the conjecture (though, of course, are not proving it).

\subsection{Manipulating Chern-Simons terms in $D=6$}
\label{ssec:cs6d}

All configurations that we analyze in this paper have four spatial dimensions compactified 
on torus $T^4$, and are uncharged under Kaluza-Klein 1-form gauge fields originating from four
compactified dimensions. Taking from the start that corresponding gauge fields 
vanish\footnote{Such truncation is expected to be consistent.} one obtains that the effective 
action is the same as in the section \ref{ssec:10Daction}, but now considering all fields and 
variables to be 6-dimensional. Effectively, one just has to replace everywhere $(10)$ with 
$(6)$ and take indices corresponding to 6-dimensional space-time, i.e., 
$M,N,\ldots = 0,1,\ldots,5$). To shorten the expressions, we immediately fix the values of
Newton constant and $\alpha'$, which in our normalization take values 
$G_6=2$ and $\alpha'=16$.

Appearance of gravitational Chern-Simons term in (\ref{hbcs}) introduces two problems. One is 
that it introduces in the action terms which are not manifestly diff-covariant, and that prevents 
direct use of Sen's entropy function formalism. A second problem is that due to (\ref{modcon}) 
and (\ref{hbcs}) Chern-Simons term is mixed in a complicated way with other 
$\alpha'$-corrections. We handle these problems by using the following two-step procedure 
(introduced in \cite{Sahoo:2006pm}).

First, we introduce an additional 3-form $K^{(6)}=dC^{(6)}$ and put a theory in a classically
equivalent form in which Lagrangian is given by\footnote{Similar dual formulations are known for 
some time, see, e.g., \cite{Nishino:1991sr}.}
\begin{eqnarray}
\sqrt{-G^{(6)}} \widetilde{\mathcal{L}}^{(6)} &=&
 \sqrt{-G^{(6)}} \mathcal{L}^{(6)}
 + \frac{1}{(24\pi)^2} \epsilon^{MNPQRS} K^{(6)}_{MNP} \overline{H}^{(6)}_{QRS}
\nonumber \\
&& + \frac{3\alpha'}{(24\pi)^2} \epsilon^{MNPQRS} 
 K^{(6)}_{MNP} \overline{\Omega}^{(6)}_{QRS} \;,
\label{gnewact}
\end{eqnarray}
and where now $\overline{H}^{(6)}_{MNP}$ should not be treated as a gauge field strength
but as an auxiliary 3-form. Antisymmetric tensor density $\epsilon^{MNPQRS}$ is defined 
by $\epsilon^{012345}=1$. As a result, Chern-Simons term is now isolated
as a single $\alpha'^1$-correction, in a way which will eventually allow us to write it in a 
manifestly covariant form.

Before passing to a second step of the procedure from \cite{Sahoo:2006pm}, we need to 
isolate in (\ref{gnewact}) ordinary Chern-Simons term $\Omega^{(6)}$ (obtained from
standard Levi-Civita connection) from the rest by using (\ref{modcon}). The result is
\cite{Chemissany:2007he}
\begin{equation} \label{bcscs}
\overline{\Omega}^{(6)}_{MNP} = \Omega^{(6)}_{MNP} + \mathcal{A}^{(6)}_{MNP}
\end{equation}
where
\begin{eqnarray} \label{a6mnp}
 \mathcal{A}^{(6)}_{MNP} &=&
 \frac{1}{4} \partial_M \left( \Gamma^{(6)R}_{NQ} \overline{H}^{(6)Q}_{RP} \right)
 + \frac{1}{8} \overline{H}^{(6)R}_{MQ} \nabla_N \overline{H}^{(6)Q}_{RP}
 - \frac{1}{4} R^{(6)\; QR}_{MN} \overline{H}^{(6)}_{PQR} \nonumber \\
&& + \frac{1}{24} \overline{H}^{(6)R}_{MQ} \overline{H}^{(6)S}_{NR} \overline{H}^{(6)Q}_{PS}
 \quad \mbox{(antisymmetrized in $M,N,P$).}
\end{eqnarray}
Notice that when (\ref{a6mnp}) is plugged in (\ref{bcscs}), and this into (\ref{gnewact}),
which is then integrated to obtain the action, contribution from the first term in 
(\ref{a6mnp}) will, after partial integration, have a factor $dK^{(6)}$ which vanishes 
because $K^{(6)}$ is by definition exact form. We now see that $\mathcal{A}^{(6)}$ gives 
manifestly covariant contribution to the action.

Now we are ready to write 6-dimensional action
\begin{equation} \label{6dea}
\mathcal{S}^{(6)} = \int dx^{6} \sqrt{-G^{(6)}} \widetilde{\mathcal{L}}^{(6)} 
\end{equation}
in the form we are going to use extensively in the paper.
Using (\ref{10dtl}) and the above analysis, Lagrangian can be written in the following form
\begin{equation} \label{6dtl}
\widetilde{\mathcal{L}}^{(6)} = \widetilde{\mathcal{L}}^{(6)}_0
 + \widetilde{\mathcal{L}}^{(6)\prime}_1 + \widetilde{\mathcal{L}}^{(6)\prime\prime}_1
 + \Delta \mathcal{L}^{(6)}_{\rm CS} + \mathcal{L}^{(6)}_{\rm other} \;.
\end{equation}
First term is lowest order ($\alpha'^0$) contribution given by (\ref{l100})
\begin{equation} \label{6dtl0}
\widetilde{\mathcal{L}}^{(6)}_0 = \frac{e^{-2\Phi^{(6)}}}{32\pi} \left[ R^{(6)} +
 4\left(\partial \Phi^{(6)} \right)^2
 - \frac{1}{12} \overline{H}^{(6)}_{MNP} \overline{H}^{(6)MNP} \right]
 + \frac{\epsilon^{MNPQRS}}{(24\pi)^2 \sqrt{-G^{(6)}}} K^{(6)}_{MNP} \overline{H}^{(6)}_{QRS}
\end{equation}
For later convenience we have separated first-order terms in three parts. One is given by
\begin{equation} \label{6dtl1p}
\widetilde{\mathcal{L}}^{(6)\prime}_1 = \frac{\epsilon^{MNPQRS}}{12\pi^2 \sqrt{-G^{(6)}}}
 K^{(6)}_{MNP} \left( \frac{1}{8} \overline{H}^{(6)U}_{QT} \nabla_R \overline{H}^{(6)T}_{US}
 - \frac{1}{4} R^{(6)\; TU}_{QR} \overline{H}^{(6)}_{STU}
 + \frac{1}{24} \overline{H}^{(6)U}_{QT} \overline{H}^{(6)V}_{RU} \overline{H}^{(6)T}_{SV} \right)
\end{equation}
The second part, which contains gravitational Chern-Simons term and is not manifestly 
covariant, is given by
\begin{equation} \label{6dtl1pp}
\widetilde{\mathcal{L}}^{(6)\prime\prime}_1 = \frac{\epsilon^{MNPQRS}}{12\pi^2 \sqrt{-G^{(6)}}}
 K^{(6)}_{MNP} \Omega^{(6)}_{QRS} \;.
\end{equation}
Finally, the third part is contained in $\Delta \mathcal{L}^{(6)}_{\rm CS}$ (\ref{l101}). In 
\cite{Sahoo:2006pm} it was shown how to rewrite (\ref{6dtl1pp}) in the manifestly covariant form 
for the particular type of the backgrounds which includes those we shall analyze in this paper.

All in all, we shall start from the reduced action with Lagrangian given by
\begin{equation} \label{6dtlred}
\widetilde{\mathcal{L}}^{(6)}_{\rm red} = \widetilde{\mathcal{L}}^{(6)}_0
 + \widetilde{\mathcal{L}}^{(6)\prime}_1 + \widetilde{\mathcal{L}}^{(6)\prime\prime}_1 \;,
\end{equation}
and check if the near horizon solutions satisfy the condition $\overline{R}_{MNPQ} = 0$. If
this is satisfied, it follows immediately that they are also solutions of the action with Lagrangian
\begin{equation} \label{6dtlsusy}
\widetilde{\mathcal{L}}^{(6)}_{\rm susy} = \widetilde{\mathcal{L}}^{(6)}_{\rm red}
 + \Delta\mathcal{L}^{(6)}_{\rm CS} \;,
\end{equation}
and, under the above mentioned assumption on $\mathcal{L}^{(6)}_{\rm other}$, of the
full heterotic action (\ref{6dtl}).

\subsection{Issues with magnetic charges: $H_{MNP}$ vs. $\overline{H}_{MNP}$}
\label{ssec:chdef}

Generally, when some configuration carries magnetic charge $Q_m$, it can be obtained from 
magnetic flux of a corresponding $(p-1)$-form gauge field $A_{p-1}$
\begin{equation} \label{magch2}
Q = \frac{1}{\Omega_p} \oint_{\mathcal{S}} F_p \,.
\end{equation}
Here $F_p$ is $p$-form gauge field strength of the gauge field $A_{p-1}$, and $\mathcal{S}$ is a 
$p$-dimensional closed surface enclosing the object in question. To be sure that we are catching
the total charge $Q_\infty$ carried by the configuration, we have to take surface 
$\mathcal{S}$ to be in "asymptotic infinity" (we denote it $\mathcal{S}_\infty$). 

The problem is that we know solutions exactly only in near-horizon region (this is what entropy
function formalism is able) we can only evaluate integral in (\ref{magch2}) when surface 
$\mathcal{S}$ is in this region (we call such surface $\mathcal{S}_h$). The obtained "horizon 
charge" $Q_h$ is generally different from $Q_\infty$, and, for the topologies relevant to us, the
difference can be calculated from
\begin{equation} \label{difmch}
Q_\infty - Q_h = \frac{1}{\Omega_p} \int dF_p \,.
\end{equation}
Integration is over $(p+1)$-dimensional surface (e.g., fixed-time surface) which is bounded by 
$\mathcal{S}_\infty$ and $\mathcal{S}_h$. This integration includes an "intermediate region", 
in which exact solution is typically not known (for complicated higher-derivative theory). But,
if the gauge field strength is (at least on-shell) closed, so $dF_p = 0$ (which is certainly fulfilled 
when $F_d = dA_{p-1}$), then $Q_h = Q_\infty$ and we can safely use near-horizon charges
to represent the total charge of the configuration.\footnote{For the near-horizon backgrounds 
treated in this review, we have $dF_d=0$ for all gauge fields which means that $Q_h$ is 
well-defined (not depending on a particular choice for $\mathcal{S}_h$.}

Let us apply the above analysis to configurations which are magnetically charged under 
Kalb-Ramond 2-form gauge field $B_{MN}$. Now, first question is which of the 3-form strengths
should be used in (\ref{magch2}), $\overline{H}_{MNP}$ (defined in (\ref{hbcs})-(\ref{modcon})) or 
$H_{MNP}$ (defined in (\ref{hcs})-(\ref{cgcs}))? From (\ref{bcscs}) follows that they are related by
\begin{equation} \label{hhbar}
H_{MNP} = \overline{H}_{MNP} + 3 \alpha' \mathcal{A}_{MNP} \;.
\end{equation}
where $\mathcal{A}_{MNP}$ is given in (\ref{a6mnp}). Because the difference is higher-order in
derivatives (and manifestly diff-covariant after elimination of first term which will not contribute to 
integral in (\ref{magch2})), if the integral in (\ref{magch2}) is performed on $\mathcal{S}_\infty$ we 
are sure that either we use $\overline{H}_{MNP}$ or $H_{MNP}$ the result (total magnetic charge 
$N_\infty = \overline{N}_\infty \equiv Q_5$) will be the same.
But, when we move the integration surface in the near-horizon region 
($\mathcal{S}=\mathcal{S}_h$) there is no reason to expect this agreement any more, and indeed 
we shall see explicitly on some examples that (\ref{magch2}) leads to different results for 
$\overline{H}_{MNP}$ or $H_{MNP}$. To understand the situation we have to analyze 
the right-hand side of (\ref{difmch}) in both cases.

For $H_{MNP}$ from (\ref{hcs}) follows the well-known result
\begin{equation} \label{dH}
dH = \frac{3}{8} \alpha' \mbox{tr} (R \wedge R) \,.
\end{equation}
An important point is that on the right-hand side we have topological density, so we can calculate 
integral in (\ref{difmch}) by using any configuration which belongs to the same topology class as
the solution in question. For example, in the case of large black hole backgrounds, we can use
lowest-order ($\alpha'$-uncorrected) solution which is normally known in the whole space-time so
we can easily calculate integral in (\ref{difmch}). Especially simple case is the background of 
5-dimensional large black holes of section \ref{sssec:3cbh} - solution is in the trivial topology class, 
i.e., it can be continuously deformed into flat background (in the region of integration, which is 
between the horizon and asymptotically flat infinity), which means that integral in (\ref{difmch}) 
trivially vanishes and so $N_h = Q_5$.

As for $\overline{H}_{MNP}$, from (\ref{hbcs}) follows
\begin{equation} \label{dHbar}
d\overline{H} = \frac{3}{8} \alpha' \mbox{tr} (\overline{R} \wedge \overline{R}) \,.
\end{equation}
Though for all of our near-horizon solutions right-hand side of (\ref{dHbar}) vanishes, this is not 
the case in intermediate region and we expect that integral in (\ref{difmch}) does not vanish and so
$\overline{N}_h \ne Q_5$. We shall show this explicitly in our examples. The outcome is that 
$\overline{H}_{MNP}$ is generally not convenient for calculation of (total) magnetic charge. 
However, due to its meaning in supersymmetry algebra, we shall show in our examples that the corresponding horizon charge $\overline{N}_h$ is directly connected to levels of affine algebra of symmetries on worldsheet.

Let us emphasize that in the dual scheme described in Sec. \ref{ssec:cs6d} there are no such
uncertainties because instead of $B_{MN}$ we have as a gauge field 2-form $C_{MN}$ with the
3-form strength $K_{MNP}$ which is by definition closed. As this is also true also for 1-form gauge 
fields obtained by Kaluza-Klein compactification, we can safely use it to calculate charges in 
near-horizon region.

Let us mention that for topologically more complex geometries, presence of Chern-Simons terms
can introduce additional problems in the proper definitions of electric and magnetic charges, e.g., connected with large gauge transformations (see, e.g., \cite{deWit:2009de}). As this is not 
affecting analyses of any of examples treated in this review, we shall ignore it.

\subsection{Compactification to $D<6$}
\label{ssec:Dcomp}

Our main interest are black holes in $D=5$ and $D=4$ dimensions, so we 
consider further compactification on $(6-D)$ circles $S^1$. Using the standard
Kaluza-Klein compactification we obtain $D$-dimensional fields
$G_{\mu\nu}$, $C_{\mu\nu}$, $\Phi$, $\widehat{G}_{mn}$,
$\widehat{C}_{mn}$ and $A_\mu^{(i)}$ ($0\le\mu,\nu\le D-1$,
$D\le m,n\le 5$, $1\le i\le 2(6-D)$): 
\begin{eqnarray} \label{d6dD}
&& \widehat{G}_{mn} = G^{(6)}_{mn}\,, \quad
 \widehat{G}^{mn} = (\widehat{G}^{-1})^{mn} \,, \quad
 \widehat{C}_{mn} = C^{(6)}_{mn}\,,
\nonumber \\
&& A^{(m-D+1)}_\mu = \frac{1}{2} \widehat{G}^{nm} G^{(6)}_{n\mu} \,,
 \quad
 A^{(m-2D+7)}_\mu = \frac{1}{2} C^{(6)}_{m\mu}
 - \widehat{C}_{mn} A^{(n-D+1)}_\mu \, ,
\nonumber \\
&& G_{\mu\nu} = G^{(6)}_{\mu\nu} - \widehat{G}^{mn} G^{(6)}_{m\mu}
 G^{(6)}_{n\nu}\, ,
\nonumber \\
&& C_{\mu\nu} = C^{(6)}_{\mu\nu} - 4 \widehat{C}_{mn}
 A^{(m-D+1)}_\mu A^{(n-D+1)}_\nu
 - 2(A^{(m-D+1)}_\mu A^{(m-2D+7)}_\nu
 - A^{(m-D+1)}_\nu A^{(m-2D+7)}_\mu)
\nonumber \\
&& \Phi = \Phi^{(6)} - \frac{1}{2} \ln \mathcal{V}_{6-D} \,,
\end{eqnarray}
There is also (now auxiliary) field $\overline{H}^{(6)}_{MNP}$ which produces
$D$-dimensional fields $\overline{H}_{\mu\nu\rho}$, $\overline{H}_{\mu\nu m}$, 
$\overline{H}_{\mu mn}$ and $\overline{H}_{mnp}$. 
As in \cite{Sahoo:2006pm}, we take for the circle coordinates 
$0\le x^m<2\pi\sqrt{\alpha'}=8\pi$, so that the volume 
$\mathcal{V}_{6-D}$ is
\begin{equation} \label{volum}
\mathcal{V}_{6-D} = (8\pi)^{6-D} \sqrt{\widehat{G}} \;.
\end{equation}
The gauge invariant field strengths associated with
$A_\mu^{(i)}$ and $C_{\mu\nu}$ are
\begin{equation} \label{eag2a}
F^{(i)}_{\mu\nu} =
 \partial_\mu A^{(i)}_\nu - \partial_\nu A^{(i)}_\mu \, ,
 \qquad 1\le i,j\le 2(6-D) \, ,
\end{equation}
\begin{equation} \label{eag2b}
K_{\mu\nu\rho} = \left( \partial_\mu C_{\nu\rho}
 + 2 A_\mu^{(i)} L_{ij} F^{(j)}_{\nu\rho} \right)
 + \hbox{cyclic permutations of $\mu$, $\nu$, $\rho$} \, ,
\end{equation}
where
\begin{equation} \label{edefl}
L = \pmatrix{ 0 & I_{6-D} \cr I_{6-D} & 0} \, ,
\end{equation}
$I_{6-D}$ being a $(6-D)$-dimensional identity matrix.

For the black holes we are interested in, we have
\begin{equation}
A_\mu^{(i)} L_{ij} F_{\nu\rho}^{(j)} = 0 \;.
\end{equation}

Normally, the next step would be to perform the Kaluza-Klein reduction on the 6-di\-men\-sional 
low-energy effective action to obtain a $D$-dimensional effective action, which can be quite 
complicated. We shall follow a more efficient procedure \cite{Sahoo:2006pm} -- we go to $D$ 
dimensions just to use the symmetries of the action to construct an ansatz for the background 
($AdS_2\times S^{D-2}$ in our case), and then perform an uplift to 6 dimensions (by inverting 
(\ref{d6dD})) where the action is simpler and calculations are easier.

\subsection{4-dimensional 4-charge black holes in heterotic theory}
\label{ssec:4chbh}

Now we want to use the formalism we developed above for studying $\alpha'$-corrections to the
near-horizon geometry of extremal 4-dimensional 4-charge black holes appearing in the 
heterotic string theory compactified on $T^4\times S^1\times \widehat{S}^1$ introduced in 
section \ref{sssec:4cbh}. One can obtain an effective 4-dimensional theory by putting $D=4$ in 
(\ref{d6dD}) (using the formulation of the 6-dimensional action from section \ref{ssec:cs6d}) 
and taking as non-vanishing only the following fields: string metric $G_{\mu\nu}$, dilaton 
$\Phi$, moduli $T_1=(\widehat{G}_{44})^{1/2}$ and $T_2=(\widehat{G}_{55})^{1/2}$, four 
Kaluza-Klein gauge fields $A_\mu^{(i)}$ ($0\le\mu,\nu\le3$, $1\le i\le 4$) coming from 
$G^{(6)}_{MN}$ and 2-form potential $C^{(6)}_{MN}$, and two auxiliary 2-forms 
$D_{\mu\nu}^{(n)}$ ($n=1,2$) coming from $\overline{H}^{(6)}_{MNP}$ (which is now, as 
explained in section \ref{ssec:cs6d}, an auxiliary field).

The black holes we are interested in are charged purely electrically with respect to 
$A_\mu^{(1)}$ and $A_\mu^{(3)}$, and purely magnetically with respect to $A_\mu^{(2)}$ and 
$A_\mu^{(4)}$. As discussed before, from the heterotic string theory viewpoint, these black 
holes should correspond to 4-charge states in which, beside fundamental string wound around 
the $S^1$ circle (with coordinate $x^4$), and with nonvanishing momentum on it, there are 
also Kaluza-Klein and H-monopoles (NS5-branes) wound around $T^4\times S^1$ (with a "nut" 
on $\widehat{S}^1$).

For extremal black holes one expects $AdS_2\times S^2$ near-horizon geometry 
\cite{Kunduri:2007vf,Astefanesei:2007bf,Goldstein:2007km} which in the present case is given 
by:
\begin{eqnarray} \label{d6d4}
&& ds^2 \equiv G_{\mu\nu} dx^\mu dx^\nu = v_1 \left( -r^2 dt^2 + {dr^2\over r^2} \right)
 + v_2 (d\theta^2 + \sin^2\theta d\phi^2)\, , \nonumber \\
&& e^{-2\Phi} = u_S \,,\qquad T_1 = u_1 \,,\qquad T_2 = u_2 \,, \nonumber \\
&& F^{(1)}_{rt} = \widetilde{e}_1, \qquad F^{(3)}_{rt}
 = \frac{\widetilde{e}_3}{16} \,, \qquad
 F^{(2)}_{\theta\phi} = \frac{\widetilde{p}_2}{4\pi} \sin\theta \,,
 \qquad F^{(4)}_{\theta\phi} = \frac{\widetilde{p}_4}{64\pi}\sin\theta \,,\nonumber \\
&& D^{(1)\,rt} = \frac{2\,u_1^2\, h_1}{v_1 v_2 u_S} \,, \qquad
 D^{(2)\,\theta\phi} = -\frac{8\pi\, u_2^2\,h_2}{v_1 v_2 u_S \sin\theta} \,.
\end{eqnarray}
Here $v_1$, $v_2$, $u_S$, $u_n$, $\widetilde{e}_i$ and $h_n$ ($n=1,2$, $i=1,\ldots,4$) 
are unknown variables fixed by equations of motion and values of electric charges 
$\widetilde{q}_{1,3}$. Somewhat unusual normalization for $h_{1,2}$ is introduced for later 
convenience.

Once we have a background obeying the full group of symmetries of AdS$_2 \times S^2$ space,
we can use Sen's entropy function formalism reviewed in section \ref{ssec:senef}. We could 
calculate the entropy function from
\begin{equation} \label{entf}
\mathcal{E} = 2\pi \left( \sum_{I} \widetilde{q}_I \, \widetilde{e}_I
 - \int_{S^2} \sqrt{-G} \, \widetilde{\mathcal{L}} \right) \;,
\end{equation}
where $\widetilde{q}_I$ are electric charges, and $\widetilde{\mathcal{L}}$ is the 
effective Lagrangian in four dimensions. For this, we would need to calculate 
$\widetilde{\mathcal{L}}$ by doing dimensional reduction from six to four dimensions, which
would give us quite a complicated effective Lagrangian.

Instead of this, it is much easier to perform calculation of the entropy function 
$\mathcal{E}$ directly in six dimensions were we already know the action. For this, we have to 
lift the background to six dimensions, which for (\ref{d6d4}) gives
\begin{eqnarray} \label{d4d6}
&& ds_6^2 \equiv G^{(6)}_{MN} dx^M dx^N
 = ds^2 + u_1^2 \left( dx^4 + 2\widetilde{e}_1 rdt \right)^2
 + u_2^2 \left( dx^5 - \frac{\widetilde{p}_2}{2\pi} \cos\theta\, d\phi \right)^2 \,,
\nonumber \\
&& K^{(6)}_{tr4} = \frac{\widetilde{e}_3}{8} \,, \qquad
 K^{(6)}_{\theta\phi 5} = -\frac{\widetilde{p}_4}{32\pi} \sin\theta \,,
\nonumber \\
&& \overline{H}^{(6)tr4} = \frac{4\,h_1}{v_1 v_2 u_S} \,, \qquad
 \overline{H}^{(6)\theta\phi 5} = \frac{16\pi\,h_2}{v_1 v_2 u_S \sin\theta} \,,
\nonumber \\
&& e^{-2\Phi^{(6)}} =  \frac{u_S}{64\pi^2\, u_1 u_2} \,.
\end{eqnarray}
Instead of $\widetilde{\mathcal{L}}$ and $G$ we now use in (\ref{entf}) the six dimensional 
Lagrangian $\widetilde{\mathcal{L}}^{(6)}$ given in (\ref{6dtl})-(\ref{6dtl1pp}) and the 
determinant $G^{(6)}$
\begin{equation} \label{entf6d}
\mathcal{E} = 2\pi \left( \sum_{I} \widetilde{q}_I \, \widetilde{e}_I
 - \int_{S^2 \times S^1 \times \widehat{S}^1} \sqrt{-G^{(6)}} \,
 \widetilde{\mathcal{L}}^{(6)} \right) \;.
\end{equation}
This is obviously equivalent to (\ref{entf}). Equations of motion turn into 
extremization of the entropy function (\ref{entf}) over variables 
$\{\varphi_a\}=\{v_1,v_2, u_S, u_n,\widetilde{e}_i,h_n\}$,
\begin{equation} \label{geneom}
0 = \frac{\partial\mathcal{E}}{\partial\varphi_a} \,
 \Bigg|_{\varphi=\bar{\varphi}} \;.
\end{equation}
The black hole entropy is given by the value of the entropy function at the extremum
\begin{equation} \label{bhgen}
S_{\rm bh} = \mathcal{E}(\bar{\varphi}) \;,
\end{equation}
which is a function of electric and magnetic charges only.

As we discussed in sections \ref{ssec:10Daction} and \ref{ssec:cs6d}, we concentrate on the part 
of the action connected by 10-dimensional supersymmetry with Chern-Simons term (obtained by 
neglecting $\mathcal{L}^{(6)}_{\rm other}$ in (\ref{6dtl})). For the moment we also neglect 
$\Delta\mathcal{L}^{(6)}_{\rm CS}$, for which we show a posteriori that it does not contribute to 
the near-horizon solutions and the entropies. This means that we start with the reduced Lagrangian 
$\widetilde{\mathcal{L}}^{(6)}_{\rm red}$ defined by (\ref{6dtlred}), (\ref{6dtl0}), (\ref{6dtl1p}) and 
(\ref{6dtl1pp}). Putting (\ref{d4d6}) in (\ref{6dtlred}), and then this into the entropy function 
(\ref{entf6d}), we obtain
\begin{equation} \label{entfcs}
\mathcal{E} = \mathcal{E}_0 + \mathcal{E}'_1 + \mathcal{E}''_1 \,,
\end{equation}
where
\begin{eqnarray} \label{ef4c0}
\mathcal{E}_0 &=& 2\pi \left[ \widetilde{q}_1 \widetilde{e}_1
 + \widetilde{q}_3 \widetilde{e}_3
 - \int d\theta\, d\phi\, dx^4 dx^5 \sqrt{-G^{(6)}} \widetilde{\mathcal{L}}^{(6)}_0 \right]
\nonumber \\
 &=& 2\pi \left[ \widetilde{q}_1 \widetilde{e}_1 + \widetilde{q}_3 \widetilde{e}_3
  - \frac{1}{8} v_1 v_2 u_S \left( -\frac{2}{v_1} + \frac{2}{v_2}
 + \frac{2 u_1^2\, \widetilde{e}_1^2}{v_1^2}
 + \frac{128\pi^2 u_2^2 h_2 (2\widetilde{e}_3 - h_2)}{v_1^2\, u_S^2}
 \right. \right.
\nonumber \\
&& \left. \left. \qquad
 - \frac{u_2^2\, \widetilde{p}_2^2}{8\pi^2 v_2^2}
 - \frac{8 u_1^2 h_1 (2\widetilde{p}_4 - h_1)}{v_2^2\, u_S^2} \right)
 \right] \,,
\end{eqnarray}
and
\begin{eqnarray} \label{ef4c1}
\mathcal{E}'_1 &=& -2\pi \int d\theta\, d\phi\, dx^4 dx^5 \sqrt{-G^{(6)}}
 \widetilde{\mathcal{L}}^{(6)\prime}_1
\nonumber \\
&=& -4\pi v_1 v_2 u_S \left( \frac{8192\pi^4 u_2^4 \widetilde{e}_3 h_2^3}{v_1^4\, u_S^4}
 + \frac{8 u_2^4 \widetilde{e}_3 h_2 \widetilde{p}_2^2}{v_1^2\, v_2^2\, u_S^2} 
 - \frac{128\pi^2 u_2^2 \widetilde{e}_3 h_2}{v_1^2\, v_2\, u_S^2} \right.
\nonumber \\
&& \qquad\qquad\qquad \left.  + \frac{32 u_1^4 \widetilde{p}_4 h_1^3}{v_2^4\, u_S^4}
 + \frac{8 u_1^4 \widetilde{e}_1^2 h_1 \widetilde{p}_4}{v_1^2\, v_2^2\, u_S^2}
 - \frac{8 u_1^2 \widetilde{p}_4 h_1}{v_1\, v_2^2\, u_S^2}
 \right) \,.
\end{eqnarray}
With $\mathcal{E}''_1$, defined by
\begin{equation} \label{ef4c12d}
\mathcal{E}''_1 = -2\pi \int d\theta\, d\phi\, dx^4 dx^5 \sqrt{-G^{(6)}}
 \widetilde{\mathcal{L}}^{(6)\prime\prime}_1 \,,
\end{equation}
the situation is a bit tricky because of the presence of Chern-Simons
density in (\ref{6dtl1pp}). This means that $\widetilde{\mathcal{L}}^{(6)\prime\prime}_1$ is 
not manifestly diffeomorphism covariant, and one cannot apply directly Sen's entropy function 
formalism. Fortunately, this problem was solved in \cite{Sahoo:2006pm} where it was shown how 
for the class of the metrics, to which (\ref{d4d6}) belongs, one can write $\mathcal{E}''_1$ 
in a manifestly covariant form. 

Next, notice that the background (\ref{d5d6}) has a form of a product of two 3-dimensional 
backgrounds, the first one is on $(t,r,x^5)$ space (AdS$_2\times S^1$) and the second one on 
$(\theta,\phi,x^4)$ space ($S^2\times\widehat{S}^1$). From this follows
\begin{equation} \label{ef4c1ppf}
\mathcal{E}_1'' = \frac{1}{6\pi} \int d\theta d\phi dx^4 dx^5 \epsilon^{ijk} \epsilon^{abc}
 \left( K^{(6)}_{ijk} \Omega^{(6)}_{abc} - \Omega^{(6)}_{ijk} K^{(6)}_{abc} \right) \,,
\end{equation}
where $\{a,b,c\}=\{t,r,4\}$ and $\{i,j,k\}=\{\theta,\phi,5\}$, and the convention for the antisymmetric 
tensor densities is
\begin{equation}
\epsilon^{tr4} = 1 \,, \qquad \epsilon^{\theta\phi5} = 1 \,.
\end{equation}
Furthermore, Kaluza-Klein compactification is performed on $x^4$ and $x^5$ which
leaves us with 4-dimensional effective space. So, for our purposes it would be enough to have 
result which is manifestly covariant in two reduced 2-dimensional spaces (AdS$_2$ and $S^2$).

In three dimensions it is known \cite{Guralnik:2003we,Sahoo:2006vz} that for the metrics of 
the "Kaluza-Klein form"
\begin{equation} \label{3KK2}
ds^2  = \phi(x)
 \left[ g_{mn}(x) dx^m dx^n + \left(dy + 2A_m(x) dx^m\right)^2 \right]
 \,, 
\end{equation}
where $0\le m,n\le1$, we have (modulo total derivative terms)
\begin{equation} \label{CS3cov}
\epsilon^{\alpha\beta\gamma} \Omega_{\alpha\beta\gamma} = 
 \frac{1}{2} \epsilon^{mn} \left[ R^{(2)} F_{mn}
 + 4 g^{m'p'} g^{q'q} F_{mm'} F_{p'q'} F_{qn} \right] \;,
\end{equation}
where $F_{mn}=\partial_m A_n - \partial_n A_m$, $\epsilon^{mn}$ is antisymmetric with 
$\epsilon^{01}=1$, and $R^{(2)}$ is a Ricci scalar obtained from $g_{mn}$. (\ref{CS3cov}) 
gives us the desired \emph{manifestly covariant form} (in the reduced 2-dimensional space) for 
the gravitational Chern-Simons term.

Using (\ref{CS3cov}) for AdS$_2\times S^1$ and $S^2\times\widehat{S}^1$ separately, it is now 
easy to obtain \cite{Sahoo:2006pm}
\begin{equation} \label{ef4c12}
\mathcal{E}''_1 = - (8\pi)^2 \left[ \frac{\widetilde{p}_4}{4\pi}
 \left( \frac{u_1^2}{v_1} \widetilde{e}_1 - 2 \frac{u_1^4}{v_1^2}\,\widetilde{e}_1^3 \right)
 + \widetilde{e}_3 \, \left( \frac{u_2^2}{v_2} \, \frac{\widetilde{p}_2}{4\pi}
 - 2 \frac{u_2^4}{v_2^2} \, \left( \frac{\widetilde{p}_2}{4\pi} \right)^3 \right) \right] \,.
\end{equation}

We are now ready to find near-horizon solutions, by solving the system (\ref{geneom}), and 
black hole entropy from (\ref{bhgen}). As we want to compare the results with the statistical
entropy obtained in string theory by counting of microstates, it is convenient to express 
charges $(\widetilde{q},\widetilde{p})$ in terms of (integer valued) charges naturally 
appearing in the string theory. By comparing the lowest-order solution (one uses just 
(\ref{ef4c0}), which is an easy exercise) with the near-horizon solution (\ref{4c4d0}) we 
obtain
\begin{equation} \label{chnorm}
\widetilde{q}_1 = \frac{n}{2} \,,\qquad \widetilde{p}_2 = 4\pi \widehat{N} \,,\qquad
 \widetilde{q}_3 = - 4\pi \widehat{W} \,,\qquad \widetilde{p}_4 = - \frac{w}{2} \,,
\end{equation}
where $n$ and $w$ are momentum and winding number of string wound along circle $S^1$, and 
$\widehat{N}$ and $\widehat{W}$ are Kaluza-Klein monopole and H-monopole charges associated 
with the circle $\widehat{S}^1$. As all gauge field strengths are closed, $\alpha'$-corrections will not introduce corrections in relations (\ref{chnorm})?

Using (\ref{entfcs})-(\ref{chnorm}) in (\ref{geneom}), we obtain quite a complicated algebraic
system, naively not expected to be solvable analytically. Amazingly, we have found analytic 
near-horizon solutions for all values of charges, corresponding to BPS and non-BPS black 
holes.\footnote{The way we constructed solutions was indirect - we managed to conjecture them 
from perturbative calculations (which we did up to $\alpha'^4$), and then checked them by putting 
into exact equations. For some special sets of charges we then numerically checked that there are 
no other physically acceptable solutions.} While in BPS case analytic solutions are expected 
because one can use BPS conditions to drastically simplify calculations, in non-BPS case in 
theories which involve higher-derivative corrections analytic solutions are typically not 
known\footnote{However, one exception can be found in \cite{Cvitan:2007en}.}.

We mentioned in section \ref{sssec:4cbh} that from the supersymmetry point of view there are 
three types of solutions, differing in relative signs of products $nw$ and 
$\widehat{N}\widehat{W}$. We now analyze them separately, by writing $\alpha'$-exact 
near-horizon solutions explicitly for representative cases of each type. We nore that 
expressions for entropies are representative independent.

The type-1 consists of supersymmetric 1/4-BPS large black holes for which $nw>0$, 
$\widehat{N}\widehat{W}>0$. For clarity of presentation, we take $n,w,\widehat{N}\widehat{W}>0$ 
as a representative of this type. The near-horizon solution is
then given by
\begin{eqnarray} \label{4solb}
&& v_1 = v_2 = 4(\widehat{N}\widehat{W}+2) \,,\qquad u_S
 = \sqrt{\frac{nw}{\widehat{N}\widehat{W}+4}} \,,
\nonumber \\
&& u_1 = \sqrt{\frac{n(\widehat{N}\widehat{W}+2)}{w(\widehat{N}\widehat{W}+4)}} \,,\qquad
 u_2 = \sqrt{\frac{\widehat{W}}{\widehat{N}} \left(1+\frac{2}{\widehat{N}\widehat{W}}\right)}
 \,, \\
&& \widetilde{e}_1 = \frac{1}{n} \sqrt{nw(\widehat{N}\widehat{W}+4)} \,,\qquad
 \widetilde{e}_3 = h_2 = - \frac{\widehat{N}}{8\pi} \sqrt{\frac{nw}{\widehat{N}\widehat{W}+4}}
 \,,\qquad
 h_1= - \frac{w}{2} \,.
\nonumber
\end{eqnarray}
For the entropy of type-1 black holes we obtain
\begin{equation} \label{4entb}
S_{\rm bh}^{\rm BPS} = 2\pi \sqrt{nw(\widehat{N}\widehat{W}+4)} \,, \qquad
 nw>0 \,, \quad \widehat{N}\widehat{W}>0
\end{equation}
This is exactly what one obtains by microstate counting in string theory (\ref{micro4}), in
the limit $nw \gg \widehat{N}\widehat{W}$, which corresponds to tree-level approximation on 
gravity side.

The type-2 consists of non-supersymmetric large black holes for which $nw<0$, 
$\widehat{N}\widehat{W}<0$. For clarity of presentation, we take $n,\widehat{N}<0$ and 
$w,\widehat{W}>0$ as a representative of this type. The near-horizon solution is then given 
by\footnote{While this review was in preparation this solution was presented in 
\cite{Bellucci:2009nn}.}
\begin{eqnarray} \label{4soln2}
&& v_1 = v_2 = 4|\widehat{N}\widehat{W}| \,,\qquad
 u_S = \sqrt{\left| \frac{nw}{\widehat{N}\widehat{W}} \right|} \,, \qquad
 u_1 = \sqrt{\left| \frac{n}{w} \right|} \,,\qquad
 u_2 = \sqrt{\bigg| \frac{\widehat{W}}{\widehat{N}} \bigg|}
 \,, \nonumber \\
&& \widetilde{e}_1 = \frac{1}{n} \sqrt{|nw\widehat{N}\widehat{W}|} \,,\qquad
 \widetilde{e}_3 = h_2 = - \frac{\widehat{N}}{8\pi} \sqrt{\left| \frac{nw}{\widehat{N}\widehat{W}}\right|}
 \,,\qquad h_1= - \frac{w}{2} \,.
\end{eqnarray}
For the entropy of type-2 black holes we obtain
\begin{equation} \label{4entn2}
S_{\rm bh}^{\rm non-BPS} = 2\pi \sqrt{|nw\widehat{N}\widehat{W}|} \,, \qquad
 nw<0 \,, \quad \widehat{N}\widehat{W}<0 \,.
\end{equation}
Again, agreement with statistical calculation in string theory (\ref{micro4n2}) is exact in $\alpha'$.
Note that here both the near-horizon solution (\ref{4soln2}) and the black hole entropy 
(\ref{4entn2}) are $\alpha'$-\emph{uncorrected}. We shall comment this later.

The type-3 consists of non-supersymmetric large black holes for which 
$nw\widehat{N}\widehat{W}<0$. For clarity of presentation, we take $n<0$, 
$w,\widehat{N}\widehat{W}>0$ as a representative of this type. The near-horizon solution is
then given by
\begin{eqnarray} \label{4soln}
&& v_1 = v_2 = 4(\widehat{N}\widehat{W}+2) \,,\qquad
 u_S = \sqrt{\frac{|n|w}{\widehat{N}\widehat{W}+2}} \,,
\nonumber \\
&& u_1 = \sqrt{\frac{|n|}{w}} \,,\qquad\qquad\qquad\quad
 u_2 = \sqrt{\frac{\widehat{W}}{\widehat{N}} \left(1+\frac{2}{\widehat{N}\widehat{W}}\right)}
 \,, \\
&& \widetilde{e}_1 = \frac{1}{n} \sqrt{|n|w(\widehat{N}\widehat{W}+2)} \,,\qquad
 \widetilde{e}_3 = h_2 = - \frac{\widehat{N}}{8\pi} \sqrt{\frac{|n|w}{\widehat{N}\widehat{W}+2}}
 \,,\qquad h_1= - \frac{w}{2} \,.
\nonumber
\end{eqnarray}
For the entropy of type-3 black holes we obtain
\begin{equation} \label{4entn}
S_{\rm bh}^{\rm non-BPS} = 2\pi \sqrt{|nw|(|\widehat{N}\widehat{W}|+2)} \,, \qquad
 nw\widehat{N}\widehat{W}<0
\end{equation}
Again, agreement with statistical calculation in string theory (\ref{cft4n}) is exact in $\alpha'$.

Now we have to check that $\overline{R}^{(6)}_{MNPQ}$ vanishes when evaluated on our 
solutions. From (\ref{modcon}) one gets
\begin{equation} \label{modrie}
\overline{R}^{(6)M}_{\qquad NPQ}
 = R^{(6)M}_{\qquad NPQ} + \nabla_{[P} \overline{H}^{(6)M}_{\qquad Q]N}
 - \frac{1}{2} \overline{H}^{(6)M}_{\qquad R[P} \overline{H}^{(6)R}_{\qquad Q]N} \;.
\end{equation}
It is easy to show that all three solutions, (\ref{4solb}), (\ref{4soln}) and (\ref{4soln}), 
when used in 6-dimensional background (\ref{d4d6}) give
\begin{equation} \label{modrie0}
\overline{R}^{(6)}_{MNPQ} = 0 \;.
\end{equation}
As explained in section \ref{ssec:10Daction}, from this follows that inclusion of the term 
$\Delta\mathcal{L}^{(6)}_{\rm CS}$ does not change neither the near-horizon solutions 
(\ref{4solb}), (\ref{4soln2}) and (\ref{4soln}) nor the corresponding black hole entropies 
(\ref{4entb}), (\ref{4entn2}) and (\ref{4entn}), which means that all our results would be
obtained if we started with the more complicated supersymmetric Lagrangian (\ref{6dtlsusy}), 
constructed by supersymmetrizing gravitational Chern-Simons term.

It is interesting to calculate "horizon charge" $\overline{W}_h$ obtained from magnetic flux of 
$B^{(6)}_{MN}$ by using for strength 3-form $\overline{H}^{(6)}_{MN}$ 
\begin{equation} \label{Wbdef}
\overline{W}_h \equiv - \frac{1}{64\,\pi^2} \oint_{S^2 \times S^1} \overline{H}
\end{equation}
we obtain for type-1 and type-3 solutions
\begin{equation} \label{WbWh13}
\overline{W}_h =  \widehat{W} \left(1 + \frac{2}{|\widehat{N}\widehat{W}|}\right) \,,
\end{equation}
while for type-2 we obtain simply
\begin{equation} \label{WbWh2}
\overline{W}_h =  \widehat{W}  \,.
\end{equation}
By using $\overline{W}_h$ instead of $\widehat{W}$ we see that in non-BPS cases (type-2 and 
type-3) solutions and entropies have $\alpha'$-uncorrected form. As for the BPS (type-1) case, 
black hole entropy is
\begin{equation} \label{4entbbar}
S_{\rm bh}^{\rm BPS} = 2\pi \sqrt{nw(\widehat{N}\overline{W}_h+2)} \,.
\end{equation}
Comparing the entropy formulas with AdS/CFT results \cite{Kutasov:1998zh}, one can read the 
meaning of $overline{W}_h$ - it is giving the total level $k$ of the world-sheet affine algebra 
$\widehat{SL(2)}$ in the supersymmetric sector through a relation $k = \widehat{N}\overline{W}_h$.

\subsection{5-dimensional 3-charge black holes in heterotic theory}
\label{ssec:3chbh}

Here we consider the 5-dimensional spherically symmetric 3-charge extremal 
black holes which appear in the heterotic string theory compactified on
$T^4\times S^1$. One can obtain an effective 5-dimensional theory by putting $D=5$ in
(\ref{d6dD}) (again using the formulation of the 6-dimensional action from section 
\ref{ssec:cs6d}) and taking as non-vanishing only the following fields: string metric
$G_{\mu\nu}$, dilaton $\Phi$, modulus $T=(\widehat{G}_{55})^{1/2}$, two Kaluza-Klein
gauge fields $A_\mu^{(i)}$ ($0\le\mu,\nu\le4$, $1\le i\le 2$) coming from 
$G^{(6)}_{MN}$ and 2-form potential $C^{(6)}_{MN}$, the 2-form potential $C_{\mu\nu}$
with the strength $K_{\mu\nu\rho}$, one Kaluza-Klein auxiliary two form $D_{\mu\nu}$
coming from $\overline{H}^{(6)}_{MNP}$, and auxiliary 3-form $\overline{H}_{\mu\nu\rho}$. 

The black holes we are interested in are charged purely electrically with respect to 
$A_\mu^{(i)}$, and purely magnetically with respect to $K_{\mu\nu\rho}$. From the 
heterotic string theory viewpoint, these black holes should correspond to 3-charge
states in which, beside fundamental string wound around $S^1$ circle with nonvanishing
momentum on it, there are NS5-branes wrapped around $T^4\times S^1$.

For extremal black holes we now expect\footnote{In $D=5$ there is no explicit proof that 
extremal asymptotically flat black holes must have AdS$_2\times S^3$ near-horizon geometry.
However, for the large black holes analyzed here one knows that lowest order solutions, which
were fully constructed, have such near-horizon behavior, and from continuity one expects the
same when $\alpha'$-corrections are included. Again, the situation is not that clear for 
small black holes, which we shall discuss later.} AdS$_2\times S^3$ near-horizon geometry 
which in the present case is given by:
\begin{eqnarray} \label{d6d5}
&& ds^2 \equiv G_{\mu\nu} dx^\mu dx^\nu
 = v_1\left(-r^2 dt^2 + {dr^2\over r^2}\right) + v_2 d\Omega_3 \,, 
\nonumber \\
&& F^{(1)}_{rt} = \widetilde{e}_1, \qquad
 F^{(2)}_{rt} = \frac{\widetilde{e_2}}{4} \,, \qquad
K_{234} = \frac{\widetilde{p}}{4} \sqrt{g_3} \,,
\nonumber \\
&& D^{rt} = \frac{2 u_T^2 h_1}{v_1 v_2^{3/2} u_S} \,,\qquad
 \overline{H}^{234} = - \frac{8 h_2}{v_1 v_2^{3/2} u_S \sqrt{g_3}} \,,
\nonumber \\
&& e^{-2\Phi} = u_S \,,\qquad\qquad T = u_T \,.
\end{eqnarray}
Here $g_3$ is a determinant of the metric on the unit 3-sphere $S^3$
(with coordinates $x^i$, $i=2,3,4$).

We follow the procedure from section \ref{ssec:4chbh}. Lift of (\ref{d6d5}) to six 
dimensions gives 
\begin{eqnarray} \label{d5d6}
&& ds_6^2 \equiv G^{(6)}_{MN} dx^M dx^N
 = ds^2 + u_T^2 \left( dx^5 + 2\widetilde{e}_1 rdt \right)^2 \,,
\nonumber \\
&& K^{(6)}_{tr5} = \frac{\widetilde{e}_2}{2} \,, \qquad\qquad\qquad
 K^{(6)}_{234} = K_{234} = \frac{\widetilde{p}}{4} \sqrt{g_3} \,,
\nonumber \\
&& \overline{H}^{(6)tr5} = \frac{4 h_1}{v_1 v_2^{3/2} u_S} \,, \qquad
 \overline{H}^{(6)234} = - \frac{8 h_2}{v_1 v_2^{3/2} u_S \sqrt{g_3}} \,,
\nonumber \\
&& e^{-2\Phi^{(6)}} =  \frac{u_S}{8\pi\, u_T} \,.
\end{eqnarray}
Now $v_1$, $v_2$, $u_S$, $u_T$, $\widetilde{e}_1$, $\widetilde{e}_2$, $h_1$
and $h_2$ are unknown variables whose solution is to be found by extremizing
the entropy function for the fixed values of electric and magnetic charges 
$\widetilde{q}_{1,2}$ and $\widetilde{p}$. Entropy function is now given by
\begin{eqnarray} \label{entf5d}
\mathcal{E} &=& 2\pi \left( \sum_{i=1}^2 \widetilde{q}_i \, \widetilde{e}_i
 - \int_{S^3} \sqrt{-G} \, \widetilde{\mathcal{L}} \right)
= 2\pi \left( \sum_{i=1}^2 \widetilde{q}_i \, \widetilde{e}_i
 - \int_{S^3} \sqrt{-G^{(6)}} \, \widetilde{\mathcal{L}}^{(6)} \right)
\nonumber \\
&=& \mathcal{E}_0 + \mathcal{E}'_1 + \mathcal{E}''_1 \,,
\end{eqnarray}
where
\begin{eqnarray} \label{ef50}
\mathcal{E}_0 &=& 2\pi \left[ \widetilde{q}_1\widetilde{e}_1
 + \widetilde{q}_2 \widetilde{e}_2 - \frac{\pi}{16} v_1 v_2^{3/2} u_S
 \left( -\frac{2}{v_1} + \frac{6}{v_2}
 + \frac{2 u_T^2 \widetilde{e}_1^2}{v_1^2}
 + \frac{32\, h_2 (2\widetilde{e}_2 - h_2)}{v_1^2\, u_S^2}
 \right. \right.
\nonumber \\
&& \left. \left. \qquad
 - \frac{8 u_T^2 h (2\widetilde{p} - h)}{v_2^3\, u_S^2} \right)
 \right] \,,
\end{eqnarray}

\begin{equation} \label{ef51p}
\mathcal{E}_1' = - 2\pi^2 v_1 v_2^{3/2} u_S
 \Bigg[ \frac{512\, e_2 h_2^3}{v_1^4 u_S^4}
 + \frac{32\, u_T^4 \widetilde{p}\, h_1^3}{v_2^6 u_S^4}
 + \frac{8 u_T^4 \widetilde{p}\, h_1 \widetilde{e}_1^2}{v_1^2 v_2^3 u_S^2}
 - \frac{8 u_T^2 \widetilde{p}\, h_1}{v_1 v_2^3 u_S^2}
 - \frac{96\, \widetilde{e}_2 h_2}{v_1^2 v_2 u_S^2} \Bigg] \,,
\end{equation}

\begin{equation} \label{ef51pp}
\mathcal{E}_1'' = - 8\pi^2 \widetilde{p} \left(
 \frac{u_T^2}{v_1} \widetilde{e}_1
 - 2 \frac{u_T^4}{v_1^2} \widetilde{e}_1^3 \right) \,.
\end{equation}
Again, to obtain (\ref{ef51pp}) we had to deal with gravitational Chern-Simons term, which
is done following the procedure reviewed in section \ref{ssec:4chbh}.\footnote{For the 3-sphere 
the Chern-Simons term vanishes
\begin{displaymath}
\int_{S^3} \epsilon^{ijk} \Omega_{ijk} = 0 \,.
\end{displaymath}
This is obvious if one calculates $\Omega_{ijk}$ using standard non-covariant formula (\ref{gcs}).
As sphere does not have boundaries, the inclusion of boundary total-derivative terms (which
"covariantize" CS term) cannot change the result.}

We are now ready to find near-horizon solutions, by solving the system (\ref{geneom}), and 
black hole entropy from (\ref{bhgen}). As we want to compare the results with the statistical
entropy obtained in string theory by counting of microstates, it is convenient to express 
charges $(\widetilde{q},\widetilde{p})$ in terms of (integer valued) charges naturally 
appearing in the string theory. By comparing the lowest-order solution (one uses just 
(\ref{ef50}), which is an easy exercise) with the near-horizon solution (\ref{3c5d0}) we 
obtain
\begin{equation} \label{3chnorm}
\widetilde{q}_1 = \frac{n}{2} \,,\qquad \widetilde{q}_2 = - 16\pi m
\,,\qquad \widetilde{p} = - \frac{w}{\pi} \,.
\end{equation}
Here $n$ and $w$ are momentum and winding number of string wound around $S^1$. We 
expect that $m$ denotes number of NS5-branes wrapped around $T^4\times S^1$. 

Again, we were able to find analytic solutions to algebraic system for all values of charges. As 
discussed in section \ref{sssec:3cbh}, from the supersymmetry viewpoint there are two types of 
black hole solutions which differ in sign of product $nw$. For clarity of presentation, we restrict to 
$w,m>0$. Then $n>0$ ($n<0$) correspond to 1/4-BPS (non-BPS) black holes.

In the BPS case ($n,w,m>0$) near-horizon solutions for 3-charge black holes
are given by
\begin{eqnarray} \label{3solb}
&& v_1 = 4(m+1) \,,\qquad v_2 = 4 v_1 \,,\qquad
 u_S = \frac{1}{8\pi} \sqrt{\frac{nw}{(m+1)(m+3)}} \,,\qquad
 u_T = \sqrt{\frac{n(m+1)}{w(m+3)}} \,,
\nonumber \\
&& \widetilde{e}_1 = \frac{1}{n} \sqrt{nw(m+3)} \,,\qquad
 \widetilde{e}_2 = h_2 = - \frac{1}{32\pi} \sqrt{\frac{nw}{m+3}} \,,\qquad
 h_1 = - \frac{w}{\pi} \,.
\end{eqnarray}
For the entropy we obtain
\begin{equation} \label{3entb}
S_{\rm bh}^{\rm BPS} = 2\pi \sqrt{nw(|m|+3)} \,, \qquad nw>0 \,,
\end{equation}
which agrees with microscopic statistical result obtained in \cite{Castro:2008ys}.

In the non-BPS case ($n<0$, $w,m>0$) we obtain
\begin{eqnarray} \label{3soln}
&& v_1 = 4(m+1) \,,\qquad v_2 = 4 v_1 \,,\qquad
 u_S = \frac{\sqrt{|n|w}}{8\pi(m+1)} \,,\qquad
 u_T = \sqrt{\frac{|n|}{w}} \,,
\nonumber \\
&& \widetilde{e}_1 = \frac{1}{n} \sqrt{|n|w(m+1)} \,,\qquad
 \widetilde{e}_2 = h_2 = - \frac{1}{32\pi} \sqrt{\frac{|n|w}{m+1}} \,,\qquad
 h_1 = - \frac{w}{\pi} \,.
\end{eqnarray}
For the entropy we obtain
\begin{equation} \label{3entn}
S_{\rm bh}^{\rm non-BPS} = 2\pi \sqrt{|nw|(|m|+1)} \,, \qquad nw<0 \,.
\end{equation}
This is exactly equal to the result conjectured in \cite{Cvitan:2007hu} (on the basis of
$\alpha'^3$-order perturbative results).

If someone is suspicious that magnetic charge $m$, obtained from the dual 3-form $K$, is indeed 
NS5-brane charge (which should be equal to the number of NS5-branes) which we denote by 
$Q_5$. As we discussed in Sec. \ref{ssec:chdef} $Q_5$ can be obtained from the magnetic flux 
calculated from 3-form $H^{(6)}$ in the near-horizon region. From solutions (\ref{3solb}) and 
(\ref{3soln}), and using (\ref{hhbar}) and (\ref{a6mnp}), we obtain
\begin{equation} \label{Hmc}
Q_5 = \frac{1}{64\pi^2} \oint_{S^3} H^{(6)} = m \,,
\end{equation}
which confirms the claim.

Again, it is interesting to calculate the horizon charge $\overline{N}_h$ obtained from 3-form 
strength $\overline{H}^{(6)}$. From solutions (\ref{3solb}) and (\ref{3soln}) we easily get
\begin{equation} \label{N5m}
\overline{N}_h \equiv \frac{1}{64\pi^2} \oint_{S^3} \overline{H}^{(6)} =  m+\frac{m}{|m|}\,.
\end{equation}
If we use $\overline{N}_h$ instead of $m=Q_5$, we see that in non-BPS case both solution and
entropy receive $\alpha'$-uncorrected form. In the BPS case, entropy is
\begin{equation} \label{3entbp}
S_{\rm bh}^{\rm BPS} = 2\pi \sqrt{nw(|\overline{N}_h|+2)} \,.
\end{equation}
Comparison of entropy formulas with AdS/CFT calculation \cite{Kutasov:1998zh} (presented in
(\ref{c3ch})-(\ref{cft3n})) is giving us the meaning of $\overline{N}_h$ - it is equal to the total 
level $k$ of the affine world-sheet symmetry algebra $\widehat{SL(2)}$ in the supersymmetric 
(right-moving) sector.  

Finally, it is easy to check that both BPS and non-BPS near-horizon solutions presented in this 
section satisfy 6-dimensional relation (\ref{modrie0}), which again means that inclusion of 
$\Delta\mathcal{L}^{(6)}_{\rm CS}$ in the action would not change our solutions and entropies 
(so they are also solutions of the action (\ref{6dtlsusy})).

\subsection{Comments on $\alpha'$-exact calculation using full effective action}
\label{ssec:comalpha}

\begin{itemize}
\item
All non-BPS solutions have $\alpha'$-\emph{un}corrected form in our scheme when we use, 
instead of NS5-brane charges, a horizon charges obtained from 3-form $\overline{H}$. Now, it 
was shown \cite{Tseytlin:1996as,Cvetic:1995bj} that lowest-order BPS solutions are $\alpha'$-exact solutions from the sigma model calculations. As we use different scheme, our solutions cannot be 
directly compared to sigma model ones. 
\item
The expressions for black hole entropies are in agreement with those obtained from AdS/CFT correspondence, using the results for central charges calculated in \cite{Kutasov:1998zh} (see 
section \ref{ssec:adscft} for more details).\footnote{A word of caution concerning a notation used 
in \cite{Kutasov:1998zh} is necessary here. In Sec. 3.3. of \cite{Kutasov:1998zh} total 
$\widehat{SL(2)}$ on the worldsheet in the right-moving (supersymmetric) sector is denoted by 
$k$ (which is equal to our $\overline{N}_h$), which is in \cite{Kutasov:1998zh} also used to denote 
the number of NS5-branes (which we denote $Q_5=m$). However, we have shown at the end of 
Sec. \ref{ssec:3chbh} that $\overline{N}_h$ and $m$ are not equal but connected through the 
relation (\ref{Hmc}).}   
\item
In the BPS cases results for entropies are agreeing $\alpha'$-exactly with statistical entropies 
obtained by direct microstate counting \cite{Sen:2007qy,Castro:2008ys}. 
\item
Important consequence of our calculation is that $\alpha'$-corrections to entropies and 
near-horizon solutions is solely coming from Chern-Simons term, for all values of charges. Now,
results from \cite{Kraus:2005vz,David:2007ak,Kaura:2008us} show that this should be expected 
for black holes which are connected to backgrounds which lead to ($\mathcal{N}=2$) 
supersymmetric AdS$_3$ gravities (this AdS$_3$ comes from AdS$_2 \times S^1$). But, we see 
in our examples that it works for all signs of the charges, \emph{even for those which are connected 
to nonsupersymmetric AdS$_3$ gravities}. It would be interesting to see how far one can extend 
the results from \cite{Kraus:2005vz,David:2007ak,Kaura:2008us}.
\item
Near-horizon solutions of 5-dimensional 3-charge black holes are exactly equal to the
solutions of 4-dimensional 4-charge black holes when there is one Kaluza-Klein monopole with
charge $\widehat{N}=\pm1$, if we take $\widehat{W} = m\mp1$. This is expected, 
because, on one hand for $\widehat{N}=\pm1$ $S^2 \times S^1$ part of the near-horizon geometry 
can be identified with $S^3$ manifold (for general $\widehat{N}$ with $S^3/Z_{\widehat{N}}$), and 
on the other hand, as we already noted, Kaluza-Klein monopole generates $(\mp1)$-unit of 
NS5-brane charge (remember that $m=Q_5$ is the number of NS5-branes). It is interesting to
track where this $(-1)$-shift between 4-dimensional and 5-dimensional NS5-brane charge appears
in our near-horizon calculation - it comes from the evaluation of gravitational Chern-Simons term, 
which in $S^2 \times S^1$ (4-dimensional black hole) case produces an additional term compared 
with $S^3$ (5-dimensional black hole) case (to see this just compare (\ref{ef4c12}) with 
(\ref{ef51pp})). 
\item
Finally, we note that results agree with perturbative calculations up to $\alpha'^2$-order obtained 
in \cite{Sahoo:2006pm,Cvitan:2007hu} by using the 4-derivative effective action derived in 
\cite{Metsaev:1987zx}. This is expected, as it was shown in \cite{Chemissany:2007he} that this 
action is equivalent up to $\alpha'^1$ with the action used in calculations above. We note that this 
action by itself is not working beyond $\alpha'^2$ order, which means that in the field redefinition 
scheme used in \cite{Metsaev:1987zx} one has to take into account also terms with more than four derivatives to obtain $\alpha'$-exact entropies \cite{Cvitan:2007hu}. 
\end{itemize}

\subsection{Black holes in type-II string theories}
\label{ssec:typeII}

If we considered type-II string theories, instead of heterotic, compactified in the same way, with
charges restricted to NS-NS sector, there are large black holes corresponding to all cases we
discussed up until now. In fact, it happens that $\alpha'$-exact results can be obtained 
immediately with no effort. This is because structure of NS-NS sector of tree-level low energy 
effective action of type-II theories -  it differs from the heterotic effective action in that there are no 
Chern-Simons terms (present in \ref{hbcs}), and correspondingly, in notation from \ref{10dtl}, one 
has $\Delta\mathcal{L}_{CS}=0$ (from this follows that there are no 4-derivative, together with 
6-derivative, terms in the action at all). As we saw in the heterotic case that Chern-Simons term 
was solely responsible for $\alpha'$-corrections, we can immediately conclude that entropies and 
near-horizon solutions remain $\alpha'$-uncorrected for such black holes.

In particular, for type-II 4-charge extremal black holes in $D=4$ (compactification on $S^1 \times
S^1 \times T^4$, NS-NS charged only) the entropy is
\begin{equation} \label{lbhII4d}
S_{\rm bh}^{\rm (II)} = 2\pi \sqrt{|nw\widehat{N}\widehat{W}|},
\end{equation}
while for corresponding 3-charge extremal black holes in $D=5$ (compactification on 
$S^1 \times T^4$, NS-NS charged only) it is
\begin{equation} \label{lbhII5d}
S_{\rm bh}^{\rm (II)} = 2\pi \sqrt{|nwm|}.
\end{equation}
The meaning of charges is the same as in the heterotic case, except that $\widehat{W}$ is now 
also the number of NS5-branes (KK-monopoles are not "NS5-brane charged").\footnote{This is 
again a consequence of lack of presence of mixed Chern-Simons term in tree-level action of 
type-II theories compactified on torii.}. These results are expected from microscopic point of view.

\section{$R^2$ actions with $\mathcal{N}=2$ off-shell SUGRA}
\label{sec:offshsusy}

\subsection{Calabi-Yau compactifications of M-theory}
\label{ssec:susy5}

Let us start with M-theory description of string theory, whose low energy effective action is
11-dimensional $N=1$ SUGRA which has maximal supersymmetry (32 generators). It can be consistently reduced to $D=5$ dimensions by Kaluza-Klein compactification on 6-dimensional 
Calabi-Yau 3-fold. One obtains $N=2$ 5-dimensional SUGRA with the bosonic part of 
lowest-order effective action given by
\begin{eqnarray} \label{l0susy}
4\pi^2\mathcal{L}_0 &=& 2 \partial^a \mathcal{A}^\alpha_i \partial_a
\mathcal{A}_\alpha^i + \mathcal{A}^2 
\left(\frac{D}{4}-\frac{3}{8}R-\frac{v^2}{2}\right)
+ \mathcal{N} \left(\frac{D}{2}+\frac{R}{4}+3v^2\right)
+ 2 \mathcal{N}_I v^{ab} F_{ab}^I \nonumber \\ 
&& + \mathcal{N}_{IJ} \left( \frac{1}{4} F_{ab}^I F^{Jab} 
 + \frac{1}{2} \partial_a M^I \partial^a M^J \right)
+ \frac{e^{-1}}{24} c_{IJK} A_a^I F_{bc}^J F_{de}^K \epsilon^{abcde}
\end{eqnarray}
$R$ is Ricci scalar, $\mathcal{A}^2 = \mathcal{A}^\alpha_{i}
\mathcal{A}_\alpha^{i}$ and $v^2 = v_{ab}v^{ab}$. $i=1,2$ is $SU(2)$,
and $\alpha=1,2$ is $USp(2)$ index. Also,
\begin{equation}
\mathcal{N} = \frac{1}{6} c_{IJK} M^I M^J M^K , \quad
\mathcal{N}_I = \partial_I \mathcal{N} = \frac{1}{2} c_{IJK} M^J M^K
, \quad 
\mathcal{N}_{IJ} = \partial_I \partial_J \mathcal{N} = c_{IJK} M^K
\end{equation}
$M^I$ are moduli (volumes of $(1,1)$-cycles), and constants $c_{IJK}$ as intersection numbers 
of Calabi-Yau space. Condition $\mathcal{N}=1$ is a condition of real special geometry.

The bosonic field content of the theory is the following. We have Weyl multiplet which contains 
the f\"{u}nfbein $e_\mu^a$, the two-form auxiliary field $v_{ab}$, and the scalar auxiliary field 
$D$. There are $n_V$ vector multiplets enumerated by $I=1,\ldots,n_V$, each containing the 
one-form gauge field $A^I$ (with the two-form field strength $F^I=dA^I$), and the scalar $M^I$. 
Scalar fields $\mathcal{A}_\alpha^i$, which are belonging to the hypermultiplet, can be gauge 
fixed and the convenient choice is given by
$\mathcal{A}^2 = -2$, $\partial_a \mathcal{A}^\alpha_i = 0$.

Action (\ref{l0susy}) is invariant under supersymmetry variations, which when acting on the 
purely bosonic configurations are given by
\begin{eqnarray} \label{svar}
\delta\psi_\mu^i &=& \mathcal{D}_\mu\varepsilon^i + \frac{1}{2}v^{ab}
 \gamma_{\mu ab}\varepsilon^i - \gamma_\mu\eta^i \nonumber \\
\delta\xi^i &=& D\varepsilon^i 
 - 2\gamma^c\gamma^{ab}\varepsilon^i\mathcal{D}_a v_{bc}
 - 2\gamma^a\varepsilon^i\epsilon_{abcde}v^{bc}v^{de} 
 + 4\gamma\cdot v\eta^i \nonumber \\
\delta\Omega^{Ii} &=& - \frac{1}{4}\gamma\cdot F^{I}\varepsilon^i
 - \frac{1}{2}\gamma^a\partial_a M^{I}\varepsilon^i - M^{I}\eta^i
 \nonumber \\
\delta\zeta^{\alpha} &=& 
 \left(3\eta^j-\gamma\cdot v\varepsilon^j\right)\mathcal{A}_j^\alpha
\end{eqnarray}
where $\psi_\mu^i$ is gravitino, $\xi^i$ auxiliary Majorana spinor (Weyl multiplet), 
$\delta\Omega^{Ii}$ gaugino (vector multiplets), and $\zeta^{\alpha}$ is a fermion field from hypermultiplet. 

In \cite{Hanaki:2006pj} a four-derivative part of the action was constructed by supersymmetric 
completion of the mixed gauge-gravitational Chern-Simons term $A \land \textrm{tr} (R \land R)$. 
The bosonic part of the action is
\begin{eqnarray} \label{l1susy}
4\pi^2\mathcal{L}_1 &=& \frac{c_{I}}{24} \left\{ \frac{e^{-1}}{16}
\epsilon_{abcde} A^{Ia} C^{bcfg} C^{de}_{\;\;\;\,fg} 
+ M^I \left[ \frac{1}{8} C^{abcd} C_{abcd} + \frac{1}{12} D^2 
 - \frac{1}{3} C_{abcd} v^{ab} v^{cd} 
\right. \right. \nonumber \\ &&
 + 4 v_{ab}v^{bc} v_{cd} v^{da} - (v_{ab}v^{ab})^2
 + \frac{8}{3} v_{ab} \hat{\mathcal{D}}^b \hat{\mathcal{D}}_c v^{ac}
 + \frac{4}{3} \hat{\mathcal{D}}^a v^{bc} \hat{\mathcal{D}}_a v_{bc}
 + \frac{4}{3} \hat{\mathcal{D}}^a v^{bc} \hat{\mathcal{D}}_b v_{ca}
\nonumber \\ && \left. 
 - \frac{2}{3} e^{-1} \epsilon_{abcde} v^{ab} v^{cd}
   \hat{\mathcal{D}}_f v^{ef} \right] 
+ F^{Iab} \left[ \frac{1}{6} v_{ab} D - \frac{1}{2} C_{abcd} v^{cd}
 + \frac{2}{3} e^{-1} \epsilon_{abcde} v^{cd} 
   \hat{\mathcal{D}}_f v^{ef} 
\right. \nonumber \\ && \left. \left.
 + e^{-1} \epsilon_{abcde} v^{c}_{\;f} \hat{\mathcal{D}}^d v^{ef}
 - \frac{4}{3} v_{ac}v^{cd} v_{db} - \frac{1}{3} v_{ab} v^2 \right]
\right\}
\end{eqnarray}
where $c_I$ are constant coefficients connected to second Chern class of Calabbi-Yau space, 
$C_{abcd}$ is the Weyl tensor. $\hat{\mathcal{D}}_a$ is the conformal covariant derivative, which
when appearing linearly in (\ref{l1susy}) can be substituted with ordinary covariant derivative 
$\mathcal{D}_a$, but when taken twice produces additional curvature contributions
\begin{equation}
v_{ab} \hat{\mathcal{D}}^b \hat{\mathcal{D}}_c v^{ac} = 
 v_{ab} \mathcal{D}^b \mathcal{D}_c v^{ac} 
 + \frac{2}{3} v^{ac} v_{cb} R_a^b + \frac{1}{12} v^2 R \,.
\end{equation}

We are interested in extremal black hole solutions of the action obtained by combining 
(\ref{l0susy}) and (\ref{l1susy}):\footnote{Our conventions in this section are different from the rest 
of the text. We take for Newton constant $G_5=\pi^2/4$ and for the string tension $\alpha'=1$.}
\begin{equation} \label{lsusy}
\mathcal{A} = \int dx^5 \sqrt{-g} \mathcal{L} 
 = \int dx^5 \sqrt{-g} (\mathcal{L}_0 + \mathcal{L}_1)
\end{equation}

The action (\ref{lsusy}) is quartic in derivatives and generally too complicated for finding complete analytical black hole solutions even in the simplest spherically symmetric case. Again, we shall concentrate just on near-horizon behavior and apply Sen's entropy function formalism. For 
spherically symmetric extremal black holes near-horizon geometry is expected to be 
$AdS_2\times S^3$, which has $SO(2,1)\times SO(4)$ symmetry. If the Lagrangian can be written 
in a manifestly diffeomorphism covariant and gauge invariant way, it is expected that near the 
horizon the complete background should respect this symmetry. In our case it means that 
near-horizon geometry should be given by
\begin{eqnarray} \label{efhere}
&& ds^2 = v_1 \left( -x^2 dt^2 + \frac{dx^2}{x^2} \right)
 + v_2\,d\Omega_{3}^2 \nonumber \\
&& F^{I}_{tr}(x) = -e^I \;,\qquad v_{tr}(x) = V \;,\qquad
 M^I(x) = M^I \;, \qquad D(x) = D
\end{eqnarray}
where $v_{1,2}\,$, $e^I$, $M^I$, $V$, and $D$ are constants. All covariant derivatives are 
vanishing. Following standard procedure we define the entropy function
\begin{equation} \label{entfun5}
\mathcal{E} = 2\pi \left( q_I \, e^I - \mathcal{F} \right) \,,
\end{equation}
where $q_I$ are electric charges. $\mathcal{F}$ is given by
\begin{equation}\label{fdef}
\mathcal{F} = \int_{S^3} dy^3 \sqrt{-g} \, \mathcal{L} \;,
\end{equation}
where right hand side is evaluated on the background (\ref{efhere}). Explicitly,
\begin{eqnarray} \label{f0susy}
\mathcal{F}_0 &=& \frac{1}{4} \sqrt{v_2} \left[ \left( \mathcal{N}+3 \right) \left( 3v_1-v_2 \right)-
 4\,V^2 \left( 3\mathcal{N}+1 \right) \frac{v_2}{v_1} \right.
\nonumber \\ && \left. 
 \qquad\quad + 8\,V \mathcal{N}_i\,e^i \,\frac{ v_2}{v_1}
- \mathcal{N}_{ij}\,e^i e^j\,\frac{v_2}{v_1} + D\,(\mathcal{N}-1)\,v_1 v_2 \right] 
\end{eqnarray}
and
\begin{eqnarray} \label{f1susy}
\mathcal{F}_1 &=& v_1  v_2^{3/2} \left\{ \frac{c_I e^I}{48} 
\left[ - \frac{4 V^3}{3 v_1^4} + \frac{D\,V}{3 v_1^2} +
\frac{V}{v_1^2} \left( \frac{1}{v_1} - \frac{1}{v_2} \right) \right]
\phantom{\left(\frac{1}{v_1}\right)^2} \right.
\nonumber \\
&& + \left. 
\frac{c_I M^I}{48} 
\left[\frac{D^2}{12}+\frac{4\,V^4}{v_1^4}+
\frac{1}{4} \left(\frac{1}{v_1}-\frac{1}{v_2}\right)^2-
\frac{2\,V^2}{3\,v_1^2}\left(\frac{5}{v_1}+\frac{3}{v_2}\right)\right]
\right\} \;,
\end{eqnarray}
Complete function $\mathcal{F}$ is a sum
\begin{equation} \label{fsusy}
\mathcal{F} = \mathcal{F}_0 + \mathcal{F}_1 \,.
\end{equation}
Notice that for the background (\ref{efhere}) all terms containing $\varepsilon_{abcde}$ tensor 
vanish, \emph{including} the mixed Chern-Simons term. This means that we do not have to worry
about violation of manifest diffeomorphism invariance, and so we can use straightforwardly Sen's
entropy function formalism. 

Also notice that the entropy function $\mathcal{E}$ is invariant on the transformation defined with
\begin{equation} \label{cpt}
q_I \to - q_I \,, \qquad e^I \to - e^I \,, \qquad V \to - V \,, 
\end{equation}
with other variables remaining the same. This symmetry follows from CPT invariance. We can use 
it to obtain new solutions which have opposite signs of all charges from the given one.

Equations of motion are obtained from extremization of $\mathcal{E}$
\begin{equation} \label{seom}
0 = \frac{\partial \mathcal{E}}{\partial v_1} \;, \quad
0 = \frac{\partial \mathcal{E}}{\partial v_2} \;, \quad
0 = \frac{\partial \mathcal{E}}{\partial M^I} \;, \quad
0 = \frac{\partial \mathcal{E}}{\partial V} \;, \quad
0 = \frac{\partial \mathcal{E}}{\partial D} \;, \quad
0 = \frac{\partial \mathcal{E}}{\partial e^I} \;.
\end{equation}
while the black hole entropy is extremal value of $\mathcal{E}$, i.e., 
\begin{equation} \label{entropy}
S_{\rm bh} = \mathcal{E} \big|_{\rm EOM} \,.
\end{equation}

It is immediately obvious that though the system (\ref{seom}) is algebraic, it is in generic 
case too complicated to be solved in direct manner. One idea is to try to find some 
additional information. Such additional information can be obtained from supersymmetry. It is 
known that there should be 1/2 BPS black hole solutions, for which it was shown in 
\cite{Chamseddine:1996pi} that near the horizon supersymmetry is enhanced fully. This means 
that in this case we can put all variations in (\ref{svar}) to zero, which one can use to 
express all unknowns in terms of one. As we have off-shell supersymmetry, variations 
(\ref{svar}) do not receive $\alpha'$-corrections and so the results obtained in this way are 
the same as in the lowest-order calculation. Vanishing of $\delta\zeta^{\alpha}$ in (\ref{svar})
fixes the spinor parameter $\eta$ to be
\begin{equation} \label{etacond}
\eta^j = \frac{1}{3}(\gamma\cdot v)\varepsilon^j
\end{equation}
Using this, and the condition that $\varepsilon^i$ is (geometrical) Killing spinor, in the remaining equations one gets the following conditions
\begin{equation} \label{bpscond}
v_2 = 4v_1 \;, \qquad M^I = \frac{e^I}{\sqrt{v_1}} \;, \qquad 
D = -\frac{3}{v_1} \;, \qquad V = \frac{3}{4}\sqrt{v_1}
\end{equation}
As moduli $M^I$ are all by definition positive, from the second equation follows that all $e^I$ must
also be positive in this solution. We see that conditions for full supersymmetry are so constraining 
that they fix everything except one unknown, which we took above to be $v_1$. To fix it, we just 
need one equation from (\ref{seom}). In our case the simplest is to take equation for $D$, which gives
\begin{equation} \label{v1bps}
v_1^{3/2} = \frac{1}{6} c_{IJK} e^I e^J e^K - \frac{c_{I} \, e^I}{48}
\end{equation}  

We note that higher derivative corrections violate real special geometry condition, i.e., we have 
now $\mathcal{N}\neq1$.\footnote{We emphasize that one should be cautious in geometric interpretation of this result. Higher order corrections generally change relations between fields in 
the effective action and geometric moduli, and one needs field redefinitions to restore the 
relations. Then correctly defined moduli may still satisfy condition for real special geometry.}

As is typical, we are interested in expressing the results in terms of charges, not field strengths. 
The results can be put in particularly compact form by defining scaled moduli
\begin{equation}
\bar{M}^I \equiv \sqrt{v_1} M^I \;.
\end{equation}
It can be shown \cite{Castro:2007hc} that solution for them is implicitly given by
\begin{equation} \label{mbareq}
8\,c_{IJK} \bar{M}^J \bar{M}^K = q_I + \frac{c_{I}}{8}
\end{equation}
and that the black hole entropy (\ref{entropy}) becomes
\begin{equation} \label{sentm}
S_{\rm bh}^{\rm (BPS)} = \frac{8\pi}{3} c_{IJK}\bar{M}^I \bar{M}^J \bar{M}^K 
\end{equation}
A virtue of this presentation is that if one is interested only in entropies, then it is enough to
consider just equations (\ref{mbareq}) and (\ref{sentm}). Unfortunately, for generic intersection 
numbers $c_{IJK}$ it appears impossible to solve (\ref{mbareq}) explicitly and we do not have
analytic expression for black hole entropy as function of electric charges $q_I$.

As for non-BPS solutions, it is not known how to preceede in generic case. However, as we show
next, even in the two-derivative approximation (lowest order in $\alpha'$) solutions have been constructed only for some special compactifications.

\subsection{Connection to heterotic string theory on $T^4 \times S^1$}
\label{ssec:r2het}

It is known that M-theory compactified on $K3 \times T^2$ manifold is equivalent to heterotic
string theory compactified on $T^5$ manifold. As $K3 \times T^2$ space has $SU(2)$ holonomy 
it breaks less supersymmetry then generic Calabi-Yau (which has $SU(3)$ holonomy) and that is 
why in this case we obtain $N=4$ SUSY in $D=5$ dimensions. In this case the non-vanishing
components of $c_{IJK}$ (up to permutation symmetry of indices) and $c_I$ at tree-level are
\begin{equation} \label{k3t2}
c_{1ij} \equiv c_{ij} \;, \qquad c_1 = 24 \, \qquad i,j=2,\ldots,23 \;,
\end{equation}
where $c_{ij}$ is a regular constant matrix whose inverse we denote as $c^{ij}$. The 
prepotential is now given by
\begin{equation} \label{pk3t2}
\mathcal{N} = \frac{1}{2}M^1 c_{ij} \, M^iM^j \;, \qquad i,j=2,\ldots,23 \,.
\end{equation}
In this case, it is easy to show that the black hole entropy corresponding to the (now 1/4-) BPS solution defined by (\ref{bpscond}) and (\ref{mbareq}) is
\begin{equation} \label{seK3}
S_{\rm bh}^{\rm (BPS)} = 2\pi\sqrt{\frac{1}{2}(q_1+3) \, q_i \, c^{ij} q_j} \;.
\end{equation}

What is new here is that for the simpler form of prepotential (\ref{pk3t2}) we can also treat
(at least some) non-BPS solutions and obtain closed form expression for entropies \cite{Cvitan:2007en}. As we cannot use BPS conditions here, the question is what can we use instead to simplify complicated system of equations. The idea comes from the observation that the BPS conditions implied that relations for $v_2$, $D$ and $V$ in (\ref{bpscond}) are uncorrected by higher-derivative terms in the action. Now, it can be shown that there are lowest order non-BPS 
near-horizon solutions which satisfy
\begin{equation} \label{ans1}
v_2 = 4 v_1 \;, \qquad D = -\frac{1}{v_1} \;, \qquad V = \frac{1}{4}\sqrt{v_1} \,.
\end{equation}
Let us now \emph{assume} that all relations in (\ref{ans1}) are unchanged under higher-derivative 
$\alpha'$-corrections, and take them as an Ansatz.\footnote{Inspection of equations of motion shows that at least some of relations between $M^I$ and $e^I$ receive $\alpha'$-corrections and so we
exclude them from Ansatz.} After using Ansatz (\ref{ans1}), and some additional manipulations, the equations of motion can be reduced to the following system (for general Calabi-Yau 
compactification)
\begin{eqnarray}
&& 0 = c_{IJK}\left(\bar{M}^J-e^J\right)\left(\bar{M}^K-e^K\right)
\label{a1eom1} \\
&& \frac{c_I\bar{M}^I}{12} = c_{IJK} \left(\bar{M}^I+e^I\right)
 \bar{M}^J e^K
\label{a1eom3} \\
&& v_1^{3/2} = \frac{c_Ie^I}{144} - (e)^3
\label{a1eom4} \\
&& q_I - \frac{c_I}{72} = -2\,c_{IJK}e^{J}e^{K} \,.
\label{a1eom5} 
\end{eqnarray}
The above system is apparently overdetermined as there is one equation more than the number of unknowns. More precisely, Eqs.\ (\ref{a1eom1}) and (\ref{a1eom3}) should be compatible, and this 
is not happening for generic choice of parameters. So, we do not expect that our Ansatz will work
for general Calabi-Yau compactifications (i.e., generic $c_{IJK}$ and $c_I$).\footnote{Indeed, 
our efforts to find numerical solutions of the above system for random choices of $c_{IJK}$, $c_I$ and $e^I$ have all failed.}

However, there are cases in which the system is regular and there are physically acceptable solutions. Important examples are prepotentials of the type (\ref{pk3t2}). In this case 
(\ref{a1eom1}) becomes
\begin{equation} \label{k3t2eq1}
0 = \left(\bar{M}^1-e^1\right)\left(\bar{M}^i-e^i\right) \;, \qquad
0 = \left(\bar{M}^i-e^i\right)c_{ij}\left(\bar{M}^j-e^j\right)
\end{equation}
It is obvious that there is now at least one consistent solution obtained by taking 
$\bar{M}^i=e^i$ for all $i$. With this choice all equations in (\ref{k3t2eq1}) are satisfied,
and the remaining unknown scaled modulus $\bar{M}^1$ is fixed by Eq. (\ref{a1eom3}). For the
black hole entropy we obtain
\begin{equation}
S_{\rm bh}^{\rm (non-BPS)} = 2\pi\sqrt{\frac{1}{2}|\hat{q}_1|c^{ij}\hat{q}_i\hat{q}_j} \;,
\qquad \hat{q_I} = q_I - \frac{c_I}{72}
\end{equation}
Again, the influence of higher-order supersymmetric correction is just to shift electric charges 
$q_I\to\hat{q}_I$, but with the different value for the shift constant than in BPS case.

For our exemplary case of 3-charge black holes of heterotic string theory compactified on 
$T^4 \times S^1$ we can obtain analytic expressions for near-horizon solutions and entropies for \emph{all} sets of charges corresponding to large black holes. There is a basis in which 
(tree-level) prepotential (\ref{pk3t2}) has the form
\begin{equation} \label{pk3t2i}
\mathcal{N} = M^1 M^2 M^3 + \frac{1}{2} M^1 c_{ab} \, M^a M^b \;, \qquad a,b=4,\ldots,23 \,.
\end{equation}
We remind the reader that parameters $c_I$ are given in (\ref{k3t2}). To obtain 3-charge
solutions we take $q_a=0$ for $a\ge4$, which by using (\ref{a1eom5}) gives $e^a=0$ for $a\ge4$.
Analysis of systems of equations ((\ref{mbareq}) in BPS case, and (\ref{a1eom1})-(\ref{a1eom3}) in 
non-BPS case) shows that equations for moduli $M^{2,3}$ decouple from those for $M^a$, $a\ge4$  (which make singular system, regularized by quantum corrections of prepotential). It follows that
for above 3-charge configurations we can effectively work with truncated theory where index 
$I=1,2,3$ and prepotential now has simple so called $STU$-form
\begin{equation} \label{pstu}
\mathcal{N} = M^1 M^2 M^3 \,.
\end{equation}
For this prepotential in \cite{Cvitan:2007en} we constructed full set of near-horizon solutions of 
$N=2$ supersymmetric $R^2$ action. We now present the results. As the theory is symmetric
under exchange $I=2$ and $I=3$ indices, 

BPS near-horizon solutions, with $q_1\ge0$, $q_{2,3}>0$ (which satisfy \ref{bpscond})), are given 
by
\begin{eqnarray}
&& v_1 = \frac{1}{4}
 \left|\frac{q_2q_3(q_1+\zeta)^2}{q_1+3\zeta}\right|^{1/3}
\label{hetbpsv} \\
&& \frac{e^1}{\sqrt{v_1^3}}\left(q_1+3\zeta\right)
 = \frac{e^2q_2}{\sqrt{v_1^3}} = \frac{e^3q_3}{\sqrt{v_1^3}} = 4\frac{q_1+3\zeta}{q_1+\zeta}
\label{hetbpse} \\
&& \frac{M^1\sqrt{v_1}}{e^1} = \frac{M^2\sqrt{v_1}}{e^2} = \frac{M^3\sqrt{v_1}}{e^3} = 1 \,,
\label{hetbpsm}
\end{eqnarray}
together with (\ref{bpscond}). Solutions with $q_1\le0$, $q_{2,3}<0$, which are also BPS, are 
easily obtained by applying transformation (\ref{cpt}). The entropy formula, valid for both
type of BPS solutions, is given by
\begin{equation} \label{sent34}
S_{\rm bh}^{\rm (BPS)} = 2\pi \sqrt{|q_2 q_3| (|q_1|+3)} \,.
\end{equation}

For non-BPS solutions we have 6 possible combinations for picking signs of charges. By using
transformation (\ref{cpt}), and the symmetry of theory under exchange of indices $I=2$ and $I=3$,
we are left with only two independent choices. We choose those which satisfy (\ref{ans1}).

Non-BPS solutions with $q_{1,2}>0$, $q_3<0$ are given by
\begin{eqnarray}
&& v_1 = \frac{1}{4} \left|\frac{q_2q_3(q_1+\zeta/3)^2}{q_1-\zeta/3}\right|^{1/3} 
\\
&& \frac{e^1}{\sqrt{v_1^3}}\left(q_1-\frac{\zeta}{3}\right)
 = \frac{e^2q_2}{\sqrt{v_1^3}} = \frac{e^3q_3}{\sqrt{v_1^3}}
 = 4\frac{q_1-\zeta/3}{q_1+\zeta/3}
\\
&& \frac{M^3\sqrt{v_1}}{e^3} = - \frac{q_1+\zeta}{q_1-\zeta/3}
 \;, \qquad \frac{M^1\sqrt{v_1}}{e^1} = \frac{M^2\sqrt{v_1}}{e^2} = 1
\end{eqnarray}

In the non-BPS case $q_{2,3}>0$, $q_1<-1$ the only difference from solution above is
\begin{equation}
\frac{M^1\sqrt{v_1}}{e^1} = - \frac{q_1-\zeta/3}{q_1+\zeta}
 \;, \qquad \frac{M^2\sqrt{v_1}}{e^2}
 = \frac{M^3\sqrt{v_1}}{e^3} = 1
\end{equation}

For both cases of above non-BPS solutions the black hole entropy is given by
\begin{equation} \label{sent12}
S_{\rm bh}^{\rm (non-BPS)} = 2\pi \sqrt{|q_2 q_3 (q_1-1/3)|} \,.
\end{equation}

Before commenting our solutions, it is necessary to make connection to notation used in
previous sections. It is known (see, e.g., section 5.1 of \cite{Cvitan:2007en}) at lowest order
(two-derivative) supersymmetric action (\ref{l0susy}) with $STU$ prepotential (\ref{pstu})
can be put (by making Poincare duality transformation on $A^1$ gauge field, and then going 
from Einstein- to string-frame metric) in the form of the heterotic effective action (\ref{ea5d}),
where\footnote{Note that relations in (\ref{hetmod}) satisfy real geometry condition 
$\mathcal{N}=M^1M^2M^3=1$, which we showed to be violated by higher-derivative terms.
This means that interpretations of fields in effective action as geometric moduli of 
compactification manifold are receiving corrections anf one has to be careful in such 
identifications.}
\begin{equation} \label{hetmod}
M^1 = S^{2/3} \;,\qquad M^2 = S^{-1/3}T^{-1} \;,\qquad M^3 = S^{-1/3}T
\end{equation}
In addition, connection between gauge fields in two formulations is such that
\begin{equation} \label{qinwm}
q_1 = m \,,\qquad q_2 = n \,,\qquad q_3 = w \,,
\end{equation}
Using (\ref{qinwm}) in the expressions for black hole entropies (\ref{sent34}) and (\ref{sent12}),
we obtain in the BPS cases
\begin{equation} \label{sent34i}
S_{\rm bh}^{\rm (BPS)} = 2\pi \sqrt{|nw| (|m|+3)} \,,
\end{equation}
while in the non-BPS cases
\begin{equation} \label{sent12i}
S_{\rm bh}^{\rm (non-BPS)} = 2\pi \sqrt{|nw (m-1/3)|} \,.
\end{equation}

For BPS black holes entropy (\ref{sent34i}) \emph{exactly} matches our previous result (\ref{3entb}) 
obtained from full heterotic effective action (which matches microscopic result \cite{Castro:2008ys}). 
Not only that, but, if we use (\ref{hetmod}) to identify\footnote{It is interesting that choice 
$T = (M^3/M^2)^{1/2}$ (which gives different result when higher-derivative corrections are 
included) leads to $u_T = \sqrt{n/w}$, which means that this could be a correct identification with 
heterotic string compactification modulus (radius of $S^1$) \cite{Cvitan:2007pk}.} 
$T = (M^1)^{-1/2} (M^2)^{-1}$ and $S = (M^1)^{3/2}$, after taking care of differences in conventions 
and normalizations we obtain that complete near-horizon solution is in fact \emph{equal} to the 
solution given in (\ref{3solb}).

However, in non-BPS sector things are completely different. Comparison of (\ref{sent12i}) with
the corresponding entropy obtained from full heterotic effective action given in (\ref{3entn}),
shows complete disagreement (starting already at first order in $\alpha'$).

\subsection{$R^2$ supersymmetric actions in $D=4$}
\label{ssec:r2d4}

In $D=4$ dimensions it is also known how to construct off-shell $N=2$ supersymmetric action
with $R^2$ terms. This action can be used to find analytic near-horizon solutions for BPS 
spherically symmetric extremal black holes. Instead of going through the details of calculation, 
this time we shall simply present results for entropies for object of our interest, i.e., 4-charge black 
holes in theory with prepotential corresponding to tree-level heterotic string theory compactified on 
$T^4 \times \widehat{S}^1 \times S^1$ (whose lowest order solution is discussed in Sec. 
\ref{sssec:4cbh}. The details can be found in review \cite{Sen:2007qy}. Again, it should be 
emphasized that the heterotic theory has larger $N=4$ supersymmetry.

For BPS black holes, whose charges satisfy $nw>0$, $\widehat{N}\widehat{W}\ge0$, the black 
hole entropy is
\begin{equation} \label{er2susyb}
S_{\rm bh}^{\rm (BPS)} = 2\pi \sqrt{nw(\widehat{N}\widehat{W}+4)} \,,
\end{equation}
which agrees with result obtained from full heterotic action and with microscopic entropy.

In the case of non-BPS black holes, no analytic results are known. For the case $n<0$, 
$w,\widehat{N},\widehat{W}>0$, perturbative calculation was performed with the result 
\cite{Sahoo:2006rp}
\begin{equation} \label{er2susyn}
S_{\rm bh}^{\rm (non-BPS)} = 2\pi \sqrt{|n|w\widehat{N}\widehat{W}} \left(
 1 +  \frac{5}{8}\frac{1}{\widehat{N}\widehat{W}} - 29 \frac{1}{(\widehat{N}\widehat{W})^2}
 - 1904 \frac{1}{(\widehat{N}\widehat{W})^3} + \ldots \right) \,,
\end{equation}
Comparison with result obtained from full heterotic action (which agrees with microscopic result 
(\ref{cft4n})) shows disagreement already at lowest $\alpha'^{1}$-correction (instead of $5/8$ it
should be 1).

\subsection{Comments on $R^2$ supersymmetric actions}

\begin{enumerate}
\item
In the case of $R^2$ supersymmetric actions in $D=4$ and $D=5$ full solutions (in the whole space, 
not just near-horizon) for BPS black holes where constructed explicitly up to one function (which is satisfying ordinary differential equation). However, the method of construction is not working in
non-BPS cases \cite{Shah:2009xc}.
\item
One can obtain closed form results also for generic $c_I$'s.
\item
Why such $R^2$ actions are working $\alpha'$-exactly for BPS black holes, when it is evident from perturbative results that these actions are incomplete already at 4-derivative ($\alpha'^1$) order?
It has been frequently claimed that this is a consequence of AdS$_3$/CFT$_2$ duality, and a 
property that in three-dimensional $N=2$ supersymmetric gravities in AdS$_3$ the only non-trivial
higher-derivative corrections are Chern-Simons terms. However, this is insufficient to prove the
statement.
\end{enumerate}

\section{Lovelock-type actions}
\label{sec:lovelock}

\subsection{Pure Gauss-Bonnet correction}
\label{ssec:GBact}

Stimulated by successes of supersymmetric $R^2$-truncated actions in describing BPS black
holes in $D=4$ and 5 in $\alpha'$-exact manner, the natural question is can the same be obtained
with some even more simple actions. In $D=4$ the simplest choice is to take for higher-derivative
correction just the pure Gauss-Bonnet density. This means that we start with "toy" effective action
in D-dimensions given by
\begin{equation} \label{aGB}
\mathcal{A} = \mathcal{A}_0 + \mathcal{A}_{\rm GB} \,,
\end{equation}
where $\mathcal{A}_0$ is a corresponding lowest-order (2-derivative) action, and 
$\mathcal{A}_{\rm GB}$ is a 4-derivative $\alpha'^1$-correction given by\footnote{The conventions
here are (again) $G_N=2$, $\alpha'=16$.}
\begin{equation} \label{l1gb}
\mathcal{A}_{GB} = \frac{1}{32\pi}\frac{1}{8} \int d^Dx \, \sqrt{-G} \, S
 \left( R_{\mu\nu\rho\sigma} R^{\mu\nu\rho\sigma}
  - 4 \, R_{\mu\nu} R^{\mu\nu} + R^2 \right)
\end{equation}
This choice has some notable properties, of which we mention:
\begin{enumerate}
\item
Such term appears in heterotic effective action on 4-derivative ($\alpha'1$) level (but note there are 
also other 4-derivative terms).
\item
By itself Gauss-Bonnet density is topological in $D=4$, which means that it is giving contribution to
equations of motion just because it is multiplied by the dilaton field $S$ in the Lagrangian. Because 
of this it gives the simplest contribution to entropy function compared with other 4-derivative 
possibilities.
\item
It produces normal second-order field equations in all dimensions.
\end{enumerate}

Property 1 suggests that the 4-dimensional case is probably simplest to treat, so we start with
our 4-charge near-horizon geometries first. In this case we already know that $\mathcal{A}_0$ is
given in (\ref{ea4d}).  We are interested in near-horizon geometry\footnote{The extension of small 
black hole solutions to the whole space-time in the case of Gauss-Bonnet-type action was 
analyzed in \cite{Chen:2009rv}.} for which already learned it has to have AdS$_2 \times S^2$ 
form, i.e.,
\begin{eqnarray} \label{eb1}
&& ds^2  = v_1 \left( -r^2 dt^2 + \frac{dr^2}{r^2} \right)  + v_2 (d\theta^2 + \sin^2\theta d\phi^2)\, ,
\nonumber \\ 
&& S =u_S \,, \qquad T = u_1 \,, \qquad  \widehat{T} = u_2 \,, 
\nonumber \\
&& F^{(1)}_{rt} = e_1 \,, \quad F^{(3)}_{rt} = e_3 \,, \quad
 F^{(2)}_{\theta\phi} = \frac{p_2}{4\pi} \,, \quad F^{(4)}_{\theta\phi} = \frac{p_4}{4\pi} \,.    
\end{eqnarray}
The function $\mathcal{F}$ has two contributions
\begin{eqnarray} \label{ey0}
\mathcal{F}_0(\vec{u}, \vec{v}, \vec{e}, \vec{p})
 &\equiv& \int d\theta d\phi \, \sqrt{-G} \, \mathcal{L}_0
\nonumber \\ 
&=& \frac{1}{8} \, v_1 \, v_2 \, u_S \left[ - \frac{2}{v_1} + \frac{2}{v_2} + 
 \frac{2 \, u_1^2 \, e_1^2}{v_1^2} + \frac{2 \, e_3^2}{u_1^2 \, v_1^2}
 - \frac{u_2^2 \, p_2^2}{8\pi^2 v_2^2} - \frac{u_2^{-2} \, p_4^2}{8\pi^2 v_2^2} \right] \,,
\end{eqnarray}
and
\begin{equation} \label{ey1}
\mathcal{F}_{\rm GB}(\vec{u}, \vec{v}, \vec{e}, \vec{p}) \equiv
 \int d\theta d\phi \, \sqrt{-G} \, \mathcal{L}_0
 = - 2 \, u_S \,.
\end{equation}
By using formalism developed in section \ref{ssec:senef} one easily obtains near-horizon solutions 
\cite{Sen:2005iz}
\begin{eqnarray} \label{eb4}
&& v_1 = v_2 = 4 |\widehat{N}\, \widehat{W}| + 8 \,, \qquad 
 u_S = \sqrt{\frac{|nw|}{|\widehat{N} \widehat{W}| + 4}}
\nonumber \\ 
&& u_1 = \sqrt{\left| \frac{n}{w} \right|} \,, \qquad 
 u_2 = \sqrt{\bigg| \frac{\widehat{W}}{\widehat{N}} \bigg|}
\nonumber \\
&& e_1 = \frac{1}{n} \sqrt{|nw|(|\widehat{N}\widehat{W}| + 4)} \,, \qquad
 e_3 = \frac{1}{w} \sqrt{|nw|(|\widehat{N}\widehat{W}| + 4)} \,.
\end{eqnarray}
Results are expressed in terms of charges
\begin{equation} \label{erel}
n = 2\, q_1 \,, \quad w = 2\, q_3 \,, \quad \widehat{N} = 4\pi\, p_2 \,, \quad
\widehat{W} = 4\pi\, p_4 \, .
\end{equation}
where (see appendix \cite{Sen:2005iz}), $w$ and $n$ are winding and momentum numbers along
$S^1$, and $\widehat{N}$ and $\widehat{W}$ are charges corresponding to Kaluza-Klein 
monopoles and NS5-branes wrapped around $T^4 \times S^1$, respectively.

For the black hole entropy one obtains
\begin{equation} \label{entGB4}
S_{\rm bh} = 2\pi \sqrt{|nw|(|\widehat{N}\widehat{W}| + 4)}
\end{equation}
We see that for BPS black holes satisfying $nw>0$, $\widehat{N}\widehat{W}\ge0$ Wald entropy
(\ref{entGB4}) $\alpha'$-exactly matches the microscopic statistical entropy (\ref{micro4}). That this 
is not coincidental shows comparison of the near-horizon solution (\ref{eb4}) with the corresponding 
solution obtained from $\alpha'$-complete heterotic effective action presented in (\ref{4solb}) - the 
only differences are the expressions for two radii $u_{1,2}$, which in (\ref{eb4}) manifestly satisfy 
known T-dualities. Thus we expect that the solutions are equivalent, and the difference can be 
attributed to field redefinitions.

For non-BPS black holes entropy (\ref{entGB4}) obviously differs from microscopic statistical
results.

What about 3-charge heterotic black holes in $D=5$? As shown in section 7 of Ref. 
\cite{Cvitan:2007en}, for large black holes (all three charges $n,w,N$ nonvanishing) one gets the 
entropy which differs from statistical results, though in BPS case ($n,w,m>0$) the result gives
correct first $\alpha'$-correction. For small BPS black hole, given by $m=0$ and $nw>0$, we again 
obtain Wald entropy agreeing with statistical result 
\begin{equation} \label{smallbps}
S_{\rm stat}^{\rm (BPS)} = 4\pi\sqrt{nw} \,.
\end{equation}

Now, the real question here is \emph{why the sole Gauss-Bonnet correction, which is not complete 
correction to effective actions even at $\alpha'^1$-order, in some cases gives the results which are 
$\alpha'$-exact?} This question still begs for an answer.

\subsection{Small black holes in general dimensions}
\label{ssec:sbhlove}

We have observe in previous subsections that adding just the Gauss-Bonnet term as 
higher-derivative correction to lowest order heterotic effective actions is producing correct results
for the entropy of small BPS black holes in $D=4$ and 5. What about $D>5$? We start from
heterotic string compactified on $S^1 \times T^{9-D}$ and wounded on $S^1$. Taking as 
nonvanishing charges only winding $w$ and momentum $n$ (both on $S^1$) we obtain, in analogy 
to (\ref{ea9d}), the truncated $D$-dimensional effective action which is at lowest order given by
\begin{equation} \label{eaDs}
\mathcal{A}_0 = {1\over 32} \int d^D x \, \sqrt{-G} \, S \, 
\left[ R + S^{-2}\, (\partial_\mu S)^2  -  T^{-2} \, (\partial_\mu T)^2
 - T^2 \, (F^{(1)}_{\mu\nu})^2 - T^{-2} \,  (F^{(2)}_{\mu\nu})^2 \right] \, ,
\end{equation}
where $T$ is the modulus of $S^1$. 

For the higher-derivative terms we could try again with pure Gauss-Bonnet correction, i.e., with 
action of the form (\ref{aGB})-(\ref{l1gb}). However, it was shown in \cite{Sen:2005kj} that this works 
just for four- and five-dimensional small BPS black holes, while for $D>5$ one gets Wald entropy 
different from (\ref{smallbps}). But, in $D=6$ we note that another Euler density appears, 
which is of 6-derivative type. More general, in $D$ dimensions there are $[D/2]$ different generalized
Euler densities\footnote{Sometimes called extended Gauss-Bonnet densities.} $E_k$
\begin{equation}\label{lgbm}
E_k = \frac{1}{2^k}
\, \delta_{\mu_1\nu_1\ldots\mu_k\nu_k}^{\rho_1\sigma_1\ldots
\rho_k\sigma_k} \, {R^{\mu_1\nu_1}}_{\rho_1\sigma_1}\cdots
{R^{\mu_k\nu_k}}_{\rho_k\sigma_k}\;, \qquad k=1,\ldots,[D/2]
\end{equation}
where $\delta_{\alpha_1\ldots\alpha_j}^{\beta_1\ldots\beta_j}$ is totally antisymmetric product of 
$j$ Kronecker deltas, normalized to take values 0 and $\pm 1$, and $[x]$ denote integer part of 
$x$. Normalization in (\ref{lgbm}) is such that $E_1=R$ and $E_2 = (R_{\mu\nu\rho\sigma})^2 -
4(R_{\mu\nu})^2 + R^2$. Euler densities $E_k$ are in many respects generalization of the Einstein 
term. Especially, $E_k$ is a topological density in $D=2k$ dimensions. Also note that $E_k$
vanish identically for $k>[D/2]$.

We now see that in $D>5$ instead of pure Gauss-Bonnet type action it is more "natural" to consider 
more general Lovelock type action\footnote{Originally Lovelock \cite{Lovelock:1971yv} considered 
pure gravity actions without dilaton field. It is easy to check that multiplication by scalar field does 
not change any of important properties of Lovelock actions, except that term containing $E_{D/2}$ 
(for $D$ even) becomes non-topological.}
\begin{equation} \label{treeea}
\mathcal{A} = \mathcal{A}_0 + \sum_{k=2}^{[D/2]} \lambda_k \int d^Dx \sqrt{-g} \,S \, E_k \,.
\end{equation}
where $\lambda_k$ are some so far undetermined coefficients, and $\mathcal{A}_0$ is given in 
(\ref{eaDs}). It is obvious that $k$-th term consists of $2k$-derivative terms, i.e., it describes  
$\alpha'^{k-1}$-correction. In $D=4$ and 5 (\ref{treeea}) reduces to the form (\ref{aGB}).  Of course, 
the action (\ref{treeea}) is not (truncated) effective action of heterotic string theory. We shell return to 
this point later. Actions (\ref{treeea}) have many attractive and notable properties, e.g., they lead to 
normal second order equations of motion.

In \cite{Prester:2005qs} it was shown that there is \emph{unique fixed} choice of coefficients 
$\lambda_k$ which does the job we are seeking to -- if we take
\begin{equation}\label{coup}
\lambda_k = \frac{1}{4^{k-1} k!} \,, 
\end{equation}
in action (\ref{treeea}) then small extremal 2-charge black hole solutions have the Wald entropy 
given by \cite{Prester:2005qs}
\begin{equation} \label{smalllove}
S_{\rm bh} = 4\pi\sqrt{|nw|} \,,
\end{equation}
\emph{for all number of dimensions} $D$. Evidently, (\ref{smalllove}) agrees with the microscopic
statistical result of string theory in BPS case $nw>0$ (\ref{smallbps}) (and again differs in non-BPS 
cases), but now for all $D$.

Let us also present near-horizon solutions. In $D$ dimensions we expect geometry to be of the 
AdS$_2 \times S^{D-2}$ form, so we make the following Ansatz 
\begin{eqnarray} \label{nhsmall}
&& ds^2  = v_1 \left( -r^2 dt^2 + \frac{dr^2}{r^2} \right)  + v_2\, d\Omega_{D-2} \, ,
\nonumber \\ 
&& S =u_S \,, \quad T = u_T \,, \quad F^{(1)}_{rt} = e_1 \,, \quad F^{(2)}_{rt} = e_2  \,.    
\end{eqnarray}
where $d\Omega_k$ denotes standard metric on the unit $k$-dimensional sphere, and $v_{1,2}$,
$u_{S,T}$, and $e_{1,2}$ are constants to be determined from equations of motion. Using as
before entropy function formalism we obtain the following near-horizon solution
\begin{eqnarray} \label{sollove}
&& v_1 = 8 \,, \qquad v_2 = f(D) \,, \qquad 
 u_S = \frac{16\pi}{\Omega_{D-2} f(D)}\sqrt{|nw|}
\nonumber \\ 
&& u_T = \sqrt{\left| \frac{n}{w} \right|} \,, \qquad 
 e_1 = \frac{2}{n} \sqrt{|nw|} \,, \qquad
 e_2 = \frac{2}{w} \sqrt{|nw|} \,,
\end{eqnarray}
where $f(D)$ is the real positive root\footnote{We numerically checked up to $D=9$ that there is
unique real positive root.} of a particular $[D/2]$-th order polynomial. We note that the AdS$_2$
radius is $\ell_A \equiv \sqrt{v_1} = \sqrt{\alpha'/2}$ for all $D$.

\subsection{Comments on Lovelock-type action}
\label{ssec:comlove}

Some questions and comments on Lovelock-type action defined by (\ref{treeea})-(\ref{coup}):
\begin{enumerate}   
\item 
Why this action works just for small BPS black holes in $D>4$ (and large and small BPS black 
holes in $D=4$ where it reduces to the pure Gauss-Bonnet type)? Again, the real question here
is why should it work for any type at all. Still unknown.
\item
What is the connection between this action and the low energy effective action of the 
heterotic string theory (HLEEA)? It is obviously much simpler then HLEEA, as, e.g., it has finite 
number of higher-derivative corrections which are purely gravitational. However, there are some 
notable similarities. HLEEA also contains Gauss-Bonnet term (i.e., second Euler density $E_2$) 
with the same coefficient $\lambda_2=1/8$. The same is also true for the 8-derivative term 
proportional to $E_4$, where coefficient is $\lambda_4=1/1526$. But, something odd is happening
with the $E_3$ term. It is completely absent in HLEEA, while in our action it appears with 
$\lambda_3=1/96$. It is amusing that this terms, with the same coefficient, appears in bosonic string theory.
\emph{It is important to keep in mind that for small black holes, which are intrinsically stringy, one
does not expect for low energy/curvature effective action to be usable.}
\item
Is this action just a trival construct to obtain (\ref{smalllove}), void of any other meaning? We believe 
not, for the following reasons. First, note that this action is unique, with the same form for all number 
of dimensions $D$, and with coefficients in front of higher-derivative terms looking stringy (as 
discussed above). Second, note that fixing of coefficients $\lambda_k$ is not "one for one",
but "one for two". Fixing of $\lambda_2$ has to work \emph{simultaneously} in $D=4$ and $D=5$,
then fixing of $\lambda_2$ has to do the job both in $D=6$ and $D=7$, etc. Also, as discussed in
section \ref{ssec:GBact}, it works also for 8-charge \emph{large} black holes in $D=4$ (where it boils
down to Gauss-Bonnet type action).
\item
What then could be the meaning of this action? One of the most attractive possibilities is that it 
describes some new type of effectiveness in string theory, present in black hole near-horizon 
analyses.
\item
What about corresponding 2-charge small black holes in type-IIA theory compactified on 
$S^1\times T^{9-D}$, which are 1/4-BPS states? In this case it is known that string microstate 
counting gives for statistical entropy
\begin{equation} \label{typeIIsmall}
S_{\rm small} = 2\sqrt{2}\pi\sqrt{|nw|} \,.
\end{equation}
Formally, one can obtain this result in all $D$ by taking for coefficients $\lambda_k$ in 
(\ref{treeea})
\begin{equation}\label{coupII}
\lambda_k^{\rm (II)} = \frac{\lambda_k^{\rm (het)}}{2} = \frac{2}{4^k k!} \,. 
\end{equation}
From this follows that AdS$_2$ radius is now $\ell_A = \sqrt{\alpha'}/2$.
Note that LEEA of type-II string theory does not contain Gauss-Bonnet, nor any other 
4-derivative term, so the meaning of this result is unknown.
\end{enumerate}

\end{document}